\documentclass[12pt]{article}
\usepackage{amsmath,amssymb,amsfonts,amsthm,bbm}
\usepackage{epic,eepic,epsfig,longtable}
\usepackage{multirow,verbatim}
\usepackage{array}
\usepackage{graphicx}
\usepackage{floatrow}
\usepackage{subcaption}
\usepackage{paralist}
\usepackage{latexsym}
\usepackage{comment}
\usepackage{epsfig}
\usepackage{setspace}
\usepackage{CJK}
\usepackage{color}
\usepackage{bbm}
\usepackage{tikz}
\usetikzlibrary{arrows,positioning}
\usepackage{rotating}
\usepackage[authoryear, round]{natbib}
\usepackage{booktabs}

\newcommand{\bftab}{\fontseries{b}\selectfont}

\newcommand{\abs}[1]{\left\lvert#1\right\rvert}
\newcommand{\norm}[1]{\left\lVert#1\right\rVert}

\makeatletter
\def\singlespace{\def\baselinestretch{1}\@normalsize}

\makeatletter
\def\singlespace{\def\baselinestretch{1}\@normalsize}

\numberwithin{equation}{section}

\renewcommand{\hat}{\widehat}

\renewcommand{\hat}{\widehat}

\newcommand{\bfsym}[1]{\ensuremath{\boldsymbol{#1}}}

\def\1{\bfsym{1}}	

 \def\htheta{\hat {\theta}}

\newcommand{\argmax}{\mathop{\mathrm{argmax}}\limits}

\DeclareMathOperator{\sgn}{sgn}

\def\newpage{\vfill\eject}

\def\today{\ifcase\month\or
  January\or February\or March\or April\or May\or June\or
  July\or August\or September\or October\or November\or December\fi
  \space\number\day, \number\year}

\newdimen\biblioindent    \biblioindent=30pt

 at 8truept

\def\sgn{\mbox{sgn}}

\newcommand{\beq}{\begin{equation}}
  \newcommand{\eeq}{\end{equation}}
\newcommand{\beqn}{\begin{eqnarray}}
  \newcommand{\eeqn}{\end{eqnarray}}
\newcommand{\beqnn}{\begin{eqnarray*}}
  \newcommand{\eeqnn}{\end{eqnarray*}}

\allowdisplaybreaks
\usepackage{verbatim}
\def\[{\left [}  \def\]{\right ]} \def\({\left (}  \def\){\right )}
 
\def\hat{\widehat}

\newtheorem{assumption}{Assumption}
\newtheorem{theorem}{Theorem}
\newtheorem{lemma}{Lemma}

\newtheorem{proposition}{Proposition}
\theoremstyle{definition}
\newtheorem{definition}{Definition}

\newtheorem{remark}{Remark}

\graphicspath{{figures/}{./images}}

\title{ Effect of the U.S.--China Trade War on Stock Markets:
A Financial Contagion Perspective}
\author{ Minseog Oh and Donggyu Kim\\
College of Business, \\
Korea Advanced Institute of Science and Technology (KAIST)\\
}

\begin{document}
\maketitle

\begin{spacing}{1.45}

\begin{abstract}
In this paper, we investigate the effect of the U.S.--China trade war on stock markets from a financial contagion perspective, based on high-frequency financial data.
Specifically, to account for risk contagion between the U.S. and China stock markets, we develop a novel jump-diffusion process.
For example, we consider three channels for volatility contagion--such as integrated volatility, positive jump variation, and negative jump variation--and each stock market is able to affect the other stock market as an overnight risk factor.  
We develop a quasi-maximum likelihood estimator for model parameters and  establish its asymptotic properties.
Furthermore, to identify contagion channels and test the existence of a structural break, we propose hypothesis test procedures.
From the empirical study, we find evidence of financial contagion from the U.S. to China  and evidence that the risk contagion channel has changed from integrated volatility to negative jump variation.
\end{abstract}


\section{Introduction} \label{SEC-1}

The U.S.--China trade war began in early 2018 and affected the global economy.
\citet{amiti2020effect} show that the tariff actions through the years 2018 and 2019 lowered the investment growth rate of U.S. companies.
In addition, \citet{fajgelbaum2020return} and \citet{amiti2019impact} find evidence of reduction in aggregate U.S. real income and the number of imported varieties.
Thus, we can conjecture that there is a structural break before and after the U.S.--China trade war. 
Our empirical study indicates that volatilities and jump sizes significantly increased after the U.S.--China trade war (see Table \ref{Table-1} in Section \ref{SEC-5}).
In other words, the risk structure of the stock markets has changed in terms of the magnitude of volatility.
Furthermore, since  the U.S. and China were directly subjected to economic sanctions from each other, we can guess that there may be a financial contagion between the U.S. and China. 
To answer these questions rigorously,  modeling the U.S. and China stock markets from a financial contagion perspective  is required.

For volatility modeling, generalized autoregressive conditional heteroskedastic (GARCH) models \citep{bollerslev1986generalized,engle1982autoregressive} are widely employed with low-frequency data, such as daily, weekly, and monthly log-returns.
GARCH models utilize squared log-returns as innovation and are able to catch low-frequency market dynamics, such as volatility clustering.
On the other hand, since high-frequency data become available,  high-frequency volatility information, such as realized volatility estimators, is employed to model volatilities.
For example, there are several well-performing realized volatility estimators, such as two-time scale realized volatility (TSRV) \citep{zhang2005tale}, multi-scale realized volatility (MSRV) \citep{zhang2006efficient}, kernel realized volatility (KRV) \citep{barndorff2008designing}, quasi-maximum likelihood estimator (QMLE) \citep{ait2010high,xiu2010quasi}, pre-averaging realized volatility (PRV) \citep{jacod2009microstructure}, and robust pre-averaging realized volatility \citep{fan2018robust, shin2021adaptive}.  
In finance practice, we often observe price jumps, and empirical studies reveal that  the decomposition of daily variation into its continuous and jump components helps explain volatility dynamics \citep{andersen2007roughing,ait2012testing,barndorff2006econometrics,corsi2010threshold}.
To decompose daily variations, \citet{mancini2004estimation} suggested the threshold method for jump detection, and \citet{fan2007multi} utilized the wavelet method to identify jumps in high-frequency data with microstructure noise.
With these realized volatility estimators, several parametric models have been developed to explain volatility dynamics.   
Examples include the realized volatility-based modeling approaches 
\citep{andersen1997heterogeneous, andersen1997intra-day, andersen1998skeptics, andersen1998deutsche,  andersen2003modeling}, 
 the realized GARCH models \citep{hansen2012realized},  the high-frequency-based volatility (HEAVY) models \citep{shephard2010realising}, the heterogeneous auto-regressive (HAR) models \citep{corsi2009simple},  and the unified GARCH-It\^o models \citep{kim2016unified, song2021volatility}.
The empirical studies on these modeling approaches showed that incorporating the realized volatility helps capture volatility dynamics.
These models usually focus on modeling the stock market during the open-to-close period.
However, when considering the U.S. and China stock markets simultaneously, the trading hours of the China stock market are during the overnight period of the U.S. market and vice versa. 
Thus, to model the financial contagion between the U.S. and China stock markets, we need to model the whole-day period and their interaction structure.

This paper proposes a novel jump-diffusion process to account for the financial contagion between the U.S. and China by embedding the volatility spillover with three channels--realized volatility, positive jump variation, and negative jump variation--into an overnight volatility process.
Specifically, China and U.S. stock markets have disjoint trading hours--from 1:30 UTC to 7:00 UTC and from 14:30 UTC to 21:00 UTC, respectively.
On the other hand, the asset prices are only observable during the open-to-close period.
These facts imply that volatility transmission occurs through overnight volatility processes.
To reflect this,  we assume that the close-to-open instantaneous volatility process in the U.S. is affected by China's open-to-close risk factors during China's open-to-close period, and vice versa. 
As risk factors, we consider integrated volatility and squared close-to-open log-return as well as jump variations. 
 \citet{patton2015good} showed that the impact of a price jump on volatility depends on the sign of the jump.
Specifically,  negative (positive) jumps lead to higher (lower) future volatility.
Furthermore, the sign of the jumps may be related with the good and bad news released during trading hours. 
To account for the different effect of good and bad news, we decompose jump variations into positive and negative jump variations and use them as risk factors. 
Consequently, there are four types of risk factors: integrated volatility, positive and negative jump varation, and squared close-to-open log-return.
We call the proposed model the contagion GARCH-It\^{o} model.
The key feature of the proposed model is that the open-to-close conditional expected volatility of the U.S. stock market is a function of the country's past risk factors as well as the China's past risk factors, and vice versa, which helps account for the financial contagion between the U.S. and China stock markets. 
Based on the structure of the conditional expected volatility, we propose a quasi-likelihood function to estimate model parameters and establish its asymptotic properties.
Further, we propose hypothesis test procedures to identify the risk contagion channels and to test the existence of a structural break.
The empirical study shows that there is no risk contagion from China to U.S. for both pre- and post-U.S.--China trade war period, and the structural break of the U.S. market mainly originates from its own structural change.
In contrast, there is a significant risk contagion channel from the U.S. to China for both periods, and the risk contagion channel changed from integrated volatility to negative jump variation.
Details regarding this are presented in Section \ref{SEC-5}.

There are several studies on volatility spillovers across international stock markets.
For example, \citet{hamao1990correlations,engle1990meteor,king1990transmission,lin1994bulls,karolyi1995multivariate} employ multivariate GARCH models with low-frequency data and show that volatility spillovers occur across international stock markets.
\citet{ait2015modeling} focus on the jump processes for financial contagion modeling.
The proposed test procedures in this paper differ from the above literature mainly in the following three respects.
First, we propose a novel multivariate continuous jump-diffusion process to utilize high-frequency data, which helps account for market dynamics in relatively short periods.
Second, the contagion GARCH-It\^o model exploits the feature of disjoint trading hours between U.S. and China stock markets.
Third, we suggest three different contagion channels--integrated volatility, positive jump variation, and negative jump variation.

The rest of this paper is organized as follows.
In Section \ref{SEC-2}, we introduce the contagion GARCH-It\^{o} model and establish its properties.
In Section \ref{SEC-3}, we propose quasi-likelihood estimation methods and investigate their asymptotic behaviors.
Based on the asymptotic behaviors, we propose the hypothesis tests for a structural break.
In Section \ref{SEC-4}, we conduct a simulation study to check the finite sample performance for the proposed estimation and hypothesis testing methods.
In Section \ref{SEC-5}, we investigate whether there is a significant structural break, cased by the U.S.--China trade war in the volatility structure with S\&P 500 and CSI 300 high-frequency data.

%
%
%
%



\setcounter{equation}{0}
\section{Contagion GARCH-It\^{o} models} \label{SEC-2}

In this section, we develop a jump-diffusion process to model the volatility processes in U.S. and China stock markets as follows.
%

\begin{definition}\label{def-model}
    Log-prices $X_1(t)$ and $X_2(t)$, $t \in \mathbb{R}_{+}$, obey a contagion GARCH-It\^{o} model if they satisfy
    \begin{eqnarray*} 
        && dX_1(t) = \mu_1dt+\sigma_{1t}(\theta)dB_{1t} + L_{1t}^{+}d\Lambda_{1t}^{+} + L_{1t}^{-}d\Lambda_{1t}^{-}, \nonumber \\
        && dX_2(t) = \mu_2dt+\sigma_{2t}(\theta)dB_{2t} + L_{2t}^{+}d\Lambda_{2t}^{+} + L_{2t}^{-}d\Lambda_{2t}^{-},  \quad  dB_{1t} dB_{2t} = \rho dt, \cr
        && \sigma_{lt}^2 (\theta) = \begin{cases} \sigma_{l[t]_l}^2 (\theta) + \frac{(t-[t]_l)}{\lambda_l}\lbrace\omega_{lH} + (\gamma_{lH} - 1)\sigma_{l[t]_l}^2 (\theta) \rbrace \\
            + \frac{\alpha_{lH}}{\lambda_l}\int^{t}_{[t]_l}\sigma_{ls}^2 (\theta) ds + \frac{\beta_{lH}^{+}}{\lambda_{  l}} \int_{[t]_l}^{t} (L_{ls}^{+})^2 d\Lambda_{ls}^{+}\\  + \frac{\beta_{lH}^{-}}{\lambda_{l}} \int_{[t]_l}^{t} (L_{ls}^{-})^2 d\Lambda_{ls}^{-} + \frac{\nu_{lH}}{\lambda_i^2}([t]_l+\lambda_l-t)Z_{lt}^2, & \text{ if } t \in ([t]_l,[t]_l+\lambda_l], \\
        \sigma_{l[t]_l+\lambda_l}^2 (\theta) + \frac{(t-[t]_l-\lambda_l)}{1-\lambda_l}\lbrace\omega_{lL} + (\gamma_{lL} - 1)\sigma_{l[t]_l+\lambda_l}^2 (\theta) \rbrace \\
 +  \frac{\alpha_{ll'}}{\lambda_{l'}}V_{l'}(t;\theta) + \frac{\beta_{ll'}^{+}}{\lambda_{l'}}J_{l'}^{+}(t) + \frac{\beta_{ll'}^{-}}{\lambda_{l'}}J_{l'}^{-}(t) \\
+ \frac{\alpha_{lL}}{1-\lambda_l}\big(\int_{[t]_l+\lambda_l}^{t}\sigma_{ls} (\theta) dB_{ls}  \big)^2, & \text{ if } t \in [[t]_l+\lambda_l,[t]_l+1],
        \end{cases}
    \end{eqnarray*}
    where $l \in \lbrace1,2\rbrace$ and $l' \in \{1,2\} \setminus \{l\} $ denote the countries, $\lambda_l$ is the time length of the country $l$'s open-to-close period, $\tau$ is the time length of the gap between the opening times of the two countries, $[t]$ denotes the integer part of $t$, $[t]_1 = [t]$,   $[t]_2 = [t-\tau]+\tau$, $Z_{lt} = \int_{[t]_l}^{t} dW_{lt}$,
    $V_l(t;\theta) = \int_{[t]_{l'}+\lambda_{l'}}^{t} \mathbbm{1}(s \in [[t]_{l}, [t]_{l} + \lambda_l]) \sigma_{ls}^2 (\theta) ds$, $J_l^{+}(t) = \int_{[t]_{l'}+\lambda_{l'}}^{t} \mathbbm{1}(s \in [[t]_{l}, [t]_{l} + \lambda_l]) (L_{ls}^{+})^{2} d\Lambda_{ls}$, $J_l^{-}(t) = \int_{[t]_{l'}+\lambda_{l'}}^{t} \mathbbm{1}(s \in [[t]_{l}, [t]_{l} + \lambda_l]) (L_{ls}^{-})^{2} d\Lambda_{ls}$, and $\mathbbm{1}(\cdot)$ is an indicator function.
    For the jump part, $\Lambda_{1t}^{+}$, $\Lambda_{1t}^{-}$, $\Lambda_{2t}^{+}$, and $\Lambda_{2t}^{-}$ are the standard Poisson processes with constant intensity $I_1^{+}$, $I_1^{-}$, $I_2^{+}$, and $I_2^{-}$, respectively.
    $L_1^{+}$, $L_1^{-}$, $L_2^{+}$, and  $L_2^{-}$ are the i.i.d. jump sizes that are independent of the Poisson and continuous diffusion processes.
    Furthermore, the jump sizes $L_1^{+}$, $L_1^{-}$, $L_2^{+}$, and $L_2^{-}$ are equal to zero for each close-to-open period.
We denote the model parameters by $\theta= (\omega_{1H},\omega_{1L},\gamma_{1H},\gamma_{1L},\alpha_{1H},\alpha_{1L},\alpha_{12},\beta_{1H}^{+},\beta_{1H}^{-},\beta_{12}^{+},\beta_{12}^{-},\nu_{1H}, \omega_{2H},\omega_{2L},\gamma_{2H},\gamma_{2L},\alpha_{2H},\alpha_{2L},\alpha_{21},$ $\beta_{2H}^{+},\beta_{2H}^{-},\beta_{21}^{+},\beta_{21}^{-},\nu_{2H} , \mu_1,\mu_2)$.
\end{definition}

The instantaneous volatility process of the contagion GARCH-It\^{o} model is continuous with respect to time.
For each open-to-close period, the instantaneous volatility process obeys the structure of the realized GARCH-It\^{o} model \citep{song2021volatility}, but it incorporates positive and negative jump variations separately, which helps identify the effect of good and bad news. 
For each close-to-open period, the instantaneous volatility process is affected by the country's own current log-return and the other country's current integrated volatility and signed jump variations.
This structure makes it possible to account for the volatility contagion. 
We introduce $Z_{lt}$ to account for the random fluctuations of the instantaneous volatilities.
In this paper, we assign 1 and 2 for China and U.S. in UTC timeline, respectively.
At the market-opening time, the instantaneous volatility processes have the following realized GARCH-type structure:
\begin{flalign*}
    \sigma_{l\tau_{l}(n)}^2 (\theta) \quad=&\quad \omega_{lL} + \gamma_{lL}\omega_{lH} + \gamma_{lL}\gamma_{lH}\sigma_{l\tau_{l}(n-1)}^2 (\theta) + \frac{\gamma_{lL}\alpha_{lH}}{\lambda_l}\int ^{\tau_{l}(n-1)+\lambda_l }_{\tau_{l}(n-1)}\sigma ^{2}_{ls} (\theta) ds &&\\
    &\quad+ \frac{\gamma_{lL}\beta_{lH}^{+}}{\lambda_l} \int ^{\tau_{l}(n-1)+\lambda_l }_{\tau_{l}(n-1)}(L^{+}_{ls})^2d\Lambda_{ls}^{+} + \frac{\gamma_{lL}\beta_{lH}^{-}}{\lambda_l} \int ^{\tau_{l}(n-1)+\lambda_l }_{\tau_{l}(n-1)}(L^{-}_{ls})^2d\Lambda_{ls}^{-} &&\\
    &\quad+ \frac{\alpha_{lL}}{1-\lambda_l}\left( \int_{\tau_{l}(n-1)+\lambda_l}^{\tau_{l}(n)}\sigma_{ls} (\theta) dB_{ls} \right)^2 + \frac{\alpha_{ll'}}{\lambda_{l'}}\int ^{[\tau_{l}(n)]_{l'}+\lambda_{l'} }_{[\tau_{l}(n)]_{l'}}\sigma ^{2}_{l's} (\theta) ds &&\\
    &\quad+ \frac{\beta_{ll'}^{+}}{\lambda_{l'}} \int ^{[\tau_{l}(n)]_{l'}+\lambda_{l'} }_{[\tau_{l}(n)]_{l'}}(L^{+}_{2s})^2d\Lambda_{2s}^{+} + \frac{\beta_{12}^{-}}{\lambda_2} \int ^{[\tau_{l}(n)]_{l'}+\lambda_{l'} }_{[\tau_{l}(n)]_{l'}}(L^{-}_{2s})^2d\Lambda_{2s}^{-}, &&
\end{flalign*}
where $n$ is an integer, $l \in \{ 1,2 \}$, $\tau_{1}(n) = n$, and $\tau_{2}(n) = n+\tau$.
Thus, the instantaneous volatility processes are some interpolations of  the realized GARCH-type structures \citep{hansen2012realized, kim2021overnight, song2021volatility} with additional signed jump innovation terms and the volatility contagion from the other country.
Owing to this structure, we can relate the low-frequency volatility dynamics and continuous diffusion process and, thus, we can harness high-frequency data to analyze low-frequency dynamics.

The proposed contagion GARCH-It\^o model has the following property, which is used to make statistical inferences.

\begin{theorem}\label{theorem1}
For $0 < \alpha_{1H} < 1$, $0 < \alpha_{2H} < 1$, $l \in \{1,2\}$, and $n \in \mathbb{N}$, we have
\begin{equation}\label{eq-thm1}
    \int ^{\tau_{l}\left ( n \right ) +\lambda_{l}}_{\tau_{l}(n)}\sigma ^{2}_{lt}\left( \theta \right) dt=\lambda_{l} h_{l,n}\left( \theta \right) + D_{l,n} \quad a.s.,
\end{equation}
where 
\begin{eqnarray*}
h_{l,n}(\theta) &=& \omega_{l}^g + \gamma_{l}h_{l,n-1}(\theta) + \dfrac {\alpha_{l}^g}{\lambda_l}\int ^{\tau_{l}(n)-1+\lambda_{l} }_{\tau_{l}(n)-1}\sigma ^{2}_{ls} (\theta) ds + \dfrac {\alpha_{ll'}^g}{\lambda_{l'} }\int ^{\left [ {\tau_{l}(n)} \right]_{l'}+\lambda_{l'}}_{\left [ \tau_{l}(n) \right]_{l'}}\sigma ^{2}_{l's} (\theta) ds \cr
&&+ \dfrac{\beta_{l+}^{g}}{\lambda_{l}} \int ^{\tau_{l}(n)-1+\lambda_{l} }_{\tau_{l}(n)-1}(L^{+}_{ls})^2d\Lambda_{ls}^{+} + \dfrac{\beta_{l-}^{g}}{\lambda_l} \int ^{\tau_{l}(n)+\lambda_l }_{\tau_{l}(n)-1}(L^{-}_{ls})^2d\Lambda_{ls}^{-} \cr
&&+ \dfrac{\beta_{ll'+}^{g}}{\lambda_{l'}}\int ^{\left [ \tau_{l}(n) \right ]_{l'}+\lambda_{l'}}_{\left [ \tau_{l}(n) \right]_{l'}}(L^{+}_{l's})^2d\Lambda_{l's}^{+} + \dfrac{\beta_{ll'-}^{g}}{\lambda_{l'}}\int ^{\left [\tau_{l}(n) \right]_{l'}+\lambda_{l'}}_{\left[ {\tau_{l}(n)} \right]_{l'}}(L^{-}_{l's})^2d\Lambda_{l's}^{-} \cr
&&+ \dfrac {\kappa_{l}^g}{1-\lambda_{l} }\left(\int ^{\tau_{l}(n)}_{\tau_{l}(n)-1+\lambda_{l}}\sigma _{ls} (\theta) dB_{1s}\right) ^{2};
\end{eqnarray*}
$D_{1,n}$ and $D_{2,n}$ are martingale differences; and $\omega_{1}^{g}, \gamma_1, \alpha_{1}^{g}, \beta_{1+}^{g}, \beta_{1-}^{g}, \kappa_{1}^{g}, \alpha_{1,2}^g, \beta_{12+}^g, \beta_{12-}^g ,  \omega_{2}^{g}, \gamma_2,$ $\alpha_{2}^{g},$ $ \beta_{2+}^{g},$ $ \beta_{2-}^{g},\kappa_{2}^{g} , \alpha_{21}^g, \beta_{21+}^g, \beta_{21-}^g$ are functions of $\theta$.
Their detailed forms are defined in Theorem \ref{theorem4}.
\end{theorem}

Theorem \ref{theorem1} indicates that the open-to-close integrated volatility for each asset can be decomposed into GARCH volatility and martingale difference.
 GARCH volatility is a function of the past open-to-close integrated volatilities and signed jump variations of two assets and its own squared close-to-open log-returns.
Under the model structure \eqref{eq-thm1}, we develop an estimation procedure for the GARCH parameter by using the integrated volatilities as proxies of the conditional GARCH volatilities $h_{1,n}(\theta)$ and $h_{2,n}(\theta)$.
Since the main purpose of this paper is to analyze the low-frequency market dynamics under the existence of the volatility contagion, the parameter of interest is the GARCH parameter $\theta ^g = (\omega_{1}^{g}, \gamma_1, \alpha_{1}^{g},$ $\beta_{1+}^{g}, \beta_{1-}^{g}, \kappa_{1}^{g}, \alpha_{1,2}^g, \beta_{12+}^g, \beta_{12-}^g ,  \omega_{2}^{g}, \gamma_2, \alpha_{2}^{g},$ $ \beta_{2+}^{g},$ $ \beta_{2-}^{g},\kappa_{2}^{g} , \alpha_{21}^g, \beta_{21+}^g, \beta_{21-}^g)$.

\section{Estimation procedure} \label{SEC-3}

\subsection{The model setup}
We assume that the underlying asset log-prices follow the contagion GARCH-It\^{o} process defined in Definition \ref{def-model}.
For the $l$th asset, the high-frequency observations during the $i$th open-to-close period are observed at $t_{l,i,j}$, $j=1,\ldots,m_{l,i}$, $l \in \{1,2\}$, where $i-1=t_{1,i,0} < t_{1,i,1} < \cdots < t_{1,i,m_{1,i}} = \lambda_1 + i - 1$ and $i-1+\tau=t_{2,i,0} < t_{2,i,1} < \cdots < t_{2,i,m_{2,i}} = \lambda_2 + i - 1 + \tau$.
For technical purpose, we define the average number of high-frequency observations--that is, $m = \frac{1}{2n}\sum_{i=1}^{n}m_{1,i} + m_{2,i}$.
In practice, high-frequency data are polluted by microstructure noise due to market inefficiencies, such as bid-ask spread and information asymmetries.
To reflect this, we assume the following additive noise structure:
\begin{equation*}
    Y_{t_{l,i,j}} = X_{t_{l,i,j}} + \epsilon_{t_{l,i,j}},
\end{equation*}
where $\epsilon_{t_{l,i,j}}$ is microstructure noise with mean zero and the log-prices and microstructure noise are independent.
The drift terms $\mu_1$ and $\mu_2$ can simply be estimated by the sample mean in terms of low-frequency data.
Moreover, the effect of the drift terms $\mu_1$ and $\mu_2$ is negligible for high-frequency realized volatility estimators.
Therefore, for simplicity, we assume $\mu_1 = \mu_2 = 0$ in Definition \ref{def-model}.

To harness high-frequency information, such as realized volatility and signed jump variations, we first need to estimate these quantities. 
In the presence of microstructure noises and price jumps, \citet{fan2007multi} proposed the jump-adjusted multiple-scale realized volatility estimator (JMSRV) to estimate realized volatility and jump variations.
The estimator uses wavelet methods to detect jumps and applies the multi-scale realized volatility estimator \citep{zhang2006efficient} to jump-adjusted data in order to estimate the continuous integrated volatility. 
They showed that both estimators for integrated volatility and jump variation have the optimal convergence rate of $m^{-1/4}$.
In this paper, we employ the JMSRV estimation procedure and  let $OV_{l,i}$, $IV_{l,i}$, and $RV_{l,i}$ be the squared close-to-open log-return, the open-to-close integrated volatility, and its estimator for country $l$ and the $i$th day, and $JV_{l,i}^{+}$, $JV_{l,i}^{-}$ be the corresponding estimators of positive and negative jump variations, respectively.
The detailed forms for estimators of realized volatility and signed jump variations are presented in Section \ref{SEC-4}.

\subsection{Estimation of GARCH parameters}

In this section, we  propose a quasi-maximum likelihood estimation procedure for making inferences on the true parameter $\theta_0 ^g = (\omega_{1}^{g}, \gamma_1, \alpha_{1}^{g}, \beta_{1+}^{g}, \beta_{1-}^{g}, \kappa_{1}^{g}, \alpha_{1,2}^g, \beta_{12+}^g, \beta_{12-}^g ,  \omega_{2}^{g}, \gamma_2, \alpha_{2}^{g},$ $ \beta_{2+}^{g},$ $ \beta_{2-}^{g},\kappa_{2}^{g} , \alpha_{21}^g, \beta_{21+}^g, \beta_{21-}^g)$.
We first define some notations.
For any given vector $v=(v_i)_{i=1,\ldots,k}$, we define $\norm{v}_{max}=\max_{i}|v_i|$.
Let $C$'s be positive generic constants whose values are independent of $\theta$, $n$, and $m_{l,i}$ and may change from occurence to occurence.

Theorem \ref{theorem1} indicates that the open-to-close integrated volatility over the $i$th period of country $l$'s asset can be decomposed into the realized GARCH volatility $h_{l,i}(\theta ^g)$ and the martingale difference $D_{l,i}$.
To employ this structure, we utilize the realized volatilities $RV_{l,i}$ as proxies of the GARCH volatilities and define the quasi-likelihood function as follows:
\begin{equation*}\label{eq-L}
    L_{n,m}(\theta ^g) = -\frac{1}{2n}\sum_{i=1}^{n}\left\lbrace\log(h_{1,i}(\theta ^g)) + \frac{RV_{1,i}/\lambda_1}{h_{1,i}(\theta ^g)} + \log(h_{2,i}(\theta ^g)) + \frac{RV_{2,i}/\lambda_2}{h_{2,i}(\theta ^g)} \right\rbrace.
\end{equation*}
Unfortunately, the true integrated volatilities and signed jump variations are not observable.
Thus, we adopt their estimators $RV_{l,i}$, $JV_{l,i}^{+}$, and $JV_{l,i}^{-}$ and use the following estimated conditional GARCH:
\begin{eqnarray*}
    \hat{h}_{1,i}(\theta ^g) &=& \omega_{1}^g + \gamma_{1}\hat{h}_{1,i-1}(\theta ^g) + \alpha_{1}^g{\lambda_1}^{-1}RV_{1,i-1} + \beta_{1+}^g{\lambda_1}^{-1}JV_{1,i-1}^{+} + \beta_{1-}^g{\lambda_1}^{-1}JV_{1,i-1}^{-}\cr
    && + \kappa_{1}^g(1-\lambda_1 )^{-1}OV_{1,i-1} + \alpha_{12}^g \lambda_2 ^{-1} RV_{2,i-1} + \beta_{12+}^g \lambda_2 ^{-1} JV_{2,i-1}^{+} + \beta_{12-}^g \lambda_2 ^{-1} JV_{2,i-1}^{-}, \cr
     \hat{h}_{2,i}(\theta ^g) &=& \omega_{2}^g + \gamma_{2}\hat{h}_{2,i-1}(\theta ^g) + \alpha_{2}^g{\lambda_2}^{-1}RV_{2,i-1} + \beta_{2+}^g{\lambda_2}^{-1}JV_{2,i-1}^{+} + \beta_{2-}^g{\lambda_2}^{-1}JV_{2,i-1}^{-} \cr
    &&   + \kappa_{2}^g(1-\lambda_2 )^{-1}OV_{2,i-1} + \alpha_{21}^g \lambda_1 ^{-1} RV_{1,i} + \beta_{21+}^g \lambda_1 ^{-1} JV_{2,i}^{+} + \beta_{21-}^g \lambda_1 ^{-1} JV_{2,i}^{-}.
\end{eqnarray*}
In the numerical study, we utilize the jump-adjusted multiple-scale realized volatility and wavelet methods to estimate realized volatilities and signed jump variations \citep{fan2007multi}, respectively. 
Then, with the realized GARCH volatility estimators $\hat{h}_{1,i}(\theta ^g)$, $\hat{h}_{2,i}(\theta ^g)$, we define the quasi-likelihood function as follows:
\begin{equation*}
    \hat{L}_{n,m}(\theta ^g) = -\frac{1}{2n}\sum_{i=1}^{n}\left\lbrace\log(\hat{h}_{1,i}(\theta ^g)) + \frac{RV_{1,i}/\lambda_1}{\hat{h}_{1,i}(\theta ^g)} + \log(\hat{h}_{2,i}(\theta ^g)) + \frac{RV_{2,i}/\lambda_2}{\hat{h}_{2,i}(\theta ^g)} \right\rbrace.
\end{equation*}
We estimate the true parameter $\theta_0$ by maximizing the quasi-likelihood function $\hat{L}_{n,m}(\theta)$ as follows:
\begin{equation*}
    \htheta ^g = \argmax_{\theta ^g \in\Theta ^g} \hat{L}_{n,m}(\theta ^g).
\end{equation*}

To establish the asymptotic properties of the proposed estimation method, the following assumptions are necessary.
\begin{assumption} \label{assumption-1}
	~
	\begin{enumerate}
		\item[(a)] Let 
		\begin{equation*}
            \begin{split}
            \Theta ^g =& \lbrace  (\omega_{1}^{g}, \gamma_1, \alpha_{1}^{g}, \beta_{1+}^{g}, \beta_{1-}^{g}, \kappa_{1}^{g}, \alpha_{1,2}^g, \beta_{12+}^g, \beta_{12-}^g  ,  \omega_{2}^{g}, \gamma_2, \alpha_{2}^{g}, \beta_{2+}^{g}, \beta_{2-}^{g},\kappa_{2}^{g} , \alpha_{2,1}^g, \beta_{21+}^g, \beta_{21-}^g):\\
		    &\omega_l < \omega _{1} ^ {g}, \omega _{2} ^ {g} < \omega_u, \gamma_l < \gamma_1 , \gamma_2 < \gamma_u < 1, \alpha_l < \alpha_{1}^{g}, \alpha_{2}^{g}, \alpha_{1,2}^{g}, \alpha_{2,1}^{g} < \alpha_u < 1,\\
            &\beta_l <\beta_{1+}^{g}, \beta_{1-}^{g}, \beta_{2+}^{g}, \beta_{2-}^{g}, \beta_{12+}^g, \beta_{12-}^g, \beta_{21+}^g, \beta_{21-}^g < \beta_u, \kappa_l <\kappa_{1}^{g},\kappa_{2}^{g} < \kappa_u   \rbrace,
            \end{split}
        \end{equation*} 
		where $\omega_ l, \omega_u,  \gamma_l, \gamma_u, \alpha_l,$ $\alpha_u,$ $\kappa_l, \kappa_u,$ $\beta_l, \beta_u$ are some constants, such that $h_n(\theta ^g) > c$ and $\hat{h}_n(\theta ^g) > c$ a.s. for all $\theta ^g \in \Theta ^g$ and some positive constant $c$.

        \item[(b)] For some positive constant $C$, we have
        \begin{equation*}
            \begin{split}
                & \sup_{i \in \mathbb{N}}E\[(X_{1,i}-X_{1,i-1+\lambda_1})^4\] \leq C, \quad \sup_{i\in \mathbb{N}}E\[(X_{1,i-1+\lambda_1}-X_{1,i-1})^4\] \leq C ,\\
                & \sup_{i\in \mathbb{N}}E\[(X_{2,i-1+\tau+\lambda_2}-X_{2,i-1+\tau})^4\] \leq C, \quad \sup_{i\in \mathbb{N}}E\[(X_{2,i+\tau}-X_{2,i-1+\tau+\lambda_2})^4\] \leq C ,\\
                & \sup_{i\in \mathbb{N}}E\[(D_{1,i})^4\] \leq C, \quad \sup_{i\in \mathbb{N}}E\[(D_{2,i})^4\] \leq C.
            \end{split}
        \end{equation*} 

        \item[(c)] For some positive constant $\eta$ and $l \in \{1,2\}$, we have
        \begin{flalign*}
            & \sup_{t\in\mathbb{R}_{+}} E\[\sigma_{l,t}^4(\theta_0)\] \leq C \quad \text{ and } \quad  \sup_{i \in \mathbb{N}} E\left[ h^{2+\eta}_{l,i}(\theta_0) \right] \leq C.
        \end{flalign*}

        \item[(d)] There exist some fixed constants $C_1$ and $C_2$, such that $C_1 m \leq m_i \leq C_2 m$, and\\
        $\sup_{1 \leq j \leq m_i} | t_{i,j} - t_{i,j-1}| = O(m^{-1})$ and $n^2 m^{-1} \rightarrow 0$ as $m,n \rightarrow \infty$.
        
		
        \item[(e)] For any $i \in \mathbb{N}$ and $l \in \{1,2\}$, we have
        \begin{equation*}
            \begin{split}
                & E \left [ (RV_{l,i}- IV_{l,i} )^2 \right ] \leq C m^{-1/2},  \quad E \left [ (JV_{l,i}^{+} - \int_{\tau_{l}(i)-1}^{\tau_{l}(i)-1+\lambda_l} (L_{ls}^{+})^{2} d\Lambda_{ls}^{+} )^2 \right ] \leq C   m^{-1/2},\cr
                & E \left [ (JV_{l,i}^{-} - \int_{\tau_{l}(i)-1}^{\tau_{l}(i)-1+\lambda_l} (L_{ls}^{-})^{2} d\Lambda_{ls}^{-} )^2 \right ] \leq C   m^{-1/2},
            \end{split}
        \end{equation*}
        where $IV_{l,i} = \int ^{\tau_{l}(i)-1+\lambda_{l} }_{\tau_{l}(i)-1}\sigma ^{2}_{ls} (\theta) ds$.
		
		\item[(f)] $\Big(D_{1,i}, D_{2,i}, IV_{1,i}, IV_{2,i}, OV_{1,i}, OV_{2,i},\int_{i-1}^{i-1+\lambda_1} (L_{1s}^{+})^{2} d\Lambda_{1s}^{+}, \int_{i-1}^{i-1+\lambda_1} (L_{1s}^{-})^{2} d\Lambda_{1s}^{-} ,$\\
        $\int_{i-1+\tau}^{i-1+\tau+\lambda_2} (L_{2s}^{+})^{2} d\Lambda_{2s}^{+}, \int_{i-1+\tau}^{i-1+\tau+\lambda_2} (L_{2s}^{-})^{2} d\Lambda_{2s}^{-} \Big)$ is a stationary ergodic process.
	\end{enumerate}
\end{assumption}

\begin{remark}\label{remark1}
    Assumption \ref{assumption-1}(a) describes the parameter space that guarantees the positive conditional GARCH volatilities.
    Assumption \ref{assumption-1}(b) and (c) are related to the finite fourth moment condition, which is required to study the second moment target parameter, such as volatilities.
    Assumption \ref{assumption-1}(d) is a typical condition in high-frequency literature.
    For the realized volatility and jump variations estimators, we take $RV_{l,i}$ as the jump-adjusted multi-scale realized volatility estimator \citep{fan2007multi}, and  we employ the jump variation estimators given in \citet{fan2007multi} that utilize wavelet methods to detect jump locations.
    The detailed forms of signed jump variation are defined in Section \ref{SEC-4}.
    \citet{tao2013fast} and \citet{kim2016asymptotic} showed that Assumption \ref{assumption-1}(e) is satisfied with these estimators.
    In order to derive an asymptotic normality for the proposed estimator, Assumption \ref{assumption-1}(f) is required.
\end{remark}

    

\begin{theorem}\label{theorem2}
    Under Assupmtion \ref{assumption-1}(a)--(e) (except for $n^2m^{-1} \rightarrow 0$ in Assumption \ref{assumption-1}(d)), we have
    \begin{equation}\label{eq-convrate}
        \norm{\htheta ^g - \theta_0 ^g}_{max} = O_p\left( m^{-1/4} + n^{-1/2} \right).    
    \end{equation}
    Furthermore, under Assumption \ref{assumption-1}, we have, as $m,n \rightarrow \infty$,
    \begin{equation}\label{eq-asymp}
        \sqrt{n}\left(\htheta ^g - \theta_{0} ^g\right) \xrightarrow{d} N(0,B^{-1}AB^{-1}),
    \end{equation}
    where $s_l = E\left[(D_{l,1})^2\big|\mathcal{F}_0\right]$ with $l \in \{1,2\}$, and 
\begin{eqnarray*}
	&&A = \frac{1}{4} E\left[\frac{\partial h_{1,1}(\theta_0 ^g)}{\partial\theta ^g}\frac{\partial h_{1,1}(\theta_0 ^g)}{\partial \theta ^{g\top} }h_{1,1}^{-4}(\theta_0 ^g)  s_1\lambda_1^{-2} + \frac{\partial h_{2,1}(\theta_0 ^g)}{\partial\theta ^g}\frac{\partial h_{2,1}(\theta_0 ^g)}{\partial\theta ^{g\top}}h_{2,1}^{-4}(\theta_0 ^g)  s_2\lambda_2^{-2} \right], \cr
    &&B = \frac{1}{2}E\left[\frac{\partial h_{1,1}(\theta_0 ^g)}{\partial\theta ^g}\frac{\partial h_{1,1}(\theta_0 ^g)}{\partial\theta ^{g\top}}h_{1,1}^{-2}(\theta_0 ^g)+ \frac{\partial h_{2,1}(\theta_0 ^g)}{\partial\theta ^g}\frac{\partial h_{2,1}(\theta_0 ^g)}{\partial\theta ^{g\top}}h_{2,1}^{-2}(\theta_0 ^g)\right].
\end{eqnarray*}
\end{theorem}

\begin{remark}
    Theorem \ref{theorem2} shows that the proposed estimator $\hat{\theta} ^g$ has the convergence rate $m^{-1/4} + n^{-1/2}$.
    The rate $m^{-1/4}$ comes from estimating integrated volatilities and signed jump variations with high-frequency data.
    This rate is known as the optimal convergence rate with the presence of market microstructure noises and jumps.
    Further, the rate $n^{-1/2}$ is the usual convergence rate for estimating low-frequency parametric structure.
    Theorem \ref{theorem2} also establishes the asymptotic normality for the proposed estimator.
    We utilize this asymptotic normality for testing the existence of a structural break.
\end{remark}

\subsection{Hypothesis tests for a structural break}

The main goal of this paper is to test whether there is a structural break.
In this section, we discuss how to conduct the hypothesis test for the structural break.
We denote the possible break point by zero. 
To test the hypothesis
\begin{equation*}
    \begin{split}
        H_0 &: \text{there is no structural break}\cr
        H_1 &: \text{there is a structural break at time 0},
    \end{split}
\end{equation*}
we first derive the asymptotic distribution of the difference between estimated GARCH parameters for two different periods, $[-n_1,0]$ and $[0,n_2]$. 
Specifically, we estimate the model parameters as follows:
$$
\htheta_1 ^g= \argmax_{\theta ^g\in\Theta ^g} \hat{L}_{1,n_1,m}(\theta ^g) \quad \text{and} \quad \htheta_2 ^g= \argmax_{\theta ^g\in\Theta ^g} \hat{L}_{2,n_2,m}(\theta ^g),
$$
where
\begin{flalign*}
    &\hat{L}_{1,n_1,m}(\theta ^g) = -\frac{1}{2n_1}\sum_{i=1}^{n_1}\left\lbrace\log(\hat{h}_{1,i-n_1}(\theta ^g)) + \frac{RV_{1,i-n_1}/\lambda_1}{\hat{h}_{1,i-n_1}(\theta ^g)} + \log(\hat{h}_{2,i-n_1}(\theta ^g)) + \frac{RV_{2,i-n_1}/\lambda_2}{\hat{h}_{2,i-n_1}(\theta ^g)} \right\rbrace, \cr
    &\hat{L}_{2,n_2,m}(\theta ^g) = -\frac{1}{2n_2}\sum_{i=1}^{n_2}\left\lbrace\log(\hat{h}_{1,i}(\theta ^g)) + \frac{RV_{1,i}/\lambda_1}{\hat{h}_{1,i}(\theta ^g)} + \log(\hat{h}_{2,i}(\theta ^g)) + \frac{RV_{2,i}/\lambda_2}{\hat{h}_{2,i}(\theta ^g)} \right\rbrace.
\end{flalign*}
Then, in the following theorem, we investigate an asymptotic distribution of $\htheta_1 ^g - \htheta_2 ^g$.

\begin{theorem}\label{theorem3}
    Under Assumption \ref{assumption-1}, as $m, n_1 \rightarrow \infty$ and $n_1 / n_2 \rightarrow r \in (0,\infty)$, we have
    \begin{equation*}
        \sqrt{n_1}\left(\htheta_1 ^g -\htheta_2 ^g - \delta \right)\xrightarrow{d} N(0,B_1^{-1}A_1B_1^{-1}+r B_2^{-1}A_2B_2^{-1}),
    \end{equation*}
    where $\delta = \theta_1 ^g - \theta_2 ^g$, $s_{1l} = E\left[(D_{l,-n_1+1})^2\big|\mathcal{F}_{-n_1}\right]$, $s_{2l} = E\left[(D_{l,1})^2\big|\mathcal{F}_0\right]$,\\
    $A_1 = \frac{1}{4} E\left[\frac{\partial h_{1,-n_1+1}(\theta_0 ^g)}{\partial\theta ^g}\frac{\partial h_{1,-n_1+1}(\theta_0 ^g)}{\partial\theta ^{g\top}}h_{1,-n_1+1}^{-4}(\theta_0 ^g)  \frac{s_{11}}{\lambda_1^{2}} + \frac{\partial h_{2,-n_1+1}(\theta_0 ^g)}{\partial\theta ^g}\frac{\partial h_{2,-n_1+1}(\theta_0 ^g)}{\partial\theta ^{g\top}}h_{2,-n_1+1}^{-4}(\theta_0 ^g)  \frac{s_{12}}{\lambda_2^{2}} \right]$,\\
    $B_1 = \frac{1}{2}E\left[\frac{\partial h_{1,-n_1+1}(\theta_0 ^g)}{\partial\theta ^g}\frac{\partial h_{1,-n_1+1}(\theta_0 ^g)}{\partial\theta ^{g\top}}h_{1,-n_1+1}^{-2}(\theta_0 ^g)+ \frac{\partial h_{2,-n_1+1}(\theta_0 ^g)}{\partial\theta ^g}\frac{\partial h_{2,-n_1+1}(\theta_0 ^g)}{\partial\theta ^{g\top}}h_{2,-n_1+1}^{-2}(\theta_0 ^g)\right]$,\\
    $A_2 = \frac{1}{4} E\left[\frac{\partial h_{1,1}(\theta_0 ^g)}{\partial\theta ^g}\frac{\partial h_{1,1}(\theta_0 ^g)}{\partial\theta ^{g\top}}h_{1,1}^{-4}(\theta_0 ^g)  s_{21}\lambda_1^{-2} + \frac{\partial h_{2,1}(\theta_0 ^g)}{\partial\theta ^g}\frac{\partial h_{2,1}(\theta_0 ^g)}{\partial\theta ^{g\top}}h_{2,1}^{-4}(\theta_0 ^g)  s_{22}\lambda_2^{-2} \right]$,\\
    $B_2 = \frac{1}{2}E\left[\frac{\partial h_{1,1}(\theta_0 ^g)}{\partial\theta ^g}\frac{\partial h_{1,1}(\theta_0 ^g)}{\partial\theta ^{g\top}}h_{1,1}^{-2}(\theta_0 ^g)+ \frac{\partial h_{2,1}(\theta_0 ^g)}{\partial\theta ^g}\frac{\partial h_{2,1}(\theta_0 ^g)}{\partial\theta ^{g\top}}h_{2,1}^{-2}(\theta_0 ^g)\right]$.
    
\end{theorem}

In order to simultaneously test the change in the GARCH parameters, we derive the Wald statistic from the asymptotic property in Theorem \ref{theorem3} and a consistent estimator of its asymptotic variance in the following proposition.

\begin{proposition}\label{proposition1}
    Under Assumption \ref{assumption-1}, as $m, n_1 \rightarrow \infty$ and $n_1/n_2 \rightarrow r \in (0,\infty)$, we have 
    \begin{equation}\label{eq-3.1}
        \mathcal{W}_{n_1,n_2}(\delta) = n_1 (\hat{\theta} ^g  _1 - \hat{\theta} ^g  _2 - \delta) ^\top \left ( \hat{B}_1^{-1} \hat{A}_1 \hat{B}_1^{-1} + r \hat{B}_2^{-1} \hat{A}_2 \hat{B}_2^{-1} \right ) ^{-1}(\hat{\theta} ^g _1 - \hat{\theta} ^g _2 - \delta) \xrightarrow{d} \chi ^2 (18),
    \end{equation}
    where
    \begin{eqnarray*}
           & \hat{A}_1  =& \frac{1}{4n} \sum_{i=-n_1+1}^{0} \frac{\partial \hat{h}_{1,i}(\htheta _1 ^g)}{\partial\theta ^g}\frac{\partial \hat{h}_{1,i}(\htheta _1 ^g)}{\partial\theta ^{g\top}}\hat{h}_{1,i}^{-4}(\htheta _1 ^g)  \left( \lambda_1 ^{-1} RV_{1,i} - \hat{h}_{1,i}(\htheta _1 ^g) \right) ^2\cr
            && \qquad  + \frac{\partial \hat{h}_{2,i}(\htheta _1 ^g)}{\partial\theta ^g}\frac{\partial \hat{h}_{2,i}(\htheta _1 ^g)}{\partial\theta ^{g\top}} \hat{h}_{2,i}^{-4}(\htheta _1 ^g) \left( \lambda_2 ^{-1} RV_{2,i} - \hat{h}_{2,i}(\htheta _1 ^g) \right) ^2 ,\cr
            &\hat{A}_2  =& \frac{1}{4n} \sum_{i=1}^{n_2} \frac{\partial \hat{h}_{1,i}(\htheta _2 ^g)}{\partial\theta ^g}\frac{\partial \hat{h}_{1,i}(\htheta _2 ^g)}{\partial\theta ^{g\top}}\hat{h}_{1,i}^{-4}(\htheta _2 ^g)  \left( \lambda_1 ^{-1} RV_{1,i} - \hat{h}_{1,i}(\htheta _2 ^g) \right) ^2\cr
            && \qquad + \frac{\partial \hat{h}_{2,i}(\htheta _2 ^g)}{\partial\theta ^g}\frac{\partial \hat{h}_{2,i}(\htheta _2 ^g)}{\partial\theta ^{g\top}} \hat{h}_{2,i}^{-4}(\htheta _2 ^g)  \left( \lambda_2 ^{-1} RV_{2,i} - \hat{h}_{2,i}(\htheta _2 ^g) \right) ^2 , \cr
            &\hat{B}_1  =& \frac{1}{2n}\sum_{i=-n_1+1}^{0}\left\lbrace \frac{\partial \hat{h}_{1,i}(\htheta_1 ^g)}{\partial\theta ^g}\frac{\partial \hat{h}_{1,i}(\htheta_1 ^g)}{\partial\theta ^{g\top}}\hat{h}_{1,i}^{-2}(\htheta_1 ^g)+  \frac{\partial \hat{h}_{2,i}(\htheta_1 ^g)}{\partial\theta ^g}\frac{\partial \hat{h}_{2,i}(\htheta_1)}{\partial\theta ^{g\top}} \hat{h}_{2,i}^{-2}(\htheta_1 ^g) \right\rbrace ,\cr
            &\hat{B}_2 =& \frac{1}{2n}\sum_{i=1}^{n}\left\lbrace \frac{\partial \hat{h}_{1,i}(\htheta_2 ^g)}{\partial\theta ^g}\frac{\partial \hat{h}_{1,i}(\htheta_2 ^g)}{\partial\theta ^{g\top}}\hat{h}_{1,i}^{-2}(\htheta_2 ^g)+  \frac{\partial \hat{h}_{2,i}(\htheta_2 ^g)}{\partial\theta ^g}\frac{\partial \hat{h}_{2,i}(\htheta_2 ^g)}{\partial\theta ^{g\top}} \hat{h}_{2,i}^{-2}(\htheta_2 ^g) \right\rbrace .
    \end{eqnarray*}
\end{proposition}

Testing the structural break is equivalent to testing whether $\delta=\theta_1 ^g - \theta_2 ^g = 0$.
Then, under the null hypothesis--that is, $\delta=0$--by Proposition \ref{proposition1}, we have $\mathcal{W}_{n_1,n_2}(0) \xrightarrow{d} \chi ^2 (18)$.
Similarly, we can test the existence of a structural break for each country, and the test statistic asymptotically follows the $\chi ^2 (9)$ distribution.
Specifically, with the chi-squared distribution as the null distribution, we can conduct a hypothesis test for the structural break.
If we reject the null hypothesis, it becomes of interest to test  which GARCH parameter is significantly changed. 
Similar to the simultaneous test, we can derive test statistics for individual tests as follows.

\begin{proposition}\label{proposition2}
    Under Assumption \ref{assumption-1}, as $m,n_1 \rightarrow \infty$ and $n_1 / n_2 \rightarrow r \in (0,\infty)$,  we have 
    \begin{equation}\label{eq-3.2}
        Z_{i,n_1,n_2}(\delta_i) = \sqrt{n_1} \hat{\Sigma}_{ii} ^{-1/2} \left ( \hat{\theta}_{1i} ^g - \hat{\theta}_{2i} ^g - \delta_i \right) \xrightarrow{d} N(0,1),
    \end{equation}
    for $i \in \lbrace 1,2, \ldots , 18 \rbrace$, where $\hat{\Sigma} = \hat{B}_1^{-1} \hat{A}_1 \hat{B}_1^{-1} + r \hat{B}_2^{-1} \hat{A}_2 \hat{B}_2^{-1}$.
\end{proposition}

By Proposition \ref{proposition2}, under $\delta_i=0$, we have the asymptotic normal distribution for each GARCH parameter. 
Then, based on the usual normal test procedure, we can conduct individual hypothesis tests.

\section{A simulation study} \label{SEC-4}
We conducted simulations to check finite sample performances of the proposed estimation procedures and verify the asymptotic results of the proposed hypothesis tests.
We generated the log-prices of two assets $X_{1,t_{1,i,j}}$ and $X_{2,t_{2,i,j}}$--the open-to-close trading hours of which do not overlap--for $n$ days, with a frequency $1/m$ for each day.
Let $t_{1,i,j} = i - 1 + \frac{\lambda}{m}j$ and $t_{2,i,j} = i - 1 + \tau + \frac{\lambda}{m}j$ for $i = 1, \ldots,n$, $j = 0, \ldots, m$; the open-to-close trading hours $\lambda_1 = \lambda_2 = 0.25$; and the difference in opening times $\tau=0.5$.
The true log stock prices follow the diffusion process defined in Definition \ref{def-model} with the following set of parameters $(\omega_{1H}, \gamma_{1H}, \alpha_{1H}, \beta_{1H}^{+}, \beta_{1H}^{-}, \nu_{1,H}, \omega_{1L}, \gamma_{1L}, \alpha_{1L}, \alpha_{12}, \beta_{12}^{+}, \beta_{12}^{-}, \omega_{2H}, \gamma_{2H}, \alpha_{2H}, \beta_{2H}^{+}, \beta_{2H}^{-},$ $\nu_{2H},$ $\omega_{2L},\quad$ $\gamma_{2L},$ $\alpha_{2L},$ $\alpha_{21}, \beta_{21}^{+}, \beta_{21}^{-}) = (0.001, 0.3, 0.7, 0.25, 0.3, 0.1, 0.0005, 0.4,
       0.1, 0.12, 0.1, 0.12, 0.0015,$ $0.4,$ $ 0.6,$ $ 0.3,
       0.4, 0.1, 0.0005, 0.4, 0.1, 0.12, 0.1, 0.1)$ and $\rho = 0$.
For the jump process, we set the intensities $I_1^{+} = 12$, $I_1^{-} = 16$, $I_2^{+} = 16$, $I_2^{-} = 12$ and generated $(L_{1t}^{+})^2$, $(L_{1t}^{-})^2$, $(L_{2t}^{+})^2$, and $(L_{2t}^{-})^2$, such that $(L_{1t}^{+})^2 = b_1^{+} + M_{1t}^{+}$, $(L_{1t}^{-})^2 = b_1^{-} + M_{1t}^{-}$, $(L_{2t}^{+})^2 = b_2^{+} + M_{2t}^{+}$, and $(L_{2t}^{-})^2 = b_2^{-} + M_{2t}^{-}$, where $b_1^{+} = b_1^{-} = b_2^{+} = b_2^{-} = 0.005$ and $M_{1t}^{+}$, $M_{1t}^{-}$, $M_{2t}^{+}$ and $M_{2t}^{-}$ follow the normal distribution with mean zero and standard deviation 0.0005.
$L_{1t}^{+}$ and $L_{2t}^{+}$ are positive and $L_{1t}^{-}$ and $L_{2t}^{-}$ are negative.
We varied the sample period from 100 to 500 and the high-frequency observations from 360 to 2160. 
The entire simulation procedure was repeated 1000 times.
We followed the procedure as described in \citet{fan2007multi} to estimate the jump locations, signed jump variations, and jump-adjusted MSRV.
Specifically, positive and negative jump variations were estimated as follows:
\begin{eqnarray*}
    && JV_{l,i}^{+} = \sum_{k=1}^{\hat{q}_{li}} \mathbbm{1} \left( \bar{Y}_{li}^{+}(\hat{\tau}_{lik}) > \bar{Y}_{li}^{-}(\hat{\tau}_{lik}) \right) \left( \bar{Y}_{li}^{+}(\hat{\tau}_{lik}) - \bar{Y}_{li}^{-}(\hat{\tau}_{lik}) \right)^{2},\\
    && JV_{l,i}^{-} = \sum_{k=1}^{\hat{q}_{li}} \mathbbm{1} \left( \bar{Y}_{li}^{+}(\hat{\tau}_{lik}) < \bar{Y}_{li}^{-}(\hat{\tau}_{lik}) \right) \left( \bar{Y}_{li}^{+}(\hat{\tau}_{lik}) - \bar{Y}_{li}^{-}(\hat{\tau}_{lik}) \right)^{2},
\end{eqnarray*}
where $\hat{\tau}_{lik} \in \left[ [i-1+\tau]_{l}, [i-1+\tau]_{l} + \lambda_l  \right]$ are jump locations estimated using wavelet method, $\hat{q}_{li}$ is the number of estimated jump locations, and $\bar{Y}_{li}^{+}(\hat{\tau}_{lik})$ and $\bar{Y}_{li}^{-}(\hat{\tau}_{lik})$ are the averages of $Y_{l}$ over $\left[ \hat{\tau}_{lik}, \hat{\tau}_{lik} + \Delta \right]$ and $[ \hat{\tau}_{lik} - \Delta, \hat{\tau}_{lik} )$ for some $\Delta > 0$.
The jump-adjusted MSRV was estimated as follows:
\begin{equation*}
    RV_{l,i} = \sum_{k=1}^{M} a_{k}RV_{l,i}^{K_k} + \zeta_{li} \left( RV_{l,i}^{K_1} - RV_{l,i}^{K_M} \right),
\end{equation*}
where
\begin{equation*}
    \begin{split}
        & RV_{l,i}^{K} = \frac{1}{K} \sum_{j=1}^{m_{l,i}-K}\left[ Y_{l}^{*}(t_{l,i,j+K}) - Y_{l}^{*}(t_{l,i,j}) \right]^2, \quad a_k = \frac{12(k+C)(k-M/2-1/2)}{M(M^2-1)},\\
        & \zeta_{li} = \frac{(M+C)(C+1)}{(m_{l,i}+1)(M-1)},
    \end{split}
\end{equation*}
$Y_{l}^{*}$ is jump-adjusted data, and $M$ and $C$ are some integers.
We selected $M=11$ and $C=4$, which are the same as that in \citet{fan2007multi}.
The details of estimators can be found in \citet{fan2007multi}.

\begin{figure}[!ht]
\centering
\includegraphics[width = 1\textwidth]{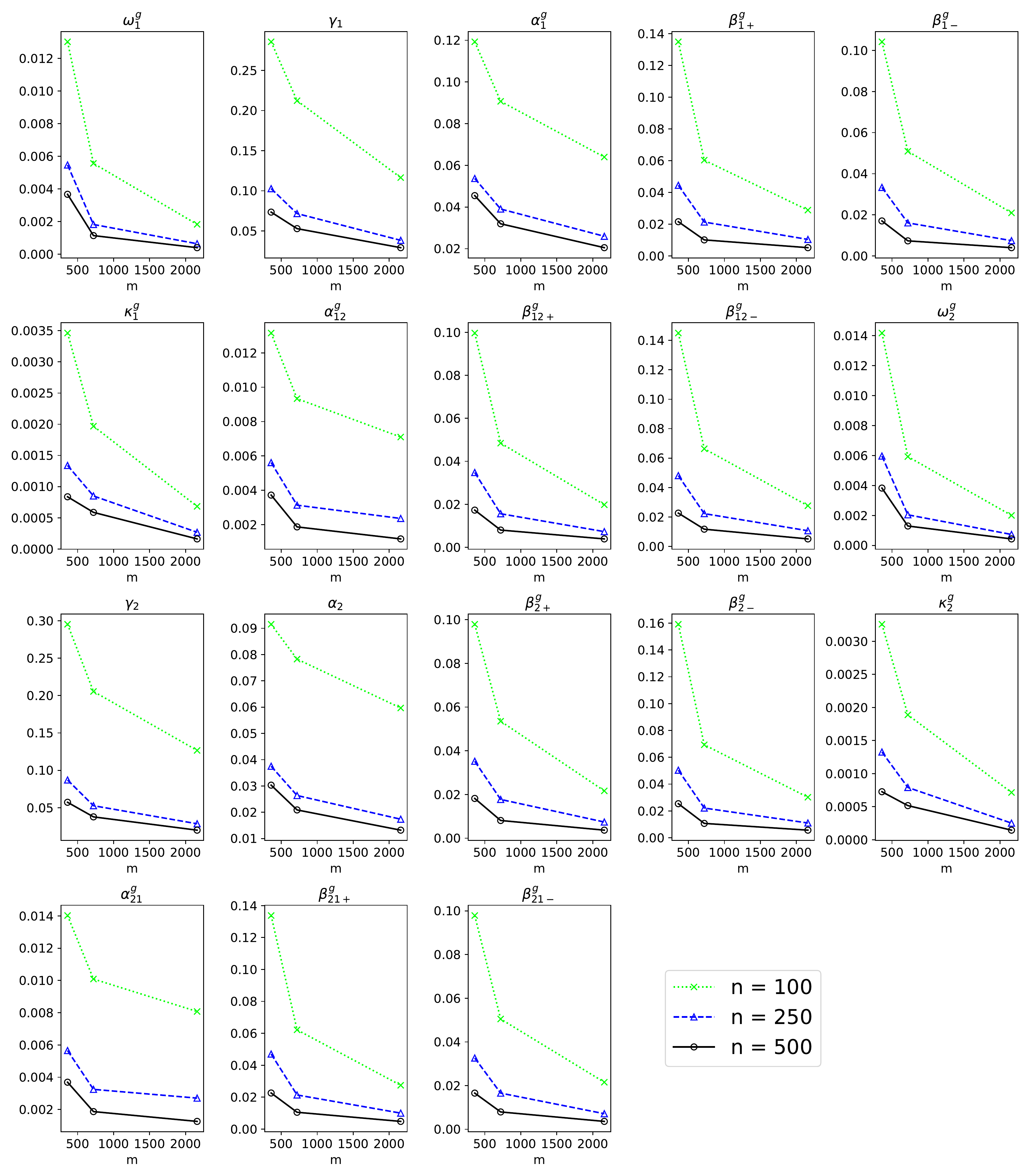}
\caption{MSEs for the proposed estimators with $n=100,250,500$ and $m=360,720,2160$.} \label{Fig-1}
\end{figure}

To verify the proposed GARCH parameter estimation procedure described in Section \ref{SEC-3}, we estimated the GARCH parameters and calculated 
mean squared errors (MSE) over 1000 simulations.  
Figure \ref{Fig-1} draws the mean squared errors (MSE) of the proposed GARCH estimators with $n = 100, 250, 500$ and $m = 360, 720, 2160$.
From Figure \ref{Fig-1}, we find that the MSEs decrease as the number of low-frequency observations or high-frequency observations increases.
This result supports the theoretical findings in Theorems \ref{theorem1} and \ref{theorem2}.

\begin{figure}[!ht]
\centering
\includegraphics[width = 1\textwidth]{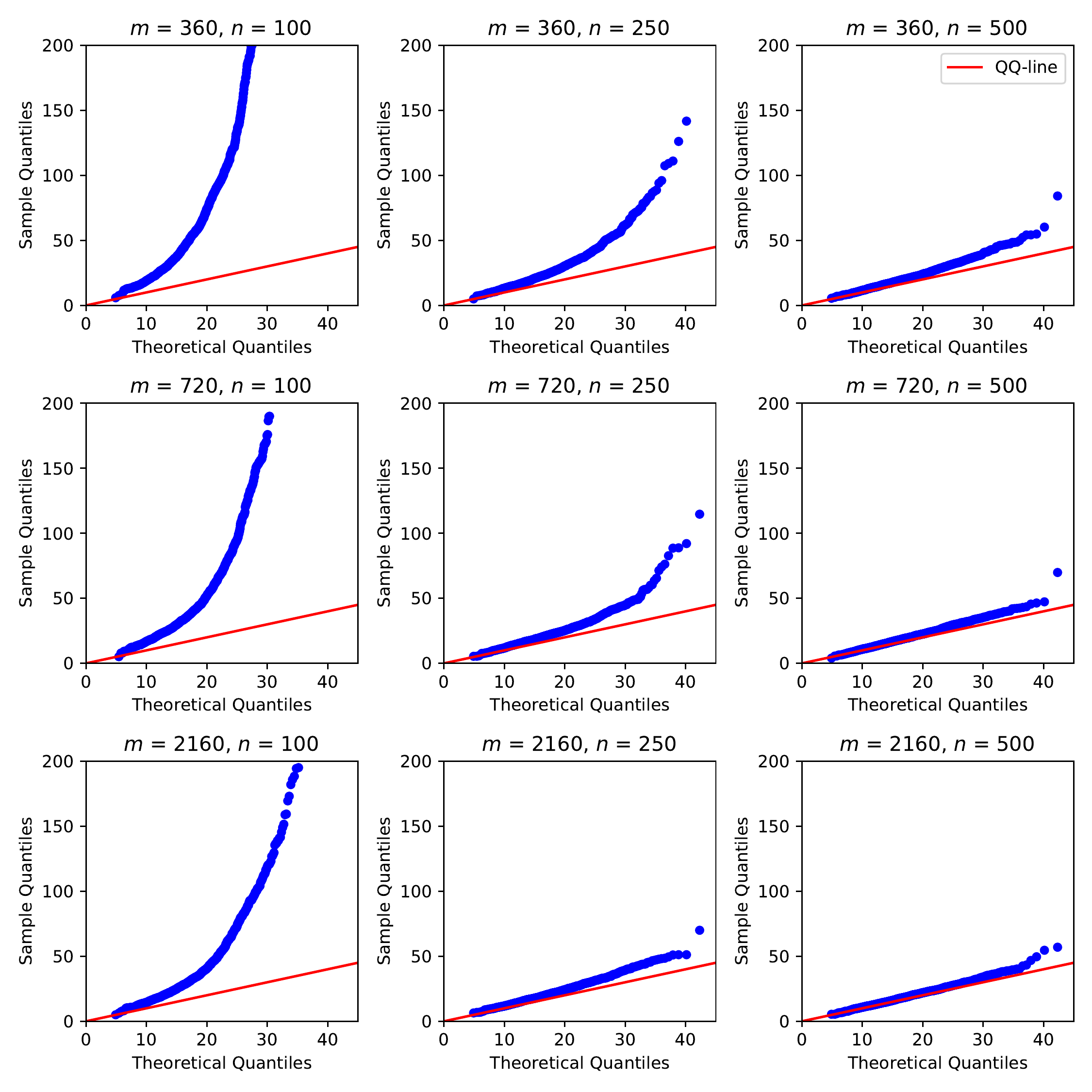}
\caption{Q-Q plots for the quantiles of the test statistic $\mathcal{W}_{n_1,n_2}(0)$ against quantiles of a $\chi^2(18)$ distribution with $n = 100,250,500$ and $m=360,720,2160$.} \label{Fig-2}
\end{figure}

\begin{figure}[!ht]
\centering
\includegraphics[width = 1\textwidth]{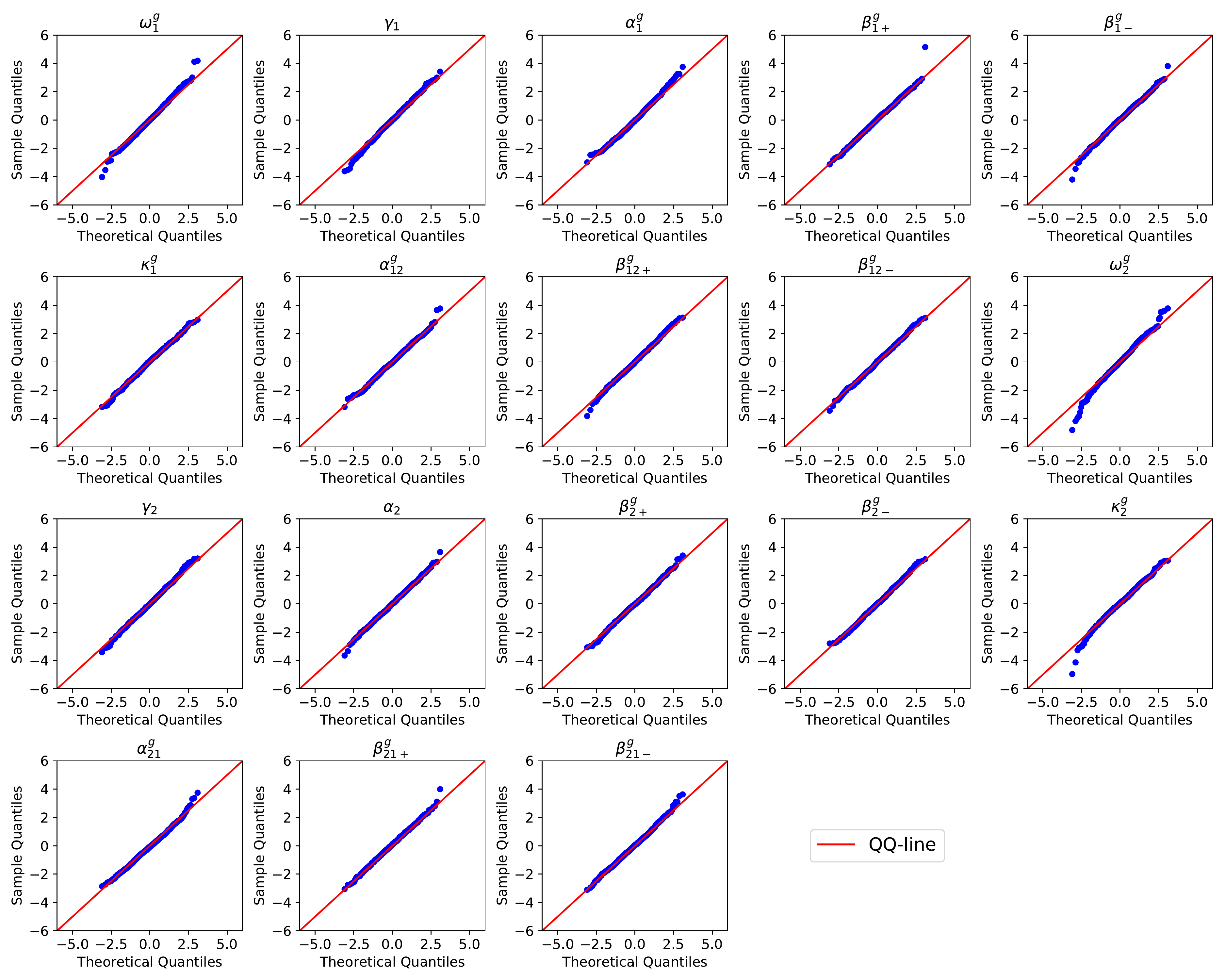}
\caption{Q-Q plots for the quantiles of the test statistic $Z_{i,n_1,n_2}(0)$ against quantiles of a $N(0,1)$ distribution with $n = 500$ and $m=2160$.} \label{Fig-3}
\end{figure}

One of the main goals of this paper is to test whether there is a structural break at the given time point.
Therefore, we checked the established asymptotic results of the proposed hypothesis tests under the null hypothesis.
We used the generated data above.
Figure \ref{Fig-2} draws the quantiles of the test statistic $\mathcal{W}_{n_1,n_2}(0)$ defined in \eqref{eq-3.1} against theoretical quantiles of a $\chi^2(18)$ distribution with $n_1 = n_2 = n$, $n=100,250,500$ and $m=360,720,2160$.
Figure \ref{Fig-3} depicts the quantiles of the test statistics $Z_{i,n_1,n_2}(0)$, $i=1,\ldots,18$, defined in \eqref{eq-3.2} against theoretical quantiles of a standard normal distribution with $n_1=n_2=500$ and $m=2160$.
The solid red lines in Figures \ref{Fig-2} and \ref{Fig-3} indicate the reference line of the asymptotic quantile-quantile (Q-Q) plots, based on the $\chi^2$ and standard normal distribution, respectively.
Figure \ref{Fig-2} shows that the test statistics $\mathcal{W}_{n_1,n_2}(0)$ become closer to the $\chi^2(18)$ distribution with the increase in the number of high-frequency observations or low-frequency observations.
From Figure \ref{Fig-3}, we find that the test statistics $Z_{i,n_1,n_2}(0)$, $i=1,\ldots,18$, with sufficiently large $n$ and $m$ almost have standard normal distribution.
These results support the theoretical findings presented in Section \ref{SEC-3}.
Thus, with the proposed test statistics, we can conduct the hypothesis tests.

\section{An empirical study} \label{SEC-5}

We applied the proposed contagion GARCH-It\^o model to the U.S. and China stock markets. 
We used the 10-seconds intra-day S\&P 500 Composite index and CSI 300 index, which represent a broad cross-section of the U.S. and China equity markets, respectively.
The indices are obtained from Tick Data Inc.
We set the break point for the U.S.--China trade war as January 22, 2018, when Trump first announced tariffs.
We selected the two periods to be one year before and after the break date--from January 22, 2017 to January 21, 2018 and from January 22, 2018 to January 21, 2019, respectively.
For non-trading dates that are trading dates for the other--for example Chinese New Year--we utilized the data from the latest trading date.
We defined the trading hours of China and U.S. stock markets from 1:30 UTC to 7:00 UTC and from 14:30 UTC to 21:00 UTC, respectively--that is, $\lambda_1 = 5.5 / 24$ and $\lambda_2 = 6.5 / 24$.
We note that the Shanghai Stock Exchange closes for lunch from 3:30 UTC to 5:00 UTC, thereby making the CSI 300 index unobservable during this period.
To handle this issue, we estimated the integrated volatilities and jump variations for the open-to-close period of the CSI 300 index as follows:
\begin{eqnarray*}
    && RV_{1,i} = RV_{AM1,i} + RV_{PM1,i} + (Y_{1,i-1+3.5/24} - Y_{1,i-1+2/24})^2,\cr
    && JV_{1,i}^{+} = JV_{AM1,i}^{+} + JV_{PM1,i}^{+}, \qquad  JV_{1,i}^{-} = JV_{AM1,i}^{-} + JV_{PM1,i}^{-},
\end{eqnarray*}
where $RV_{AM1,i}$, $JV_{AM1,i}^{+}$, and $JV_{AM1,i}^{-}$ are the jump-adjusted MSRV, positive jump variations, and negative jump variations, respectively, estimated by applying the procedure in Sections \ref{SEC-4} to morning trading session data; $RV_{PM1,i}$, $JV_{PM1,i}^{+}$, and $JV_{PM1,i}^{-}$ are estimated in the same manner, using afternoon trading session data.
$Y_{1,i-1+2/24}$ is the close log-price of the morning trading session, and $Y_{1,i-1+3.5/24}$ is the open log-price of the afternoon trading session.
We utilized the log-prices for estimating the GARCH parameters of the volatility structures.

\begin{table}[!ht]
\caption{The sample means of the $RV$, $OV$, $JV$, $JV^{+}$, and $JV^{-}$ for each period and country. The sample standard deviations are reported in parentheses. The sample means and sample standard deviations are multiplied by 10000. Further, p-values of the mean difference tests between the two periods are reported.
 }\label{Table-1}
\centering
\scalebox{1}{
\begin{tabular}{lccclccc}
\hline
\multicolumn{1}{c}{} & \multicolumn{3}{c}{China}                                                                      &  & \multicolumn{3}{c}{U.S.}                                                                         \\ \cline{2-4} \cline{6-8} 
                                  & Before   & After    & \begin{tabular}[c]{@{}c@{}}p-value\\ of the mean\\ difference test\end{tabular} &  & Before   & After    & \begin{tabular}[c]{@{}c@{}}p-value\\ of the mean\\ difference test\end{tabular} \\ \hline
$RV \times 10^4$                              & 0.3734   & 0.8924   & 0.0000                                                                   &  & 0.0823   & 0.6600   & 0.0000                                                                   \\
                                  & (0.3219) & (0.6249) &                                                                          &  & (0.0692) & (0.8845) &                                                                          \\
$OV \times 10^4$                              & 0.0633   & 0.5626   & 0.0000                                                                   &  & 0.0480   & 0.2517   & 0.0000                                                                   \\
                                  & (0.2698) & (1.4654) &                                                                          &  & (0.0900) & (0.6930) &                                                                          \\
$JV \times 10^4$                              & 0.0123   & 0.0999   & 0.0000                                                                   &  & 0.0233   & 0.0707   & 0.0000                                                                   \\
                                  & (0.0350) & (0.1151) &                                                                          &  & (0.0443) & (0.1611) &                                                                          \\
$JV^{+} \times 10^4$                          & 0.0078   & 0.0765   & 0.0000                                                                   &  & 0.0099   & 0.0335   & 0.0000                                                                   \\
                                  & (0.0199) & (0.0938) &                                                                          &  & (0.0168) & (0.0733) &                                                                          \\
$JV^{-} \times 10^4$                          & 0.0045   & 0.0234   & 0.0000                                                                   &  & 0.0134   & 0.0372   & 0.0001                                                                   \\
                                  & (0.0203) & (0.0373) &                                                                          &  & (0.0287) & (0.0943) &                                                                          \\ \hline
\end{tabular}
}
\end{table}

We first checked the magnitudes of volatilities before and after the U.S.--China trade war. 
Table \ref{Table-1} reports the sample means and standard deviations of the estimated $RV$, $OV$, $JV$, $JV^{+}$, and $JV^{-}$ for each period and country and the p-values of the mean difference tests between two periods.
From Table \ref{Table-1}, we find that the average of the realized volatilities, overnight volatilities, and signed jump variations increased after the U.S.--China trade war.
The mean difference tests for each volatility   have p-values lower than $0.001$.
Thus, we can conjecture that the U.S.--China trade war changed the volatility structure. 
We need to check the source of the structural break using the proposed contagion GARCH-It\^o model.

\begin{table}[!ht]
\caption{Estimation results and the p-values of the proposed Z-tests.
 }\label{Table-2}
\centering
\scalebox{1}{
\begin{tabular}{lrrrrrrrrr}
\hline
China                                                           & \multicolumn{1}{c}{$\omega_{1}^{g}$} & \multicolumn{1}{c}{$\gamma_1$} & \multicolumn{1}{c}{$\alpha_{1}^{g}$} & \multicolumn{1}{c}{$\beta_{1+}^{g}$} & \multicolumn{1}{c}{$\beta_{1-}^{g}$} & \multicolumn{1}{c}{$\kappa_{1}^{g}$} & \multicolumn{1}{c}{$\alpha_{12}^{g}$} & \multicolumn{1}{c}{$\beta_{12+}^{g}$} & \multicolumn{1}{c}{$\beta_{12-}^{g}$} \\ \hline
                                                                & \multicolumn{9}{c}{Before the U.S.--China trade war}                                                                                                                                                                                                                                                                                                            \\
Coefficients                                                    & 2.92e-7                              & \bftab.499                        & \bftab.518                              & \bftab-2.108                            & .207                                 & \bftab.140                              & \bftab.261                               & -.220                                 & -.468                                 \\
Standard errors                                                 & 3.84e-6                              & .066                          & .070                                & .850                                & 1.400                                & .054                                & .114                                 & .930                                 & .699                                 \\
p-values                                                      & .940                                 & .000                           & .000                                 & .013                                 & .882                                 & .009                                 & .022                                  & .813                                  & .503                                  \\
                                                                & \multicolumn{9}{c}{After the U.S.--China trade war}                                                                                                                                                                                                                                                                                                             \\
Coefficients                                                    & \bftab5.01e-5                           & \bftab.274                        & \bftab.555                              & -.284                             & -.147                                & \bftab.336                              & -.026                                 & .112                                  & \bftab.742                               \\
Standard errors                                                 & 1.45e-5                              & .072                          & .069                                & .242                                & .963                                & .107                                & .023                                 & .452                                 & .289                                 \\
p-values                                                      & .001                                 & .000                           & .000                                 & .240                                 & .879                                 & .002                                 & .272                                  & .805                                  & .010                                  \\ \hline
\begin{tabular}[c]{@{}l@{}}p-values of the\\ structural break test\end{tabular} & \bftab.001                             & \bftab.021                        & .708                                 & \bftab.039                              & .835                                 & .102                                 & \bftab.014                               & .748                                  & .109                                  \\ \hline\hline
U.S.                                                              & $\omega_{2}^{g}$                     & $\gamma_2$                     & $\alpha_{2}^{g}$                     & $\beta_{2+}^{g}$                     & $\beta_{2-}^{g}$                     & $\kappa_{2}^{g}$                     & $\alpha_{21}^{g}$                     & $\beta_{21+}^{g}$                     & $\beta_{21-}^{g}$                     \\ \hline
                                                                & \multicolumn{9}{c}{Before the U.S.--China trade war}                                                                                                                                                                                                                                                                                                            \\
Coefficients                                                    & \bftab9.79e-6                           & \bftab.181                        & \bftab.500                              & .132                                 & -.293                                & \bftab.448                              & -.010                                 & -.024                                 & .072                                  \\
Standard errors                                                 & 1.60e-6                              & .038                          & .056                                & .303                                & .216                                & .154                                & .015                                 & .083                                 & .080                                 \\
p-values                                                      & .000                                 & .000                           & .000                                 & .664                                 & .173                                 & .004                                 & .513                                  & .769                                  & .371                                  \\
                                                                & \multicolumn{9}{c}{After the U.S.--China trade war}                                                                                                                                                                                                                                                                                                             \\
Coefficients                                                    & 7.42e-6                             & \bftab.216                        & \bftab.677                              & .186                                 & .930                              & \bftab.693                              & -.008                                 & -.082                                 & -.090                                 \\
Standard errors                                                 & 6.72e-6                              & .083                          & .097                                & .478                                & .490                                & .206                                & .013                                 & .056                                 & .214                                 \\
p-values                                                      & .269                                 & .010                           & .000                                 & .698                                 & .058                                 & .001                                 & .519                                  & .142                                  & .675                                  \\ \hline
\begin{tabular}[c]{@{}l@{}}p-values of the\\ structural break test\end{tabular} & .731                                 & .697                           & .113                                 & .924                                 & \bftab.022                              & .339                                 & .936                                  & .563                                  & .480                                  \\ \hline
\multicolumn{10}{l}{\small NOTE: Significant coefficients at 5\% and p-values below 0.05 are presented in bold.}\\
\end{tabular}
}
\end{table}

To check whether the volatility structures were changed, we estimated parameters of the proposed model and conducted structural break tests.
The p-values of the simultaneous structural break tests for the GARCH parameters of China, U.S., and both are below 0.001.
Thus, we conclude that the U.S.--China trade war caused the structural break.
To further check the channel of the structural break, we conducted individual tests.
Table \ref{Table-2} reports the results of the proposed estimators and structural break tests.
In case of the Chinese GARCH parameters estimated from data before the structural break, $\gamma_{1}$, $\alpha_{1}^g$, $\beta_{1+}^g$, $\kappa_{1}^g$, and $\alpha_{12}^g$ are significant.
The negative sign of  $\beta_{1+}^g$ implies that the positive news during the Chinese trading hours reduced future risk in the Chinese market.
The positive sign of  $\alpha_{12}^g$ implies that the U.S. market risk increased future Chinese market risk.
In other words, the realized volatility of the U.S. is the channel of the risk contagion before the U.S.--China trade war. 
For estimated Chinese parameters during the trade war period, $\omega_{1}^g$, $\gamma_{1}$, $\alpha_{1}^g$, $\kappa_{1}^g$, and $\beta_{12-}^g$ are significant.
The positive sign of  $\beta_{12-}^g$ implies that bad news released in U.S. trading hours leads to an increase in future Chinese financial risk.
Unlike the result before the U.S.--China trade war, the contagion channel is the negative jump during the trade war. 
In other words, the bad news released in the U.S. trading hours significantly affect the China market. 
Furthermore, among the Chinese GARCH parameters, the p-values of the structural break tests for $\omega_{1}^g$, $\gamma_{1}^g$, $\beta_{1+}^g$, and $\alpha_{12}^g$ are below 0.05.
Decreased $\gamma_1$ implies that the effect of the risk factors decayed faster.
Further, the positive effect of good news during Chinese trade hours was weakened for the trade war.
The effect of U.S. market risk on the Chinese market was diluted for the trade war, while the effect of overnight volatility and negative news increased with p-values $0.102$ and $0.109$, respectively.
On the other hand, in case of U.S. parameters, only the GARCH parameters for the U.S. risk factors are significant. 
We note that the estimated $\beta_{2-}^g$ for the trade war period is positive and significant, with p-value $0.058$ and p-value from its structural break test $0.022$.
This implies that for the trade war period, the effect of negative news in U.S. trading hours increased future risk in the U.S. market.
From these results, we can conclude that the U.S. and China stock markets experienced a structural break from the U.S.--China trade war, and negative news announced during the U.S. trading hours is a key ingredient of these structural breaks.

\subsection{Robustness analysis}

\subsubsection{Realized volatility measures}

To check the validity of our empirical analysis, we investigated whether the findings in the previous section are robust to realized volatility measures.
In Sections \ref{SEC-4} and \ref{SEC-5}, we employed the jump-adjusted MSRV, which is the consistent and efficient estimator for integrated volatility under some bounded fourth-moment assumption in the presence of i.i.d. microstructure noises.
However, we often observe that stock return data are  heavy-tailed.
Recently, to deal with  heavy-tailed observations which have the finite $a$th moment for $a<4$, \citet{shin2021adaptive}  proposed the adaptive robust pre-averaging realized volatility (ARP) estimator.
They employed pre-averaging and truncation schemes  as follows:
\begin{equation*}
    ARP_{l,i} = \frac{1}{(m_{l,i}-K_{l,i})\vartheta_{li}}\sum_{k=1}^{m_{l,i}-K_{l,i}}\Psi_{\hat{a}_{li}}\left( \vartheta_{li} Q_{l}(\tau_{l,i,k}) \right) - \frac{\zeta}{\varphi_{li} \vartheta^{*}_{li} K_{l,i}} \sum_{k=1}^{m_{l,i}-1} \Psi_{\hat{a}_{li}} \left( \vartheta^{*}_{li} Q_{l}^{*}(\tau_{l,i,k}) \right),
\end{equation*}
where $K_{l,i} = [m_{l,i}^{1/2}]$, $Q_{l}(\tau_{l,i,k})=\frac{m_{l,i}-K_{l,i}}{\varphi_{li} K_{l,i}}\left\{\bar{Y}^{*}_{l}(\tau_{l,i,k};K_{l,i})\right\}^2$, $\varphi_{li} = \frac{1}{K_{l,i}}\sum_{k=0}^{K_{l,i} - 1} g^2(\frac{k}{K_{l,i}})$, $g(x) = x \land (1-x)$, $\bar{Y}^{*}_{l}(\tau_{l,i,k};w) = \sum_{r=0}^{w-1}g\left(\frac{r}{w}\right) \{ Y^{*}_{l}(\tau_{l,i,k+r+1}) - Y^{*}_{l}(\tau_{l,i,k+r}) \}$,
$Q^{*}_{l}(\tau_{l,i,k}) = \frac{1}{2}\{ Y^{*}_{l}(\tau_{l,i,k+1}) - Y^{*}_{l}(\tau_{l,i,k}) \}^2$,
$\Psi_{a}(x) = -\sgn(x) \log \{1 - (|x| \land t_a) + c_a (|x| \land t_a)^a \}$, $c_a = \frac{(a-1)}{a} \lor \sqrt{(2-a)/a} $, and $t_a = (1/ac_a)^{1/(a-1)}$;
$\vartheta_{li}$ and $\vartheta_{li}^{*}$ are truncation parameters; $\hat{a}_{li}$ is the estimated order of the highest finite moment for the continuous part of $Q_{l}(\tau_{l,i,k})$.
We chose the tuning parameters as follows:
\begin{equation*}
    \vartheta_{li} = c\left( \frac{K_{l,i}}{(\hat{a}_{li}-1)c_{\hat{a}_{li}}\hat{S}_{li}(m_{l,i} - K_{l,i})}\right)^{1/\hat{a}_{li}} \quad \text{and} \quad \vartheta_{li}^{*} = c\left( \frac{1}{(\hat{a}_{li}-1)c_{\hat{a}_{li}}\hat{S}^{*}_{li}(m_{l,i} - 1)}\right)^{1/\hat{a}_{li}},
\end{equation*}
where $\hat{S}_{li} = \frac{1}{m_{l,i}-K_{l,i}} \sum_{k=1}^{m_{l,i}-K_{l,i}} \left \{ |Q_{l}(\tau_{l,i,l})|^{\hat{a}_{li}}  \right \} $, $\hat{S}^{*}_{li} = \frac{1}{m_{l,i}-1} \sum_{k=1}^{m_{l,i}-1} \left \{ |Q_{l}^{*}(\tau_{l,i,l})|^{\hat{a}_{li}}  \right \}$, and $c=0.15$.
The details of estimators can be found in \citet{shin2021adaptive}.
On the other hand, \citet{li2021robust} indicated that microstructure noise in high-frequency data  has prominent intraday patterns, such as U-shape or reverse J-shape, in the scale of the noise. 
To handle these non-i.i.d. microstructure noise structures, they developed the pre-averaging--ReMeDI (PaReMeDI) estimator.
They employ the pre-averaging method and the ReMeDI approach \citep{li2021remedi} to eliminate the effect of the microstructure noise which is autocorrelated and dependent on the stock price.
Specifically, jump-adjusted PaReMeDI were estimated as follows:
\begin{equation*}
    PaReMeDI_{l,i} = \frac{1}{w_{li} \phi_{li}}\sum_{k=0}^{m_{l,i}-w_{li}} \left\{ \bar{Y}^{*}_{l}(\tau_{l,i,k};w_{li}) \right\}^2 - \frac{1}{w_{li}^2\phi_{li}}\sum_{k=k_{li}}^{m_{l,i} - w_{li}} \sum_{|\varsigma| \leq \varsigma_{li} } \bar{\phi}_{li\varsigma}\Delta_{k,\varsigma}^{k_{li}}Y_{l}^{*},
\end{equation*}
where $\Delta_{k,\varsigma}^{k_{li}}Y_{l}^{*} = \{ Y_{l}^{*}(\tau_{l,i,k+\varsigma}) - Y_{l}^{*}(\tau_{l,i,k+\varsigma + k_{li}}) \} \{Y_{l}^{*}(\tau_{l,i,k}) - Y_{l}^{*}(\tau_{l,i,k - k_{li}}) \}$,
$\phi_{li} = \frac{1}{w_{li}} \sum_{k=1}^{w_{li}} g^2(\frac{k}{w_{li}})$,
$\bar{\phi}_{li\varsigma} = w_{li} \sum_{k=\varsigma}^{w_{li}} \{ g ( \frac{k+1}{w_{li}}  ) - g ( \frac{k}{w_{li}}  )   \} \{ g ( \frac{k-\varsigma+1}{w_{li}}  ) - g ( \frac{k - \varsigma}{w_{li}}  )   \} $,
$k_{li} = 10$, $\varsigma_{li} = [ m_{l,i}^{1/7}  ]$, $w_{li} = [ \varpi_{li} m_{l,i}^{-1/2}  ] $, $\varpi_{li} = \frac{K_{\Phi}\sqrt{R_{li}}}{\sqrt{RV_{l,i}}} $,
$R_{li} = \frac{1}{m_{l,i}}  \sum_{k=k_{li}}^{m_{l,i}-\varsigma-k_{li}} \sum_{|\varsigma|\leq \varsigma_{li}} \Delta_{k,\varsigma}^{k_{li}} Y_{l}^{*} $, and $K_{\Phi} = 4.78$.
The details of estimators can be found in \citet{li2021robust}.
In sum, the ARP estimator is robust to the heavy-tailedness structure of observed data, while the  PaReMeDI  estimator is robust to the dependent structure of the microstructure noise.

\begin{table}[!ht]
\caption{Robustness analysis for the realized volatility measures, adopting the jump-adjusted ARP method as the integrated volatility estimator.
}\label{robust-GARCH-ARP}
\centering
\scalebox{1}{
\begin{tabular}{lrrrrrrrrr}
\hline
China                                                                            & \multicolumn{1}{c}{$\omega_{1}^{g}$} & \multicolumn{1}{c}{$\gamma_1$} & \multicolumn{1}{c}{$\alpha_{1}^{g}$} & \multicolumn{1}{c}{$\beta_{1+}^{g}$} & \multicolumn{1}{c}{$\beta_{1-}^{g}$} & \multicolumn{1}{c}{$\kappa_{1}^{g}$} & \multicolumn{1}{c}{$\alpha_{12}^{g}$} & \multicolumn{1}{c}{$\beta_{12+}^{g}$} & \multicolumn{1}{c}{$\beta_{12-}^{g}$} \\ \hline
                                                                                 & \multicolumn{9}{c}{Before the U.S.--China trade war}                                                                                                                                                                                                                                                                                                      \\
Coefficients                                                                     & 8.85e-7                              & \bftab.536                     & \bftab.463                           & \bftab-1.433                         & .812                                 & \bftab.085                           & \bftab.241                            & -.221                                 & -.331                                 \\
Standard errors                                                                  & 3.15e-6                              & .060                           & .062                                 & .665                                 & 1.052                                & .032                                 & .105                                  & .819                                  & .577                                  \\
p-values                                                                         & .779                                 & .000                           & .000                                 & .031                                 & .440                                 & .008                                 & .022                                  & .788                                  & .567                                  \\
                                                                                 & \multicolumn{9}{c}{After the U.S.--China trade war}                                                                                                                                                                                                                                                                                                       \\
Coefficients                                                                     & \bftab3.60e-5                        & \bftab.331                     & \bftab.518                           & -.068                                & -.527                                & \bftab.213                           & -.021                                 & .013                                  & \bftab.627                            \\
Standard errors                                                                  & 1.12e-5                              & .070                           & .063                                 & .161                                 & .605                                 & .078                                 & .020                                  & .375                                  & .239                                  \\
p-values                                                                         & .001                                 & .000                           & .000                                 & .674                                 & .384                                 & .006                                 & .277                                  & .972                                  & .009                                  \\ \hline
\begin{tabular}[c]{@{}l@{}}p-values of the\\ structural break test \end{tabular} & \bftab.003                           & \bftab.027                     & .530                                 & \bftab.046                           & .270                                 & .127                                 & \bftab.014                            & .795                                  & .126                                  \\ \hline\hline
U.S.                                                                             & $\omega_{1}^{g}$                     & $\gamma_1$                     & $\alpha_{1}^{g}$                     & $\beta_{1+}^{g}$                     & $\beta_{1-}^{g}$                     & $\kappa_{1}^{g}$                     & $\alpha_{12}^{g}$                     & $\beta_{12+}^{g}$                     & $\beta_{12-}^{g}$                     \\ \hline
                                                                                 & \multicolumn{9}{c}{Before the U.S.--China trade war}                                                                                                                                                                                                                                                                                                      \\
Coefficients                                                                     & \bftab9.88e-6                        & \bftab.150                     & \bftab.462                           & .237                                 & -.189                                & \bftab.369                           & -.012                                 & -.036                                 & .061                                  \\
Standard errors                                                                  & 1.51e-6                              & .039                           & .054                                 & .269                                 & .188                                 & .133                                 & .015                                  & .071                                  & .056                                  \\
p-values                                                                         & .000                                 & .000                           & .000                                 & .379                                 & .315                                 & .005                                 & .416                                  & .618                                  & .278                                  \\
                                                                                 & \multicolumn{9}{c}{After the U.S.--China trade war}                                                                                                                                                                                                                                                                                                       \\
Coefficients                                                                     & 5.64e-6                              & \bftab.242                     & \bftab.640                           & .283                                 & \bftab.891                           & \bftab.633                           & -.007                                 & -.069                                 & -.117                                 \\
Standard errors                                                                  & 5.33e-6                              & .080                           & .090                                 & .430                                 & .416                                 & .201                                 & .014                                  & .044                                  & .176                                  \\
p-values                                                                         & .290                                 & .002                           & .000                                 & .510                                 & .032                                 & .002                                 & .629                                  & .114                                  & .507                                  \\ \hline
\begin{tabular}[c]{@{}l@{}}p-values of the\\ structural break test \end{tabular} & .444                                 & .299                           & .089                                 & .927                                 & \bftab.018                           & .274                                 & .778                                  & .688                                  & .336                                  \\ \hline
\multicolumn{10}{l}{\small NOTE: Significant coefficients at 5\% and p-values below 0.05 are presented in bold.}\\
\end{tabular}
}
\end{table}
\begin{table}[!ht]
\caption{Robustness analysis for the realized volatility measures, adopting the jump-adjusted PaReMeDI method as integrated volatility estimator.
}\label{robust-GARCH-PaReMeDI}
\centering
\scalebox{1}{
\begin{tabular}{lrrrrrrrrr}
\hline
China                                                                            & \multicolumn{1}{c}{$\omega_{1}^{g}$} & \multicolumn{1}{c}{$\gamma_1$} & \multicolumn{1}{c}{$\alpha_{1}^{g}$} & \multicolumn{1}{c}{$\beta_{1+}^{g}$} & \multicolumn{1}{c}{$\beta_{1-}^{g}$} & \multicolumn{1}{c}{$\kappa_{1}^{g}$} & \multicolumn{1}{c}{$\alpha_{12}^{g}$} & \multicolumn{1}{c}{$\beta_{12+}^{g}$} & \multicolumn{1}{c}{$\beta_{12-}^{g}$} \\ \hline
                                                                                 & \multicolumn{9}{c}{Before the U.S.--China trade war}                                                                                                                                                                                                                                                                                                      \\
Coefficients                                                                     & 7.67e-7                             & \bftab.551                     & \bftab.453                           & \bftab-2.346                         & .152                                 & \bftab.163                           & \bftab.318                            & .281                                  & -1.090                                \\
Standard errors                                                                  & 5.86e-6                             & .071                           & .067                                 & .998                                 & 1.925                                & .064                                 & .137                                  & 1.403                                 & .944                                  \\
p-values                                                                         & .896                                 & .000                           & .000                                 & .019                                 & .937                                 & .011                                 & .021                                  & .841                                  & .248                                  \\
                                                                                 & \multicolumn{9}{c}{After the U.S.--China trade war}                                                                                                                                                                                                                                                                                                       \\
Coefficients                                                                     & \bftab6.99e-5                       & \bftab.261                     & \bftab.533                           & -.223                                & .273                                 & \bftab.497                           & -.025                                 & -.155                                 & \bftab.883                            \\
Standard errors                                                                  & 1.87e-5                             & .073                           & .074                                 & .329                                 & 1.403                                & .146                                 & .025                                  & .615                                  & .414                                  \\
p-values                                                                         & .000                                 & .000                           & .000                                 & .498                                 & .846                                 & .001                                 & .313                                  & .801                                  & .033                                  \\ \hline
\begin{tabular}[c]{@{}l@{}}p-values of the\\ structural break test \end{tabular} & \bftab.000                           & \bftab.004                     & .423                                 & \bftab.043                           & .960                                 & \bftab.036                           & \bftab.014                            & .776                                  & .056                                  \\ \hline\hline
U.S.                                                                             & $\omega_{1}^{g}$                     & $\gamma_1$                     & $\alpha_{1}^{g}$                     & $\beta_{1+}^{g}$                     & $\beta_{1-}^{g}$                     & $\kappa_{1}^{g}$                     & $\alpha_{12}^{g}$                     & $\beta_{12+}^{g}$                     & $\beta_{12-}^{g}$                     \\ \hline
                                                                                 & \multicolumn{9}{c}{Before the U.S.--China trade war}                                                                                                                                                                                                                                                                                                      \\
Coefficients                                                                     & \bftab1.37e-5                       & \bftab.110                     & \bftab.501                           & .566                                 & \bftab-.527                          & \bftab.651                           & -.009                                 & -.063                                 & .116                                  \\
Standard errors                                                                  & 1.80e-6                             & .019                           & .047                                 & .352                                 & .244                                 & .187                                 & .010                                  & .107                                  & .098                                  \\
p-values                                                                         & .000                                 & .000                           & .000                                 & .107                                 & .031                                 & .000                                 & .358                                  & .557                                  & .238                                  \\
                                                                                 & \multicolumn{9}{c}{After the U.S.--China trade war}                                                                                                                                                                                                                                                                                                       \\
Coefficients                                                                     & 8.52e-6                             & \bftab.228                     & \bftab.662                           & .114                                 & .934                                 & \bftab1.001                          & -.008                                 & -.090                                 & -.013                                 \\
Standard errors                                                                  & 9.01e-6                             & .087                           & .095                                 & .541                                 & .595                                 & .278                                 & .012                                  & .059                                  & .283                                  \\
p-values                                                                         & .344                                 & .009                           & .000                                 & .833                                 & .116                                 & .000                                 & .500                                  & .130                                  & .963                                  \\ \hline
\begin{tabular}[c]{@{}l@{}}p-values of the\\ structural break test \end{tabular} & .575                                 & .184                           & .128                                 & .484                                 & \bftab.023                           & .296                                 & .950                                  & .826                                  & .666                                  \\ \hline
\multicolumn{10}{l}{\small NOTE: Significant coefficients at 5\% and p-values below 0.05 are presented in bold.}\\
\end{tabular}
}
\end{table}

To assess robustness against realized volatility measures, we conducted our empirical analysis using jump-adjusted ARP and PaReMeDI estimators as  the integrated volatility estimator.
Tables \ref{robust-GARCH-ARP} and \ref{robust-GARCH-PaReMeDI} report the results of the estimation and the structural break test, adopting different realized volatility measures, such as jump-adjusted ARP and PaReMeDI, respectively.
From Tables \ref{robust-GARCH-ARP} and \ref{robust-GARCH-PaReMeDI}, we find that our main results in Section \ref{SEC-5} change remarkably little with the alternative realized volatility measures.
Thus, we can conclude that the empirical findings are robust to the realized volatility measures.

\subsubsection{Model specification}

While GARCH models are widely used to model volatility dynamics, the HAR models \citep{corsi2009simple} have some benefits--for example,  the HAR models can reflect the differences of agents' risk profiles and long memory.
Thus, we considered the HAR-type contagion model to assess robustness against the model specification.
Similar to the result of Theorem \ref{theorem1}, we defined the HAR-type contagion model as follows:
\begin{eqnarray*}
    H_{l,i} &=& \omega_{l} + \sum_{d\in\{1,5,22\}}\alpha_{l}^{(d)}\frac{RV_{l,i-1}^{(d)}}{\lambda_l} + \beta_{l+}\frac{JV_{l,i-1}^{+}}{\lambda_{l}} + \beta_{l-}\frac{JV_{l,i-1}^{-}}{\lambda_{l}} + \kappa_{l}\frac{OV_{l,i-1}}{1-\lambda_{l}}\cr
    && + \alpha_{ll'}\frac{RV_{l',i-1+\iota_l}}{\lambda_{l'}} + \beta_{ll'+}\frac{JV_{l',i-1+\iota_l}^{+}}{\lambda_{l'}} + \beta_{ll'-}\frac{JV_{l',i-1+\iota_l}^{-}}{\lambda_{l'}},
\end{eqnarray*}
where $H_{l,i} = E\left[ RV_{l,i}/\lambda_l | \mathcal{F}_{\tau_l(i)-1} \right]$, $RV_{l,i-1}^{(d)} = \frac{1}{d}\sum_{j=i-d}^{i-1}RV_{l,j}$, $\iota_1 = 0$, and $\iota_2 = 1$.
Similar to Section \ref{SEC-3}, we employed quasi-likelihood estimation procedures to obtain the difference between the estimated model parameters in each period and the consistent estimator of its asymptotic variance, which are required to conduct structural break tests.
Then, we analyzed the volatility dynamics based on the HAR-type model. 

\begin{table}[!ht]
\caption{Robustness analysis against the model specification with  the HAR-type contagion model,  adopting the jump-adjusted MSRV method as the integrated volatility estimator.
}\label{robust-HAR-MSRV}
\centering
\scalebox{1}{
\begin{tabular}{lrrrrrrrrrr}
\hline
China                                                                           & \multicolumn{1}{c}{$\omega_{1}$} & \multicolumn{1}{c}{$\alpha_{1}^{(1)}$} & \multicolumn{1}{c}{$\alpha_{1}^{(5)}$} & \multicolumn{1}{c}{$\alpha_{1}^{(22)}$} & \multicolumn{1}{c}{$\beta_{1+}$} & \multicolumn{1}{c}{$\beta_{1-}$} & \multicolumn{1}{c}{$\kappa_{1}$} & \multicolumn{1}{c}{$\alpha_{12}$} & \multicolumn{1}{c}{$\beta_{12+}$} & \multicolumn{1}{c}{$\beta_{12-}$} \\ \hline
                                                                                & \multicolumn{10}{c}{Before the U.S.--China trade war}                                                                                                                                                                                                                                                                                                                             \\
Coefficients                                                                    & 4.7e-6                           & \bftab.47                              & \bftab.45                              & .10                                     & \bftab-2.57                      & .49                              & \bftab.23                        & .29                               & .34                               & -1.29                             \\
Standard errors                                                                 & 8.4e-6                           & .09                                    & .18                                    & .10                                     & 1.04                             & 1.53                             & .05                              & .16                               & 1.08                              & .75                               \\
p-values                                                                        & .58                              & .00                                    & .01                                    & .33                                     & .01                              & .75                              & .00                              & .07                               & .75                               & .09                               \\
                                                                                & \multicolumn{10}{c}{After the U.S.--China trade war}                                                                                                                                                                                                                                                                                                                              \\
Coefficients                                                                    & \bftab5.0e-5                     & \bftab.41                              & \bftab.39                              & -.00                                    & -.12                             & .27                              & \bftab.39                        & -.01                              & .04                               & \bftab.65                         \\
Standard errors                                                                 & 2.0e-5                           & .07                                    & .09                                    & .06                                     & .28                              & .95                              & .12                              & .03                               & .47                               & .32                               \\
p-values                                                                        & .01                              & .00                                    & .00                                    & .99                                     & .66                              & .78                              & .00                              & .66                               & .93                               & .04                               \\ \hline
\begin{tabular}[c]{@{}l@{}}p-values of the\\ structural break test\end{tabular} & \bftab.04                        & .60                                    & .75                                    & .41                                     & \bftab.02                        & .90                              & .22                              & .06                               & .80                               & \bftab.02                         \\ \hline\hline
U.S.                                                                            & \multicolumn{1}{c}{$\omega_{2}$} & \multicolumn{1}{c}{$\alpha_{2}^{(1)}$} & \multicolumn{1}{c}{$\alpha_{2}^{(5)}$} & \multicolumn{1}{c}{$\alpha_{2}^{(22)}$} & \multicolumn{1}{c}{$\beta_{2+}$} & \multicolumn{1}{c}{$\beta_{2-}$} & \multicolumn{1}{c}{$\kappa_{2}$} & \multicolumn{1}{c}{$\alpha_{21}$} & \multicolumn{1}{c}{$\beta_{21+}$} & \multicolumn{1}{c}{$\beta_{21-}$} \\ \hline
                                                                                & \multicolumn{10}{c}{Before the U.S.--China trade war}                                                                                                                                                                                                                                                                                                                             \\
Coefficients                                                                    & \bftab9.4e-6                     & \bftab.47                              & .21                                    & .03                                     & .62                              & -.60                             & \bftab.46                        & \bftab-.01                        & -.01                              & .06                               \\
Standard errors                                                                 & 3.5e-6                           & .11                                    & .11                                    & .13                                     & .64                              & .35                              & .21                              & .01                               & .06                               & .04                               \\
p-values                                                                        & .01                              & .00                                    & .06                                    & .82                                     & .33                              & .09                              & .03                              & .03                               & .89                               & .17                               \\
                                                                                & \multicolumn{10}{c}{After the U.S.--China trade war}                                                                                                                                                                                                                                                                                                                              \\
Coefficients                                                                    & 8.8e-6                           & \bftab.70                              & \bftab.18                              & .03                                     & -.09                             & .92                              & \bftab.62                        & -.02                              & -.05                              & -.02                              \\
Standard errors                                                                 & 5.4e-6                           & .10                                    & .09                                    & .05                                     & .51                              & .52                              & .21                              & .01                               & .07                               & .18                               \\
p-values                                                                        & .10                              & .00                                    & .04                                    & .61                                     & .85                              & .08                              & .00                              & .23                               & .43                               & .92                               \\ \hline
\begin{tabular}[c]{@{}l@{}}p-values of the\\ structural break test\end{tabular} & .93                              & .12                                    & .86                                    & .99                                     & .38                              & \bftab.02                        & .60                              & .78                               & .62                               & .69                               \\ \hline
\multicolumn{10}{l}{\small NOTE: Significant coefficients at 5\% and p-values below 0.05 are presented in bold.}\\
\end{tabular}
}
\end{table}

Table \ref{robust-HAR-MSRV} reports the estimation results of the HAR-type contagion model and structural break tests, adopting the jump-adjusted MSRV as the integrated volatility estimator.
From Table \ref{robust-HAR-MSRV}, we find that the estimated $\alpha_{12}$ before the U.S.--China trade war and $\beta_{12-}$ during the trade war period have positive signs with p-values below $0.1$.
Moreover, the sign of $\beta_{2-}$ has changed from negative to positive with p-values below $0.1$.
These results imply that the channel of the risk contagion from China to the U.S. has changed from integrated volatility to negative jump variation and negative news announced during the U.S. trading hours has become a factor that increases subsequent market risk of both countries, which is consistent to the results in Section \ref{SEC-5}.
We note that similar results can be obtained with different realized measures, such as ARP and PaReMeDI methods, under the HAR-type contagion model.
See  Tables \ref{robust-HAR-ARP} and \ref{robust-HAR-PaReMeDI} in the Appendix.
These results show that our empirical findings are robust to the realized volatility measures and the model specification.

\section{Conclusions} \label{SEC-Conclusions}

In this paper, we developed a novel contagion GARCH-It\^o model to investigate the volatility contagion structure of two countries that have disjoint trading hours.
We proposed a quasi-likelihood estimation procedure and establish its asymptotic properties. 
Based on these asymptotic properties, we proposed hypothesis test procedures to check whether there was a structural break in the proposed volatility contagion model with a known single break point.
The empirical results showed that the Chinese stock market was more affected by the U.S. stock market than the reverse in terms of risk contagion.
We documented that both countries have the structural break from the U.S.--China trade war, and negative news announced during the U.S. trading hours is a key ingredient of the structural breaks.

 

\bibliography{myReferences}

\begin{thebibliography}{}

\bibitem[A{\"\i}t-Sahalia et~al., 2015]{ait2015modeling}
A{\"\i}t-Sahalia, Y., Cacho-Diaz, J., and Laeven, R.~J. (2015).
\newblock Modeling financial contagion using mutually exciting jump processes.
\newblock {\em Journal of Financial Economics}, 117(3):585--606.

\bibitem[A{\"\i}t-Sahalia et~al., 2010]{ait2010high}
A{\"\i}t-Sahalia, Y., Fan, J., and Xiu, D. (2010).
\newblock High-frequency covariance estimates with noisy and asynchronous
  financial data.
\newblock {\em Journal of the American Statistical Association},
  105(492):1504--1517.

\bibitem[A{\"\i}t-Sahalia et~al., 2012]{ait2012testing}
A{\"\i}t-Sahalia, Y., Jacod, J., and Li, J. (2012).
\newblock Testing for jumps in noisy high frequency data.
\newblock {\em Journal of Econometrics}, 168(2):207--222.

\bibitem[Amiti et~al., 2020]{amiti2020effect}
Amiti, M., Kong, S.~H., and Weinstein, D. (2020).
\newblock The effect of the us-china trade war on us investment.
\newblock Technical report, National Bureau of Economic Research.

\bibitem[Amiti et~al., 2019]{amiti2019impact}
Amiti, M., Redding, S.~J., and Weinstein, D.~E. (2019).
\newblock The impact of the 2018 tariffs on prices and welfare.
\newblock {\em Journal of Economic Perspectives}, 33(4):187--210.

\bibitem[Andersen and Bollerslev, 1997a]{andersen1997heterogeneous}
Andersen, T.~G. and Bollerslev, T. (1997a).
\newblock Heterogeneous information arrivals and return volatility dynamics:
  Uncovering the long-run in high frequency returns.
\newblock {\em The journal of Finance}, 52(3):975--1005.

\bibitem[Andersen and Bollerslev, 1997b]{andersen1997intra-day}
Andersen, T.~G. and Bollerslev, T. (1997b).
\newblock Intraday periodicity and volatility persistence in financial markets.
\newblock {\em Journal of empirical finance}, 4(2-3):115--158.

\bibitem[Andersen and Bollerslev, 1998a]{andersen1998skeptics}
Andersen, T.~G. and Bollerslev, T. (1998a).
\newblock Answering the skeptics: Yes, standard volatility models do provide
  accurate forecasts.
\newblock {\em International Economic Review}, 39(4):885--905.

\bibitem[Andersen and Bollerslev, 1998b]{andersen1998deutsche}
Andersen, T.~G. and Bollerslev, T. (1998b).
\newblock Deutsche mark-dollar volatility: Intraday activity patterns,
  macroeconomic announcements, and longer run dependencies.
\newblock {\em The journal of Finance}, 53(1):219--265.

\bibitem[Andersen et~al., 2007]{andersen2007roughing}
Andersen, T.~G., Bollerslev, T., and Diebold, F.~X. (2007).
\newblock Roughing it up: Including jump components in the measurement,
  modeling, and forecasting of return volatility.
\newblock {\em The review of economics and statistics}, 89(4):701--720.

\bibitem[Andersen et~al., 2003]{andersen2003modeling}
Andersen, T.~G., Bollerslev, T., Diebold, F.~X., and Labys, P. (2003).
\newblock Modeling and forecasting realized volatility.
\newblock {\em Econometrica}, 71(2):579--625.

\bibitem[Andrews, 1992]{andrews1992generic}
Andrews, D.~W. (1992).
\newblock Generic uniform convergence.
\newblock {\em Econometric theory}, 8(2):241--257.

\bibitem[Barndorff-Nielsen et~al., 2008]{barndorff2008designing}
Barndorff-Nielsen, O.~E., Hansen, P.~R., Lunde, A., and Shephard, N. (2008).
\newblock Designing realized kernels to measure the ex post variation of equity
  prices in the presence of noise.
\newblock {\em Econometrica}, 76(6):1481--1536.

\bibitem[Barndorff-Nielsen and Shephard, 2006]{barndorff2006econometrics}
Barndorff-Nielsen, O.~E. and Shephard, N. (2006).
\newblock Econometrics of testing for jumps in financial economics using
  bipower variation.
\newblock {\em Journal of financial Econometrics}, 4(1):1--30.

\bibitem[Bollerslev, 1986]{bollerslev1986generalized}
Bollerslev, T. (1986).
\newblock Generalized autoregressive conditional heteroskedasticity.
\newblock {\em Journal of econometrics}, 31(3):307--327.

\bibitem[Corsi, 2009]{corsi2009simple}
Corsi, F. (2009).
\newblock A simple approximate long-memory model of realized volatility.
\newblock {\em Journal of Financial Econometrics}, 7(2):174--196.

\bibitem[Corsi et~al., 2010]{corsi2010threshold}
Corsi, F., Pirino, D., and Reno, R. (2010).
\newblock Threshold bipower variation and the impact of jumps on volatility
  forecasting.
\newblock {\em Journal of Econometrics}, 159(2):276--288.

\bibitem[Engle, 1982]{engle1982autoregressive}
Engle, R.~F. (1982).
\newblock Autoregressive conditional heteroscedasticity with estimates of the
  variance of united kingdom inflation.
\newblock {\em Econometrica: Journal of the Econometric Society}, pages
  987--1007.

\bibitem[Engle et~al., 1990]{engle1990meteor}
Engle, R.~F., Ito, T., and Lin, W.-L. (1990).
\newblock Meteor showers or heat waves? heteroskedastic intra-daily volatility
  in the foreign exchange market.
\newblock {\em Econometrica}, 58(3):525--542.

\bibitem[Fajgelbaum et~al., 2020]{fajgelbaum2020return}
Fajgelbaum, P.~D., Goldberg, P.~K., Kennedy, P.~J., and Khandelwal, A.~K.
  (2020).
\newblock The return to protectionism.
\newblock {\em The Quarterly Journal of Economics}, 135(1):1--55.

\bibitem[Fan and Kim, 2018]{fan2018robust}
Fan, J. and Kim, D. (2018).
\newblock Robust high-dimensional volatility matrix estimation for
  high-frequency factor model.
\newblock {\em Journal of the American Statistical Association},
  113(523):1268--1283.

\bibitem[Fan and Wang, 2007]{fan2007multi}
Fan, J. and Wang, Y. (2007).
\newblock Multi-scale jump and volatility analysis for high-frequency financial
  data.
\newblock {\em Journal of the American Statistical Association},
  102(480):1349--1362.

\bibitem[Hall and Heyde, 2014]{hall2014martingale}
Hall, P. and Heyde, C.~C. (2014).
\newblock {\em Martingale limit theory and its application}.
\newblock Academic press.

\bibitem[Hamao et~al., 1990]{hamao1990correlations}
Hamao, Y., Masulis, R.~W., and Ng, V. (1990).
\newblock Correlations in price changes and volatility across international
  stock markets.
\newblock {\em The review of financial studies}, 3(2):281--307.

\bibitem[Hansen et~al., 2012]{hansen2012realized}
Hansen, P.~R., Huang, Z., and Shek, H.~H. (2012).
\newblock Realized garch: a joint model for returns and realized measures of
  volatility.
\newblock {\em Journal of Applied Econometrics}, 27(6):877--906.

\bibitem[Jacod et~al., 2009]{jacod2009microstructure}
Jacod, J., Li, Y., Mykland, P.~A., Podolskij, M., and Vetter, M. (2009).
\newblock Microstructure noise in the continuous case: the pre-averaging
  approach.
\newblock {\em Stochastic processes and their applications}, 119(7):2249--2276.

\bibitem[Karolyi, 1995]{karolyi1995multivariate}
Karolyi, G.~A. (1995).
\newblock A multivariate garch model of international transmissions of stock
  returns and volatility: The case of the united states and canada.
\newblock {\em Journal of Business \& Economic Statistics}, 13(1):11--25.

\bibitem[Kim and Wang, 2016]{kim2016unified}
Kim, D. and Wang, Y. (2016).
\newblock Unified discrete-time and continuous-time models and statistical
  inferences for merged low-frequency and high-frequency financial data.
\newblock {\em Journal of Econometrics}, 194:220--230.

\bibitem[Kim and Wang, 2021]{kim2021overnight}
Kim, D. and Wang, Y. (2021).
\newblock Overnight garch-it\^{o} volatility models.
\newblock {\em arXiv preprint arXiv:2102.13467}.

\bibitem[Kim et~al., 2016]{kim2016asymptotic}
Kim, D., Wang, Y., and Zou, J. (2016).
\newblock Asymptotic theory for large volatility matrix estimation based on
  high-frequency financial data.
\newblock {\em Stochastic Processes and their Applications}, 126:3527–--3577.

\bibitem[King and Wadhwani, 1990]{king1990transmission}
King, M.~A. and Wadhwani, S. (1990).
\newblock Transmission of volatility between stock markets.
\newblock {\em The Review of Financial Studies}, 3(1):5--33.

\bibitem[Li and Linton, 2021a]{li2021remedi}
Li, Z.~M. and Linton, O. (2021a).
\newblock A remedi for microstructure noise.
\newblock {\em Econometrica}, forthcoming.

\bibitem[Li and Linton, 2021b]{li2021robust}
Li, Z.~M. and Linton, O. (2021b).
\newblock Robust estimation of integrated volatility.
\newblock Technical report, Faculty of Economics, University of Cambridge.

\bibitem[Lin et~al., 1994]{lin1994bulls}
Lin, W.-L., Engle, R.~F., and Ito, T. (1994).
\newblock Do bulls and bears move across borders? international transmission of
  stock returns and volatility.
\newblock {\em Review of financial studies}, 7(3):507--538.

\bibitem[Mancini, 2004]{mancini2004estimation}
Mancini, C. (2004).
\newblock Estimation of the characteristics of the jumps of a general
  poisson-diffusion model.
\newblock {\em Scandinavian Actuarial Journal}, 2004(1):42--52.

\bibitem[Patton and Sheppard, 2015]{patton2015good}
Patton, A.~J. and Sheppard, K. (2015).
\newblock Good volatility, bad volatility: Signed jumps and the persistence of
  volatility.
\newblock {\em Review of Economics and Statistics}, 97(3):683--697.

\bibitem[Shephard and Sheppard, 2010]{shephard2010realising}
Shephard, N. and Sheppard, K. (2010).
\newblock Realising the future: forecasting with high-frequency-based
  volatility (heavy) models.
\newblock {\em Journal of Applied Econometrics}, 25(2):197--231.

\bibitem[Shin et~al., 2021]{shin2021adaptive}
Shin, M., Kim, D., and Fan, J. (2021).
\newblock Adaptive robust large volatility matrix estimation based on
  high-frequency financial data.
\newblock {\em Available at SSRN 3793394}.

\bibitem[Song et~al., 2021]{song2021volatility}
Song, X., Kim, D., Yuan, H., Cui, X., Lu, Z., Zhou, Y., and Wang, Y. (2021).
\newblock Volatility analysis with realized garch-it{\^o} models.
\newblock {\em Journal of Econometrics}, 222(1):393--410.

\bibitem[Tao et~al., 2013]{tao2013fast}
Tao, M., Wang, Y., and Chen, X. (2013).
\newblock Fast convergence rates in estimating large volatility matrices using
  high-frequency financial data.
\newblock {\em Econometric Theory}, 29(04):838--856.

\bibitem[Xiu, 2010]{xiu2010quasi}
Xiu, D. (2010).
\newblock Quasi-maximum likelihood estimation of volatility with high frequency
  data.
\newblock {\em Journal of Econometrics}, 159(1):235--250.

\bibitem[Zhang, 2006]{zhang2006efficient}
Zhang, L. (2006).
\newblock Efficient estimation of stochastic volatility using noisy
  observations: A multi-scale approach.
\newblock {\em Bernoulli}, 12(6):1019--1043.

\bibitem[Zhang et~al., 2005]{zhang2005tale}
Zhang, L., Mykland, P.~A., and A{\"\i}t-Sahalia, Y. (2005).
\newblock A tale of two time scales: Determining integrated volatility with
  noisy high-frequency data.
\newblock {\em Journal of the American Statistical Association},
  100(472):1394--1411.

\end{thebibliography}

\newpage

\appendix

\setcounter{table}{0}
\renewcommand{\thetable}{A\arabic{table}}

\section{Tables} \label{SEC-Table}

\begin{table}[!ht]
\caption{Robustness analysis against the model specification with the HAR-type contagion model, adopting the jump-adjusted ARP method as the integrated volatility estimator.
}\label{robust-HAR-ARP}
\centering
\scalebox{1}{
\begin{tabular}{lrrrrrrrrrr}
\hline
China                                                                           & \multicolumn{1}{c}{$\omega_{1}$} & \multicolumn{1}{c}{$\alpha_{1}^{(1)}$} & \multicolumn{1}{c}{$\alpha_{1}^{(5)}$} & \multicolumn{1}{c}{$\alpha_{1}^{(22)}$} & \multicolumn{1}{c}{$\beta_{1+}$} & \multicolumn{1}{c}{$\beta_{1-}$} & \multicolumn{1}{c}{$\kappa_{1}$} & \multicolumn{1}{c}{$\alpha_{12}$} & \multicolumn{1}{c}{$\beta_{12+}$} & \multicolumn{1}{c}{$\beta_{12-}$} \\ \hline
                                                                                & \multicolumn{10}{c}{Before the U.S.--China trade war}                                                                                                                                                                                                                                                                                                                             \\
Coefficients                                                                    & 7.1e-6                          & \bftab.41                             & \bftab.47                             & .11                                    & \bftab-1.73                      & 1.14                             & \bftab.10                       & .25                              & -.02                             & -.81                             \\
Standard errors                                                                 & 7.5e-6                          & .09                                   & .18                                   & .10                                    & .80                             & 1.24                             & .03                             & .15                              & .90                              & .63                              \\
p-values                                                                        & .35                             & .00                                   & .01                                   & .29                                    & .03                             & .36                             & .00                             & .10                              & .98                              & .20                              \\
                                                                                & \multicolumn{10}{c}{After the U.S.--China trade war}                                                                                                                                                                                                                                                                                                                              \\
Coefficients                                                                    & \bftab4.0e-5                    & \bftab.41                             & \bftab.40                             & .00                                    & .04                             & -.14                            & \bftab.26                       & -.01                             & -.14                             & \bftab.56                        \\
Standard errors                                                                 & 1.6e-5                          & .07                                   & .10                                   & .07                                    & .20                             & .64                             & .09                             & .03                              & .40                              & .25                              \\
p-values                                                                        & .02                             & .00                                   & .00                                   & .99                                    & .84                             & .82                             & .00                             & .79                              & .72                              & .02                              \\ \hline
\begin{tabular}[c]{@{}l@{}}p-values of the\\ structural break test\end{tabular} & .07                             & .99                                   & .70                                   & .39                                    & \bftab.03                       & .36                             & .07                             & .09                              & .90                              & \bftab.04                        \\ \hline\hline
U.S.                                                                            & \multicolumn{1}{c}{$\omega_{2}$} & \multicolumn{1}{c}{$\alpha_{2}^{(1)}$} & \multicolumn{1}{c}{$\alpha_{2}^{(5)}$} & \multicolumn{1}{c}{$\alpha_{2}^{(22)}$} & \multicolumn{1}{c}{$\beta_{2+}$} & \multicolumn{1}{c}{$\beta_{2-}$} & \multicolumn{1}{c}{$\kappa_{2}$} & \multicolumn{1}{c}{$\alpha_{21}$} & \multicolumn{1}{c}{$\beta_{21+}$} & \multicolumn{1}{c}{$\beta_{21-}$} \\ \hline
                                                                                & \multicolumn{10}{c}{Before the U.S.--China trade war}                                                                                                                                                                                                                                                                                                                             \\
Coefficients                                                                    & \bftab9.9e-6                    & \bftab.43                             & .21                                   & -.01                                   & .62                             & -.45                            & .39                             & \bftab-.01                       & -.02                             & .04                              \\
Standard errors                                                                 & 3.7e-6                          & .10                                   & .11                                   & .16                                    & .62                             & .33                             & .20                             & .01                              & .06                              & .03                              \\
p-values                                                                        & .01                             & .00                                   & .06                                   & .92                                    & .32                             & .17                             & .06                             & .02                              & .74                              & .18                              \\
                                                                                & \multicolumn{10}{c}{After the U.S.--China trade war}                                                                                                                                                                                                                                                                                                                              \\
Coefficients                                                                    & 6.6e-6                          & \bftab.67                             & \bftab.18                             & .03                                    & -.00                            & \bftab1.19                       & \bftab.55                       & -.01                             & -.05                             & -.02                             \\
Standard errors                                                                 & 5.1e-6                          & .10                                   & .09                                   & .05                                    & .51                             & .54                             & .21                             & .02                              & .06                              & .18                              \\
p-values                                                                        & .19                             & .00                                   & .04                                   & .55                                    & 1.00                             & .03                             & .01                             & .40                              & .40                              & .90                              \\ \hline
\begin{tabular}[c]{@{}l@{}}p-values of the\\ structural break test\end{tabular} & .59                             & .08                                   & .84                                   & .78                                    & .44                             & \bftab.01                       & .58                             & .99                              & .71                              & .72                              \\ \hline
\multicolumn{10}{l}{\small NOTE: Significant coefficients at 5\% and p-values below 0.05 are presented in bold.}\\
\end{tabular}
}
\end{table}

\begin{table}[!ht]
\caption{Robustness analysis against the model specification with  the HAR-type contagion model, adopting the jump-adjusted PaReMeDI method as the integrated volatility estimator.
}\label{robust-HAR-PaReMeDI}
\centering
\scalebox{1}{
\begin{tabular}{lrrrrrrrrrr}
\hline
China                                                                           & \multicolumn{1}{c}{$\omega_{1}$} & \multicolumn{1}{c}{$\alpha_{1}^{(1)}$} & \multicolumn{1}{c}{$\alpha_{1}^{(5)}$} & \multicolumn{1}{c}{$\alpha_{1}^{(22)}$} & \multicolumn{1}{c}{$\beta_{1+}$} & \multicolumn{1}{c}{$\beta_{1-}$} & \multicolumn{1}{c}{$\kappa_{1}$} & \multicolumn{1}{c}{$\alpha_{12}$} & \multicolumn{1}{c}{$\beta_{12+}$} & \multicolumn{1}{c}{$\beta_{12-}$} \\ \hline
                                                                                & \multicolumn{10}{c}{Before the U.S.--China trade war}                                                                                                                                                                                                                                                                                                                             \\
Coefficients                                                                    & 1.1e-5                          & \bftab.36                             & \bftab.52                             & .11                                    & \bftab-3.28                      & 1.01                             & \bftab.25                       & .29                              & 1.38                              & -2.02                             \\
Standard errors                                                                 & 1.4e-5                          & .10                                   & .23                                   & .11                                    & 1.39                             & 2.06                             & .07                             & .21                              & 1.78                              & 1.17                              \\
p-values                                                                        & .43                             & .00                                   & .02                                   & .30                                    & .02                             & .62                             & .00                             & .17                              & .44                              & .09                              \\
                                                                                & \multicolumn{10}{c}{After the U.S.--China trade war}                                                                                                                                                                                                                                                                                                                              \\
Coefficients                                                                    & \bftab6.7e-5                    & \bftab.39                             & \bftab.38                             & .01                                    & -.04                            & .47                             & \bftab.56                       & -.02                             & -.11                             & .83                              \\
Standard errors                                                                 & 2.7e-5                          & .08                                   & .10                                   & .07                                    & .37                             & 1.32                             & .16                             & .03                              & .62                              & .44                              \\
p-values                                                                        & .01                             & .00                                   & .00                                   & .90                                    & .91                             & .72                             & .00                             & .49                              & .85                              & .06                              \\ \hline
\begin{tabular}[c]{@{}l@{}}p-values of the\\ structural break test\end{tabular} & .07                             & .87                                   & .58                                   & .43                                    & \bftab.02                       & .83                             & .06                             & .15                              & .43                              & \bftab.02                        \\ \hline\hline
U.S.                                                                            & \multicolumn{1}{c}{$\omega_{2}$} & \multicolumn{1}{c}{$\alpha_{2}^{(1)}$} & \multicolumn{1}{c}{$\alpha_{2}^{(5)}$} & \multicolumn{1}{c}{$\alpha_{2}^{(22)}$} & \multicolumn{1}{c}{$\beta_{2+}$} & \multicolumn{1}{c}{$\beta_{2-}$} & \multicolumn{1}{c}{$\kappa_{2}$} & \multicolumn{1}{c}{$\alpha_{21}$} & \multicolumn{1}{c}{$\beta_{21+}$} & \multicolumn{1}{c}{$\beta_{21-}$} \\ \hline
                                                                                & \multicolumn{10}{c}{Before the U.S.--China trade war}                                                                                                                                                                                                                                                                                                                             \\
Coefficients                                                                    & \bftab1.1e-5                    & \bftab.45                             & .19                                   & .05                                    & .88                             & -.78                            & \bftab.69                       & \bftab-.01                       & -.04                             & .10                              \\
Standard errors                                                                 & 4.6e-6                          & .10                                   & .11                                   & .14                                    & .80                             & .44                             & .26                             & .00                              & .06                              & .05                              \\
p-values                                                                        & .01                             & .00                                   & .07                                   & .74                                    & .27                             & .08                             & .01                             & .01                              & .52                              & .06                              \\
                                                                                & \multicolumn{10}{c}{After the U.S.--China trade war}                                                                                                                                                                                                                                                                                                                              \\
Coefficients                                                                    & 8.2e-6                          & \bftab.67                             & .19                                   & .04                                    & -.24                            & 1.10                             & \bftab.89                       & -.02                             & -.04                             & .14                              \\
Standard errors                                                                 & 6.1e-6                          & .10                                   & .10                                   & .06                                    & .58                             & .65                             & .26                             & .01                              & .08                              & .26                              \\
p-values                                                                        & .18                             & .00                                   & .05                                   & .48                                    & .68                             & .09                             & .00                             & .17                              & .61                              & .58                              \\ \hline
\begin{tabular}[c]{@{}l@{}}p-values of the\\ structural break test\end{tabular} & .68                             & .12                                   & .98                                   & .98                                    & .26                             & \bftab.02                       & .59                             & .66                              & 1.00                              & .86                              \\ \hline
\multicolumn{10}{l}{\small NOTE: Significant coefficients at 5\% and p-values below 0.05 are presented in bold.}\\
\end{tabular}
}
\end{table}

\clearpage

\section{Proofs} \label{SEC-Proof}

We first define $\norm{X}_{L_p} = \left\{ E\left[ |X|^p \right] \right\}^{1/p}$, for any given random variable $X$ and $p \leq 1$.
We note that jump size can be rewritten as $L_{1t}^{2} = b_{1} + M_{1t}$ and $L_{2t}^{2} = b_{2} + M_{2t}$, where $M_{1t}$ and $M_{2t}$ are i.i.d. mean zero random variables.
The effect of the initial values $h_{1,1}(\theta_0)$ and $h_{2,1}(\theta_0)$ is of order $n^{-1}$, which is negligible (Lemma 1 in \citet{kim2016unified}).
Thus, we assume that $h_{1,1}(\theta_0)$ and $h_{2,1}(\theta_0)$ are given.

\subsection{Proof of Theorem \ref{theorem1}}

Theorem \ref{theorem1} is an immediate consequence of Theorem \ref{theorem4} below.
\begin{theorem}\label{theorem4}
    \begin{enumerate}
        \item [(a)] For $0 < \alpha_{1H} < 1$ and $n \in \mathbb{N}$, we have
        \begin{equation*}
            \int ^{n-1+\lambda_1}_{n-1}\sigma ^{2}_{1t}\left( \theta \right) dt=\lambda_1 h_{1,n}\left( \theta \right) + D_{1,n} \quad  \text{ a.s.}, 
        \end{equation*}
        where
        \begin{eqnarray*}
        h_{1,n}(\theta) &=& (\varrho_{12} - 2\varrho_{13})\nu_{1H} + \varrho_{12} (\beta_{1H}^{+} I_1^{+} b_1^{+} + \beta_{1H}^{-} I_1^{-} b_1^{-}) + \varrho_{12} \omega_{1H} \cr
        && + \left [ \varrho_{12} (\gamma_{1H} - 1) + \varrho_{11} \right ] \sigma_{1(n-1)}^2 (\theta) \cr
        &=& \omega_{1}^g + \gamma_{1}h_{1,n-1}(\theta) + \dfrac {\alpha_{1}^g}{\lambda_1}\int ^{n-2+\lambda_1 }_{n-2}\sigma ^{2}_{1s} (\theta) ds + \dfrac {\alpha_{12}^g}{\lambda_2 }\int ^{n-2+\tau+\lambda_2}_{n-2+\tau}\sigma ^{2}_{2s} (\theta) ds \cr
        &&+ \dfrac{\beta_{1+}^{g}}{\lambda_1} \int ^{n-2+\lambda_1 }_{n-2}(L^{+}_{1s})^2d\Lambda_{1s}^{+} + \dfrac{\beta_{1-}^{g}}{\lambda_1} \int ^{n-2+\lambda_1 }_{n-2}(L^{-}_{1s})^2d\Lambda_{1s}^{-} \cr
        &&+ \dfrac{\beta_{12+}^{g}}{\lambda_2}\int ^{n-2+\tau+\lambda_2}_{n-2+\tau}(L^{+}_{2s})^2d\Lambda_{2s}^{+} + \dfrac{\beta_{12-}^{g}}{\lambda_2}\int ^{n-2+\tau+\lambda_2}_{n-2+\tau}(L^{-}_{2s})^2d\Lambda_{2s}^{-} \cr
        &&+ \dfrac {\kappa_{1}^g}{1-\lambda_1 }\left(\int ^{n-1}_{n-2+\lambda_1}\sigma _{1s} (\theta) dB_{1s}\right) ^{2},
        \end{eqnarray*}
        $\gamma_1 = \gamma_{1L}\gamma_{1H}$, $\omega_{1}^g = (1-\gamma_1)\left \lbrace (\varrho_{12} - 2 \varrho_{13}) \nu_{1H} + \varrho_{12} (\beta_{1H}^{+} I_1^{+} b_1^{+} + \beta_{1H}^{-} I_1^{-} b_1^{-}) + \varrho_{12} \omega_{1H} \right \rbrace + \varrho_{1}(\omega_{1L} + \gamma_{1L} \omega_{1H})$,  $\alpha_{1}^g = \varrho_{1} \gamma_{1L} \alpha_{1H}$, $\alpha_{12}^g = \varrho_{1} \alpha_{12}$, $\kappa_{1}^g = \varrho_{1} \alpha_{1L}$, $\beta_{1+}^g = \varrho_{1} \gamma_{1L} \beta_{1H}^{+}$, $\beta_{1-}^g = \varrho_{1} \gamma_{1L} \beta_{1H}^{-}$, $\beta_{12+}^g = \varrho_{1} \beta_{12}^{+}$, $\beta_{12-}^g = \varrho_{1} \beta_{12}^{-}$, $\varrho_{11} = \alpha_{1H}^{-1} (e^{\alpha_{1H}} - 1)$, $\varrho_{12} = \alpha_{1H}^{-2} (e^{\alpha_{1H}} - 1 - \alpha_{1H})$, $\varrho_{13} = \alpha_{1H}^{-3} (e^{\alpha_{1H}} - 1 - \alpha_{1H} - \frac{\alpha_{1H}}{2})$, $\varrho_{1} = (\gamma_{1H} - 1)\varrho_{12} + \varrho_{11}$,   and
        \begin{eqnarray*} 
            D_{1,n} &=& D^{c}_{1,n} + D^{J}_{1,n}, \cr
            D^{c}_{1,n} &=& 2\nu_{1H}\alpha_{1H}^{-2}\int ^{n-1+\lambda_1}_{n-1 }\left [ \left \lbrace \frac{\alpha_{1H}}{\lambda_1}(n-1+\lambda_1-t) - 1 \right \rbrace e^{\frac{\alpha_{1H}}{\lambda_1}(n-1+\lambda_1-t)} + 1 \right ] Z_{1t} dZ_{1t}, \cr
            D^{J}_{1,n} &=& \beta_{1H}^{+} \alpha_{1H} ^ {-1} \biggl \lbrace \int_{n-1}^{n-1+\lambda_1} \left ( e^{\lambda_{1}^{-1}\alpha_{1H}(n-1+\lambda_1 -t)} -1 \right ) M_{1t}^{+}d\Lambda_{1t}^{+} \cr
            && \quad  + b_1^{+} \int_{n-1}^{n-1+\lambda_1} \left ( e^{\lambda_{1}^{-1}\alpha_{1H}(n-1+\lambda_1 -t)} -1 \right ) (d\Lambda_{1t}^{+} - I_1^{+} dt) \biggl \rbrace \cr
            && \quad + \beta_{1H}^{-} \alpha_{1H} ^ {-1} \biggl \lbrace \int_{n-1}^{n-1+\lambda_1} \left ( e^{\lambda_{1}^{-1}\alpha_{1H}(n-1+\lambda_1 -t)} -1 \right ) M_{1t}^{-}d\Lambda_{1t}^{-} \cr
            && \quad + b_1^{-} \int_{n-1}^{n-1+\lambda_1} \left ( e^{\lambda_{1}^{-1}\alpha_{1H}(n-1+\lambda_1 -t)} -1 \right ) (d\Lambda_{1t}^{-} - I_1^{-} dt) \biggl \rbrace  
        \end{eqnarray*}
        are all martingale differences.

        \item [(b)] For $0 < \alpha_{2H} < 1$ and $n \in \mathbb{N}$, we have
        \begin{equation*}
             \int ^{n-1+\tau+\lambda_2}_{n-1+\tau}\sigma ^{2}_{2t}\left( \theta \right) dt=\lambda_2 h_{2,n}\left( \theta \right) + D_{2,n} \quad \text{ a.s.}, 
        \end{equation*}
        where
        \begin{eqnarray*}
        h_{2,n}(\theta) &=& (\varrho_{22} - 2\varrho_{23})\nu_{2H} + \varrho_{22} (\beta_{2H}^{+} I_2^{+} b_2^{+} + \beta_{2H}^{-} I_2^{-} b_2^{-}) + \varrho_{22} \omega_{2H} \cr
        && + \left [ \varrho_{22} (\gamma_{2H} - 1) + \varrho_{21} \right ] \sigma_{2(n-1+\tau)}^2 (\theta) \cr
        &=& \omega_{2}^g + \gamma_{2}h_{2,n-1}(\theta) + \dfrac {\alpha_{2}^g}{\lambda_2}\int ^{n-2+\tau+\lambda_2 }_{n-2+\tau}\sigma ^{2}_{2s} (\theta) ds + \dfrac {\alpha_{21}^g}{\lambda_1 }\int ^{n-1+\lambda_1}_{n-1}\sigma ^{2}_{1s} (\theta) ds \cr
        &&+ \dfrac{\beta_{2+}^{g}}{\lambda_2} \int ^{n-2+\tau+\lambda_2 }_{n-2+\tau}(L^{+}_{2s})^2d\Lambda_{2s}^{+} + \dfrac{\beta_{2-}^{g}}{\lambda_2} \int ^{n-2+\tau+\lambda_2 }_{n-2+\tau}(L^{-}_{2s})^2d\Lambda_{2s}^{-} \cr
        &&+ \dfrac{\beta_{21+}^{g}}{\lambda_1}\int ^{n-1+\lambda_1}_{n-1}(L^{+}_{1s})^2d\Lambda_{1s}^{+} + \dfrac{\beta_{21-}^{g}}{\lambda_1}\int ^{n-1+\lambda_1}_{n-1}(L^{-}_{1s})^2d\Lambda_{1s}^{-} \cr
        &&+ \dfrac {\kappa_{2}^g}{1-\lambda_2 }\left(\int ^{n-1+\tau}_{n-2+\tau+\lambda_2}\sigma _{2s} (\theta) dB_{2s}\right) ^{2},
        \end{eqnarray*}
        $\gamma_2 = \gamma_{2L}\gamma_{2H}$, $\omega_{2}^g = (1-\gamma_2)\left \lbrace (\varrho_{22} - 2 \varrho_{23}) \nu_{2H} + \varrho_{22} (\beta_{2H}^{+} I_2^{+} b_2^{+} + \beta_{2H}^{-} I_2^{-} b_2^{-}) + \varrho_{22} \omega_{2H} \right \rbrace + \varrho_{2}(\omega_{2L} + \gamma_{2L} \omega_{2H})$,  $\alpha_{2}^g = \varrho_{2} \gamma_{2L} \alpha_{2H}$, $\alpha_{21}^g = \varrho_{2} \alpha_{21}$, $\kappa_{2}^g = \varrho_{2} \alpha_{2L}$, $\beta_{2+}^g = \varrho_{2} \gamma_{2L} \beta_{2H}^{+}$, $\beta_{2-}^g = \varrho_{2} \gamma_{2L} \beta_{2H}^{-}$, $\beta_{21+}^g = \varrho_{2} \beta_{21}^{+}$, $\beta_{21-}^g = \varrho_{2} \beta_{21}^{-}$, $\varrho_{21} = \alpha_{2H}^{-1} (e^{\alpha_{2H}} - 1)$, $\varrho_{22} = \alpha_{2H}^{-2} (e^{\alpha_{2H}} - 1 - \alpha_{2H})$, $\varrho_{23} = \alpha_{2H}^{-3} (e^{\alpha_{2H}} - 1 - \alpha_{2H} - \frac{\alpha_{2H}}{2})$, $\varrho_{2} = (\gamma_{2H} - 1)\varrho_{22} + \varrho_{21}$, and
        \begin{eqnarray*}
            D_{2,n} &=& D^{c}_{2,n} + D^{J}_{2,n}, \cr
            D^{c}_{2,n} &=& \int ^{n-1+\tau+\lambda_2}_{n-1+\tau }\left [ \left \lbrace \frac{\alpha_{2H}}{\lambda_2}(n-1+\tau+\lambda_2-t) - 1 \right \rbrace e^{\frac{\alpha_{2H}}{\lambda_2}(n-1+\tau+\lambda_2-t)} + 1 \right ] Z_{2t} dZ_{2t}  \cr 
            && \times 2\nu_{2H}\alpha_{2H}^{-2} , \cr
            D^{J}_{2,n} &=& \beta_{2H}^{+} \alpha_{2H} ^ {-1} \biggl \lbrace \int_{n-1+\tau}^{n-1+\tau+\lambda_2} \left ( e^{\lambda_{2}^{-1}\alpha_{2H}(n-1+\tau+\lambda_2 -t)} -1 \right ) M_{2t}^{+}d\Lambda_{2t}^{+} \cr
            && + b_2^{+} \int_{n-1+\tau}^{n-1+\tau+\lambda_2} \left ( e^{\lambda_{2}^{-1}\alpha_{2H}(n-1+\tau+\lambda_2 -t)} -1 \right ) (d\Lambda_{2t}^{+} - I_2^{+} dt) \biggl \rbrace \cr
            && + \beta_{2H}^{-} \alpha_{2H} ^ {-1} \biggl \lbrace \int_{n-1+\tau}^{n-1+\tau+\lambda_2} \left ( e^{\lambda_{2}^{-1}\alpha_{2H}(n-1+\tau+\lambda_2 -t)} -1 \right ) M_{2t}^{-}d\Lambda_{2t}^{-} \cr
            && + b_2^{-} \int_{n-1+\tau}^{n-1+\tau+\lambda_2} \left ( e^{\lambda_{2}^{-1}\alpha_{2H}(n-1+\tau+\lambda_2 -t)} -1 \right ) (d\Lambda_{2t}^{-} - I_2^{-} dt) \biggl \rbrace  
        \end{eqnarray*}
        are all martingale differences.
    \end{enumerate}
\end{theorem}
\textbf{Proof of Theorem \ref{theorem4}.}
Consider (a).
By It\^{o}'s lemma, we have
\begin{eqnarray*}
    R_1(k) & \equiv & \int_{n-1}^{n-1+\lambda_1} \frac{(n-1+\lambda_1 - t)^k}{k!} \sigma_{1,t}^2 (\theta) dt \cr
    &=& \sigma_{1,n-1}^2 (\theta) \biggl \lbrace \frac{\lambda_1^{k+1}}{(k+1)!} + (\gamma_{1,H} - 1) \lambda_1^{-1} \frac{\lambda_1^{k+2}}{(k+2)!} \biggl \rbrace + \omega_{1,H} \lambda_1^{-1} \frac{\lambda_1^{k+2}}{(k+2)!} \cr
    && + \beta_{1,H}b_1 I_1 \lambda_{1}^{-1}\frac{\lambda_1^{k+2}}{(k+2)!} + \nu_{1,H}\lambda_{1}^{k+1} \biggl\lbrace \frac{1}{(k+2)!} - \frac{2}{(k+2)!} \biggl\rbrace \cr
    && + \frac{\beta_{1,H}^{+}}{\alpha_{1,H}} \int_{n-1}^{n-1+\lambda_1} \frac{(n-1+\lambda_1-t)^{k+1}}{(k+1)!}\frac{\alpha_{1,H}}{\lambda_1} M_{1,t}^{+} d\Lambda_{1,t}^{+} \cr
    && + \frac{\beta_{1,H}^{-}}{\alpha_{1,H}} \int_{n-1}^{n-1+\lambda_1} \frac{(n-1+\lambda_1-t)^{k+1}}{(k+1)!}\frac{\alpha_{1,H}}{\lambda_1} M_{1,t}^{-} d\Lambda_{1,t}^{-} \cr
    && + 2\nu_{1,H}\alpha_{1,H}^{-2}\int_{n-1}^{n-1+\lambda_1}\left\lbrace \frac{1}{(k+1)!} - \frac{1}{(k+2)!} \right\rbrace (n-1+\lambda_1-t)^{k+2}\left(\frac{\alpha_{1,H}}{\lambda_1}\right)^2 Z_{1,t} dZ_{1,t} \cr
    && + \beta_{1,H}^{+}\alpha_{1,H}^{-1}b_1 \int_{n-1}^{n-1+\lambda_1} \frac{(n-1+\lambda_1-t)^{k+1}}{(k+1)!}\left( \frac{\alpha_{1,H}}{\lambda_1} \right) \left ( d\Lambda_{1,t}^{+} - I_1^{+} dt \right ) \cr
    && + \beta_{1,H}^{-}\alpha_{1,H}^{-1}b_1 \int_{n-1}^{n-1+\lambda_1} \frac{(n-1+\lambda_1-t)^{k+1}}{(k+1)!}\left( \frac{\alpha_{1,H}}{\lambda_1} \right) \left ( d\Lambda_{1,t}^{-} - I_1^{-} dt \right ) + \frac{\alpha_{1,H}}{\lambda_1} R_1(k+1).
\end{eqnarray*}
Then, algebraic manipulations show 
\begin{eqnarray*}
R_1(0) &=& \int_{n-1}^{n-1+\lambda_1} \sigma_{1,t}^2 dt \cr
&=& \lambda_1 [ (\varrho_2 - 2\varrho_3)\nu_{1,H} + \varrho_2 \beta_{1,H}^{+} I_1^{+} b_1 + \varrho_2 \beta_{1,H}^{-} I_1^{-} b_1 + \varrho_2 \omega_{1,H} + \left \lbrace \varrho_2 (\gamma_{1,H} - 1) + \varrho_1 \right \rbrace \sigma_{1,n-1}^2 ] \cr
&& + 2\nu_{1,H}\alpha_{1,H}^{-2}\int ^{n-1+\lambda_1}_{n-1 }\left [ \left \lbrace \frac{\alpha_{1,H}}{\lambda_1}(n-1+\lambda_1-t) - 1 \right \rbrace e^{\frac{\alpha_{1,H}}{\lambda_1}(n-1+\lambda_1-t)} + 1 \right ] Z_{1,t} dZ_{1,t} \cr
&& + \beta_{1,H}^{+} \alpha_{1,H} ^ {-1} \biggl \lbrace \int_{n-1}^{n-1+\lambda_1} \left ( e^{\lambda_1^{-1}\alpha_{1,H}(n-1+\lambda_1 -t)} -1 \right ) M_{1t}^{+}d\Lambda_{1,t}^{+} \cr
&& + b_1^{+} \int_{n-1}^{n-1+\lambda_1} \left ( e^{\lambda_1^{-1}\alpha_{1,H}(n-1+\lambda_1 -t)} -1 \right ) (d\Lambda_{1t}^{+} - I_1^{+} dt) \biggl \rbrace \cr
&& + \beta_{1,H}^{-} \alpha_{1,H} ^ {-1} \biggl \lbrace \int_{n-1}^{n-1+\lambda_1} \left ( e^{\lambda_1^{-1}\alpha_{1,H}(n-1+\lambda_1 -t)} -1 \right ) M_{1t}^{-}d\Lambda_{1,t}^{-} \cr
&& + b_1^{-} \int_{n-1}^{n-1+\lambda_1} \left ( e^{\lambda_1^{-1}\alpha_{1,H}(n-1+\lambda_1 -t)} -1 \right ) (d\Lambda_{1t}^{-} - I_1^{-} dt) \biggl \rbrace \cr
&=& \lambda_1 h_{1,n}(\theta) + D_{1,n}^{c} + D_{1,n}^{J} \quad \text{a.s.}
\end{eqnarray*}
The proof of Theorem \ref{theorem4}(b) is similar to (a), so we omit it.
$\blacksquare$

\subsection{Proof of Theorem \ref{theorem2}}
To easy the notations, we use $\theta$ instead of $\theta^g$.
Let $\xi$ be a positive constant such that $|x| < \xi$ for any $x \in \{\alpha_l,\alpha_u,\beta_l,\beta_u,\kappa_l,\kappa_u\}$.
Let
\begin{flalign*}
    &\hat{L}_{n,m}(\theta) = -\frac{1}{2n}\sum_{i=1}^{n}\left\lbrace\log(\hat{h}_{1,i}(\theta)) + \frac{RV_{1,i}/\lambda_1}{\hat{h}_{1,i}(\theta)} + \log(\hat{h}_{2,i}(\theta)) + \frac{RV_{2,i}/\lambda_2}{\hat{h}_{2,i}(\theta)} \right\rbrace = -\frac{1}{2n}\sum_{i=1}^{n}\hat{l}_i(\theta), \cr
    &\hat{L}_{n}(\theta) = -\frac{1}{2n}\sum_{i=1}^{n}\left\lbrace\log(h_{1,i}(\theta)) + \frac{IV_{1,i}/\lambda_1}{h_{1,i}(\theta)} + \log(h_{2,i}(\theta)) + \frac{IV_{2,i}/\lambda_2}{h_{2,i}(\theta)} \right\rbrace, \\
    &L_{n}(\theta) = -\frac{1}{2n}\sum_{i=1}^{n}\left\lbrace\log(h_{1,i}(\theta)) + \frac{h_{1,i}(\theta_0)}{h_{1,i}(\theta)} + \log(h_{2,i}(\theta)) + \frac{h_{1,i}(\theta_0)}{h_{2,i}(\theta)} \right\rbrace, \\
    &\hat{\psi}_{n,m}\left(\theta\right) = \frac{\partial\hat{L}_{n,m}\left(\theta\right)}{\partial\theta} , \quad \hat{\psi}_{n}\left(\theta\right) = \frac{\partial\hat{L}_{n}\left(\theta\right)}{\partial\theta}, \quad \psi_{n}\left(\theta\right) = \frac{\partial L_{n}\left(\theta\right)}{\partial\theta}.
\end{flalign*}

\begin{lemma}\label{lemma1}
Under Assupmtion \ref{assumption-1}(a), we have
\begin{enumerate}
    \item [(a)] $\sup_{i \in \mathbb{N}} E\left[\sup_{\theta \in \Theta} h_{1,i}(\theta)\right] < \infty$, $\sup_{i \in \mathbb{N}} E\left[\sup_{\theta \in \Theta} h_{2,i}(\theta)\right] < \infty$, $\sup_{i \in \mathbb{N}} E\left[\sup_{\theta \in \Theta} \hat{h}_{1,i}(\theta)\right] $ $< \infty$, and $\sup_{i \in \mathbb{N}} E\left[\sup_{\theta \in \Theta} \hat{h}_{2,i}(\theta)\right] < \infty$;

    \item [(b)] for any $p \geq 1$, $l \in \lbrace 1,2 \rbrace$, and $j,k,q \in \lbrace 1, \ldots, 18 \rbrace$, we have $\sup_{i \in \mathbb{N}} \norm{\sup_{\theta \in \Theta} h_{l,i}(\theta)^{-1} \frac{\partial h_{l,i}(\theta)}{\partial \theta_j}}_{L_p} \leq C$, 
    $\sup_{i \in \mathbb{N}} \norm{\sup_{\theta \in \Theta} \hat{h}_{l,i}(\theta)^{-1} \frac{\partial \hat{h}_{l,i}(\theta)}{\partial \theta_j}}_{L_p} \leq C$,
    $\sup_{i \in \mathbb{N}} \norm{\sup_{\theta \in \Theta} h_{l,i}(\theta)^{-1} \frac{\partial^2 h_{l,i}(\theta)}{\partial \theta_j \partial \theta_k}}_{L_p} \leq C$,\\
    $\sup_{i \in \mathbb{N}} \norm{\sup_{\theta \in \Theta} \hat{h}_{l,i}(\theta)^{-1} \frac{\partial^2 \hat{h}_{l,i}(\theta)}{\partial \theta_j \partial \theta_k}}_{L_p} \leq C$,
    $\sup_{i \in \mathbb{N}} \norm{\sup_{\theta \in \Theta} h_{l,i}(\theta)^{-1} \frac{\partial^3 h_{l,i}(\theta)}{\partial \theta_j \partial \theta_k \partial \theta_q}}_{L_p} \leq C$,\\
    and $\sup_{i \in \mathbb{N}} \norm{\sup_{\theta \in \Theta} \hat{h}_{l,i}(\theta)^{-1} \frac{\partial^3 \hat{h}_{l,i}(\theta)}{\partial \theta_j \partial \theta_k \partial \theta_q}}_{L_p} \leq C$, where $\theta = (\theta_1, \ldots, \theta_{18} ) =$ $(\omega_{1}^{g}, \gamma_1, \alpha_{1}^{g}, \beta_{1+}^{g},$ $ \beta_{1-}^{g}, \kappa_{1}^{g}, \alpha_{1,2}^g, \beta_{12+}^g, \beta_{12-}^g ,  \omega_{2}^{g}, \gamma_2,$ $\alpha_{2}^{g},$ $ \beta_{2+}^{g},$ $ \beta_{2-}^{g},\kappa_{2}^{g} , \alpha_{21}^g,$ $ \beta_{21+}^g,$ $ \beta_{21-}^g)$;

    \item [(c)] for any $p \geq 1$ and $j,k,q \in \lbrace 1,  \ldots ,18 \rbrace$, we have $\sup_{i \in \mathbb{N}} \norm {\sup_{\theta\in \Theta} \frac{\partial^3\hat{l}_i(\theta)}{\partial\theta_j\partial\theta_k\partial\theta_q}}_{L_p} < \infty$;

    \item [(d)] for any $l \in \{1,2\}$ and $j \in \{1,\ldots,18\}$, we have $\sup_{i \in \mathbb{N}} \norm{\sup_{\theta \in \Theta} \left( \hat{h}_{l,i}(\theta) - h_{l,i}(\theta) \right)}_{L_2} \leq Cm^{-1/4}$ and $\sup_{i \in \mathbb{N}} \norm{\sup_{\theta \in \Theta} \left(  \frac{\partial \hat{h}_{l,i}(\theta)}{\partial \theta_j} - \frac{\partial h_{l,i}(\theta)}{\partial \theta_j} \right)}_{L_2} \leq Cm^{-1/4}$.
\end{enumerate}

\end{lemma}
\textbf{Proof of Lemma \ref{lemma1}.}
Consider (a).
 Let
\begin{equation*}
    IJ_{l,i}^{+} =  \int_{\tau_{l}(i)-1}^{\tau_{l}(i)-1+\lambda_l} (L_{ls}^{+})^{2} d\Lambda_{ls}^{+} \quad \text{and} \quad IJ_{l,i}^{-} =  \int_{\tau_{l}(i)-1}^{\tau_{l}(i)-1+\lambda_l} (L_{ls}^{-})^{2} d\Lambda_{ls}^{-}.
\end{equation*}
Using the fact that $h_{1,i}(\theta) = \omega_{1}^g + \gamma_{1}h_{1,i-1}(\theta) + \alpha_{1}^g{\lambda_1}^{-1}IV_{1,i-1} + \beta_{1+}^g \lambda_1^{-1}IJ_{1,i-1}^{+} + \beta_{1-}^g \lambda_1^{-1}IJ_{1,i-1}^{-} + \kappa_{1}^g (1 - \lambda_1 )^{-1}OV_{1,i-1} + \alpha_{12}^g \lambda_{2}^{-1} IV_{2,i-1} + \beta_{12+}^g \lambda_{2}^{-1} IJ_{2,i-1}^{+} + \beta_{12-}^g \lambda_{2}^{-1} IJ_{2,i-1}^{-}$, we can show that
\begin{eqnarray*}
    E \left [ \sup_{\theta \in \Theta} h_{1,i}(\theta) \right ] &=& E \Biggl [\sup_{\theta \in \Theta} \biggl \{ \frac{\omega_{1}^g (1-\gamma_1 ^{(i-1)})}{1 - \gamma_1} + \alpha_{1}^g \sum_{j=0}^{i-1} IV_{1,i-1-j} \gamma^j + \beta_{1+}^g \sum_{j=0}^{i-1} IJ_{1,i-1-j}^{+} \gamma^j \\
    && + \beta_{1-}^g \sum_{j=0}^{i-1} IJ_{1,i-1-j}^{-} \gamma^j + \kappa_{1}^g \sum_{j=0}^{i-1} OV_{1,i-1-j} \gamma^j + \alpha_{12}^g \sum_{j=0}^{i-1} IV_{2,i-1-j} \gamma^j\\
    && + \beta_{12+}^g \sum_{j=0}^{i-1} IJ_{2,i-1-j}^{+} \gamma^j + \beta_{12-}^g \sum_{j=0}^{i-1} IJ_{2,i-1-j}^{-} \gamma^j \biggl \} \Biggl ]\\
    & \leq & \frac{\omega_{u}}{1 - \gamma_u} + C \frac{\xi}{1 - \gamma_u}.
\end{eqnarray*}
This implies that $\sup_{i \in \mathbb{N}} E\left[\sup_{\theta \in \Theta} h_{1,i}(\theta)\right] < \infty$. Similarly, we can show the bound for others.

Consider (b).
 We first investigate the first derivatives of $h_{l,i}(\theta)$ with $l=1$.
$h_{1,i}(\theta)$ does not depend on $\omega_{2}^{g}, \gamma_2, \alpha_{2}^{g}, \beta_{2+}^{g}, \beta_{2-}^{g}, \kappa_{2}^{g}, \alpha_{21}^{g}$, $\beta_{21+}^{g}$ and $\beta_{21-}^{g}$.
Since $h_{1,i}(\theta)$ is the linear function of $\omega_{1}^{g}, \alpha_{1}^{g}, \beta_{1+}^{g}, \beta_{1-}^{g}, \kappa_{1}^{g}, \alpha_{12}^{g}$, $\beta_{12+}^{g}$ and $\beta_{12-}^{g}$, we can show that
$$ h_{l,i}(\theta) ^{-1}\frac{\partial h_{l,i}(\theta)}{\partial \theta_j} \leq C \text{ a.s. for } j=1,3,4,5,6,7,8,9. $$
In case of $\gamma_1$, simple algebra shows that
\begin{eqnarray*}
    h_{1,i}(\theta) ^{-1} \frac{\partial h_{1,i}(\theta)}{\partial \gamma_1} &=& h_{1,i}(\theta) ^{-1} \left (  h_{1,i-1}(\theta) + \gamma_1 \frac{\partial h_{1,i-1}(\theta)}{\partial \gamma_1} \right )\\
    &=& h_{1,i}(\theta) ^{-1} \sum_{r=0}^{i-2} \gamma_1 ^{r} h_{1,i-1-r}(\theta) + \gamma_1 ^{i-1} h_{1,i}(\theta) ^{-1} \frac{\partial h_{1,1}(\theta)}{\partial \gamma_1} \\
    & \leq & \sum_{r=0}^{i-2} \frac{\gamma_1 ^{r} h_{1,i-1-r}(\theta)}{\omega_{1}^{g} + \gamma_1 ^{q} h_{1,i-1-r}(\theta)} + C \\
    & \leq & \sum_{r=0}^{i-2} \gamma_1 ^{rs} \left (\frac{h_{1,i-1-r}(\theta)}{\omega_{1}^g} \right )^{s} + C \\
    & \leq & C \sum_{r=0}^{i-2} \gamma_u ^{rs} \left ( h_{1,i-1-r}(\theta)\right)^{sp} + C \quad  \text{ a.s.},
\end{eqnarray*}
where the second inequality is due to the fact that $x/(1+x) \leq x^s$ for any $s \in [0,1]$ and all $x \geq 0$.
For given $ p \geq 1 $, we choose $s \in [0,1]$ such that $E\left (h_{1,i-1-r}(\theta) \right )^{sp} < \infty$.
Then, by the fact that $\gamma < 1$, we have
\begin{equation*}
    \sup_{i \in \mathbb{N}} \norm{\sup_{\theta \in \Theta} h_{1,i}(\theta) ^{-1} \frac{\partial h_{1,i}(\theta)}{\partial \gamma_1} }_{L_p} \leq C .
\end{equation*}
Similarly, we can show the bound for others.

Consider (c).
 We have
\begin{eqnarray*}
    \frac{\partial^3\hat{l}_i(\theta)}{\partial\theta_j\partial\theta_k\partial\theta_q} & = & \sum_{l=1}^{2} \left \lbrace 1 - \frac{RV_{l,i}/\lambda_l}{\hat{h}_{l,i}(\theta)} \right \rbrace
    \left \lbrace \frac{1}{\hat{h}_{l,i}(\theta)} \frac{\partial^3\hat{h}_{l,i}(\theta)}{\partial\theta_j\partial\theta_k\partial\theta_q} \right \rbrace
    \cr && +
    \left \lbrace 2 \frac{RV_{l,i}/\lambda_l}{\hat{h}_{l,i}(\theta)} -1  \right \rbrace
    \left \lbrace \frac{1}{\hat{h}_{l,i}(\theta)} \frac{\partial \hat{h}_{l,i}(\theta)}{\partial \theta_j} \right \rbrace
    \left \lbrace \frac{1}{\hat{h}_{l,i}(\theta)} \frac{\partial^2\hat{h}_{l,i}(\theta)}{\partial\theta_k\partial\theta_q} \right \rbrace
    \cr && +
    \left \lbrace 2 \frac{RV_{l,i}/\lambda_l}{\hat{h}_{l,i}(\theta)} -1  \right \rbrace
    \left \lbrace \frac{1}{\hat{h}_{l,i}(\theta)} \frac{\partial \hat{h}_{l,i}(\theta)}{\partial \theta_k} \right \rbrace
    \left \lbrace \frac{1}{\hat{h}_{l,i}(\theta)} \frac{\partial^2\hat{h}_{l,i}(\theta)}{\partial\theta_j\partial\theta_q} \right \rbrace
    \cr && +
    \left \lbrace 2 \frac{RV_{l,i}/\lambda_l}{\hat{h}_{l,i}(\theta)} -1  \right \rbrace
    \left \lbrace \frac{1}{\hat{h}_{l,i}(\theta)} \frac{\partial \hat{h}_{l,i}(\theta)}{\partial \theta_q} \right \rbrace
    \left \lbrace \frac{1}{\hat{h}_{l,i}(\theta)} \frac{\partial^2\hat{h}_{l,i}(\theta)}{\partial\theta_j\partial\theta_k} \right \rbrace
    \cr && +
    \left \lbrace 2 - 6 \frac{RV_{l,i}/\lambda_l}{\hat{h}_{l,i}(\theta)} \right \rbrace
    \left \lbrace \frac{1}{\hat{h}_{l,i}(\theta)} \frac{\partial \hat{h}_{l,i}(\theta)}{\partial \theta_j} \right \rbrace
    \left \lbrace \frac{1}{\hat{h}_{l,i}(\theta)} \frac{\partial \hat{h}_{l,i}(\theta)}{\partial \theta_k} \right \rbrace
    \left \lbrace \frac{1}{\hat{h}_{l,i}(\theta)} \frac{\partial \hat{h}_{l,i}(\theta)}{\partial \theta_q} \right \rbrace.
\end{eqnarray*}
Since $\hat{h}_{1,i}(\theta)$ and $\hat{h}_{2,i}(\theta)$ are stay away from 0, we have for $l \in \lbrace 1,2 \rbrace$,
\begin{eqnarray*}
    && E \left [ \sup_{\theta \in \Theta} \abs{\frac{RV_{l,i}/\lambda_l}{\hat{h}_{l,i}(\theta)} \left \lbrace \frac{1}{\hat{h}_{l,i}(\theta)} \frac{\partial^3\hat{h}_i(\theta)}{\partial\theta_j\partial\theta_k\partial\theta_q} \right \rbrace} \right ] \cr
    && \leq C \norm{RV_{l,i}}_{L_2} \norm{\sup_{\theta \in \Theta} \abs{\frac{1}{\hat{h}_{l,i}(\theta)} \frac{\partial^3\hat{h}_i(\theta)}{\partial\theta_j\partial\theta_k\partial\theta_q}}}_{L_2} \leq C,
\end{eqnarray*}
where the first and second inequalities are due to H\"{o}lder's inequality and Lemma \ref{lemma1}(b), respectively.
Similarly we can bound the rest of terms.

Consider (d).
 Simple algebra shows that
\begin{flalign*}
    \hat{h}_{1,i}(\theta) - h_{1,i}(\theta) =& \frac{\alpha_{1}^{g}}{\lambda_{1}} \sum_{j=1}^{i-1} \gamma_{1}^{j-1}(RV_{1,i-1} - IV_{1,i-1}) + \frac{\beta_{1+}^{g}}{\lambda_{1}} \sum_{j=1}^{i-1} \gamma_{1}^{j-1}(JV_{1,i-1}^{+} - IJ_{1,i-1}^{+})\\
     &+ \frac{\beta_{1-}^{g}}{\lambda_{1}} \sum_{j=1}^{i-1} \gamma_{1}^{j-1} (JV_{1,i-1}^{-} - IJ_{1,i-1}^{-})  + \frac{\alpha_{12}^{g}}{\lambda_{2}} \sum_{j=1}^{i-1} \gamma_{1}^{j-1} (RV_{2,i-1} - IV_{2,i-1})\\
     &+ \frac{\beta_{12+}^{g}}{\lambda_{2}} \sum_{j=1}^{i-1} \gamma_{1}^{j-1} (JV_{2,i-1}^{+} - IJ_{2,i-1}^{+}) + \frac{\beta_{12-}^{g}}{\lambda_{2}} \sum_{j=1}^{i-1} \gamma_{1}^{j-1} (JV_{2,i-1}^{-} - IJ_{2,i-1}^{-}).
\end{flalign*}
Using the facts that $|x| < \xi$ for any $x \in \{\alpha_{1}^{g},\beta_{1+}^{g},\beta_{1-}^{g},\alpha_{12}^{g},\beta_{12+}^{g},\beta_{12-}^{g}\}$ and $0 < \gamma_1 < 1$, and the Assumption \ref{assumption-1} (e), we can easily show
\begin{equation*}
    \sup_{i \in \mathbb{N}} \norm{\sup_{\theta \in \Theta} \left( \hat{h}_{1,i}(\theta) - h_{1,i}(\theta) \right)}_{L_2} \leq Cm^{-1/4}.    
\end{equation*}
Similarly, we can show the bound for the others.
$\blacksquare$

\begin{lemma}\label{lemma2}
    Under Assupmtion \ref{assumption-1}, $\htheta$ converges to $\theta_0$ in probability.
\end{lemma}

\textbf{Proof of Lemma \ref{lemma2}.}
First, we will show that
\begin{equation*}
    \sup_{\theta \in \Theta} | \hat{L}_{n,m} (\theta) - L _{n} (\theta) | \xrightarrow{p} 0.
\end{equation*}
By the triangular inequality, we have
\begin{equation} \label{eq-4.1}
    | \hat{L}_{n,m} (\theta) - L _{n} (\theta) | \leq | \hat{L}_{n,m} (\theta) - \hat{L} _{n} (\theta) | + | \hat{L}_{n} (\theta) - L _{n} (\theta) |. 
\end{equation}
For the first term on the right hand side of \eqref{eq-4.1}, we have
\begin{equation*}
    | \hat{L}_{n,m} (\theta) - \hat{L} _{n} (\theta) |  \leq \frac{1}{2n} \sum_{l=1}^{2} \sum_{i=1}^{n} \Big | \log(\hat{h}_{l,i}(\theta)) - \log(h_{l,i}(\theta)) \Big | +  \Biggl | \frac{RV_{l,i}/\lambda_l}{\hat{h}_{l,i}(\theta)} - \frac{IV_{l,i}/\lambda_l}{h_{l,i}(\theta)} \Biggl |.
\end{equation*}
By Lemma \ref{lemma1}(d), we have
\begin{eqnarray*}
    E \left [ \sup_{\theta \in \Theta} \frac{1}{2n}\sum_{i=1}^{n} \Big | \log(\hat{h}_{l,i}(\theta)) - \log(h_{l,i}(\theta)) \Big |  \right ] & \leq & \frac{C}{n}\sum_{i=1}^{n} E \left [ \sup_{\theta \in \Theta} \Big | \hat{h}_{l,i}(\theta) - h_{l,i}(\theta) \Big |  \right ] \\
    & \leq & C m^{-1/4},
\end{eqnarray*}
where the first inequality is due to the fact that $\hat{h}_{l,i}(\theta)$ and $h_{l,i}(\theta)$ stay away from zero.
By triangular inequality, we have
\begin{equation*}
 \frac{1}{2n}  \sum_{i=1}^{n} \Biggl | \frac{RV_{l,i}/\lambda_l}{\hat{h}_{l,i}(\theta)} - \frac{IV_{l,i}/\lambda_l}{h_{l,i}(\theta)} \Biggl |  \leq \frac{C}{n}  \sum_{i=1}^{n} \Biggl | IV_{l,i} \frac{\hat{h}_{l,i}(\theta) - h_{l,i}(\theta)}{\hat{h}_{l,i}(\theta)h_{l,i}(\theta)} \Biggl | + \frac{C}{n}  \sum_{i=1}^{n} \Biggl | \frac{RV_{l,i} - IV_{l,i}}{\hat{h}_{l,i}(\theta)} \Biggl |.
\end{equation*}
By H\"{o}lder's inequality and the fact that $\hat{h}_{l,i}(\theta)$ and $h_{l,i}(\theta)$ stay away from zero, we have
\begin{eqnarray*}
    E\left[ \sup_{\theta \in \Theta} \frac{C}{n} \sum_{i=1}^{n} \Biggl | IV_{l,i} \frac{\hat{h}_{l,i}(\theta) - h_{l,i}(\theta)}{\hat{h}_{l,i}(\theta)h_{l,i}(\theta)} \Biggl | \right ] &\leq& \frac{C}{n}\sum_{i=1}^{n}E\left[ |IV_{l,i}| \sup_{\theta \in \Theta} \left | \hat{h}_{l,i}(\theta) - h_{l,i}(\theta) \right |\right]\\
    &\leq & \frac{C}{n}\sum_{i=1}^{n}\norm{IV_{l,i}}_{L_2} \norm{\sup_{\theta \in \Theta} \left | \hat{h}_{l,i}(\theta) - h_{l,i}(\theta) \right |}_{L_2} \cr
	& \leq&  Cm^{-1/4}
\end{eqnarray*}
and
\begin{equation*}
    E\left[ \sup_{\theta \in \Theta} \frac{C}{n}  \sum_{i=1}^{n} \Biggl | \frac{RV_{l,i} - IV_{l,i}}{\hat{h}_{l,i}(\theta)} \Biggl | \right] \leq \frac{C}{n}  \sum_{i=1}^{n} \norm{RV_{l,i} - IV_{l,i}}_{L_2} \leq Cm^{-1/4},
\end{equation*}
where the third inequality is due to Lemma \ref{lemma1}(d).
This implies that
\begin{equation*}
    \sup_{\theta \in \Theta} | \hat{L}_{n,m} (\theta) - \hat{L} _{n} (\theta) | = O_p(m^{-1/4}).
\end{equation*}

For the second term on the right hand side of \eqref{eq-4.1}, we have
\begin{eqnarray*}
    | \hat{L}_{n} (\theta) - L _{n} (\theta) | & = & \left | - \frac{1}{2n} \sum_{i=1}^{n} \left \lbrace \frac{D_{1,i}}{\lambda_1 h _{1,i}(\theta)} + \frac{D_{2,i}}{\lambda_2 h _{2,i}(\theta)} \right \rbrace \right | \\
    & \leq &  \left | \frac{1}{2n} \sum_{i=1}^{n}  \frac{D_{1,i}}{\lambda_1 h _{1,i}(\theta)} \right | + \left | \frac{1}{2n} \sum_{i=1}^{n} \frac{D_{2,i}}{\lambda_2 h _{2,i}(\theta)} \right |.
\end{eqnarray*}
Since $h_{l,i}(\theta)$ is adapted to $\mathcal{F}_{i-1}$, $\frac{D_{l,i}}{h_{l,i}}(\theta)$ is also a martingale difference.
Assumption \ref{assumption-1}(b) implies that $D_{1,i}$ and $D_{2,i}$ are uniform integrable.
The fact that $\left| \frac{D_{l,i}}{h_{l,i}(\theta)} \right| \leq \frac{1}{c}\left| D_{l,i} \right|$ implies the uniform integrability of $\left| \frac{D_{l,i}}{h_{l,i}(\theta)} \right|$.
Thus, by Theorem 2.22 in \citet{hall2014martingale}, we have $| \hat{L}_{n} (\theta) - L _{n} (\theta) | \xrightarrow{p} 0$ .

To establish the uniform convergence, we define $G_n(\theta) = \hat{L}_{n} (\theta) - L _{n} (\theta)$. If $G_n(\theta)$ satisfies a weak Lipschitz condition, the uniform convergence can be obtained by Theorem 3 in \citet{andrews1992generic}.
By the mean value theorem, there exists $\theta^{*}$ between $\theta$ and $\theta '$ such that
\begin{eqnarray*}
    | G_n(\theta) - G_n(\theta ') | & = & \left | \frac{1}{2n} \sum_{l=1}^{2} \sum _{i=1} ^{n} \left [ \frac{\partial h_{l,i}(\theta^{*})}{\partial \theta} \frac{D_{l,i}}{\lambda_l h _{l,i} ^{2} (\theta^{*})}\right ] (\theta - \theta ') \right | \\
    & \leq & \frac{C}{n} \sum_{l=1}^{2} \sum _{i=1} ^{n} \norm{ \frac{\partial h_{l,i}(\theta^{*})}{\partial \theta} \frac{D_{l,i}}{\lambda_l h _{l,i} ^{2} (\theta^{*})} }_{\max} \norm{(\theta - \theta ')}_{\max}.
\end{eqnarray*}
By H\"{o}lder's inequality and Lemma \ref{lemma1}(b), we have for $j \in \lbrace 1,2, \ldots , 18 \rbrace$,
\begin{eqnarray*}
    \norm{ \frac{\partial h_{l,i}(\theta^{*})}{\partial \theta_j} \frac{D_{l,i}}{\lambda_1 h _{l,i} ^{2} (\theta^{*})} }_{L_1} & \leq & \norm{\sup_{\theta^{*} \in \Theta} \frac{\partial h_{l,i}(\theta^{*})}{\partial \theta_j} \frac{D_{l,i}}{\lambda_1 h _{l,i} ^{2} (\theta^{*})} }_{L_1} \\
    & \leq & C \norm{\sup_{\theta^{*} \in \Theta} h_{l,i}(\theta^{*}) ^{-1} \frac{\partial h_{l,i}(\theta^{*})}{\partial \theta_j} D_{l,i}}_{L_1} \\
    & \leq & C \norm{\sup_{\theta^{*} \in \Theta} h_{l,i}(\theta^{*}) ^{-1} \frac{\partial h_{l,i}(\theta^{*})}{\partial \theta_j}}_{L_2} \norm{ D_{l,i}}_{L_2} \\
    & \leq & C < \infty.
\end{eqnarray*}
Thus, $G_n(\theta)$ satisfies a weak Lipschitz condition which implies that $G_n(\theta)$ is stochastic equicontinuous and thus it uniformly converges to zero.

Now, we need to show the uniqueness of the maximizer of $L_n(\theta)$. Simple algebra shows that
\begin{equation*}
    \max_{\theta \in \Theta} L_n(\theta) \leq -\frac{1}{2n} \sum_{l=1}^{2} \sum _{i=1} ^{n} \min_{\theta_{l,i} \in \Theta} \left ( \log (h_{l,i}(\theta_{l,i})) + \frac{h_{l,i}(\theta_0)}{h_{l,i}(\theta_{l,i})} \right ).
\end{equation*}
Then $\theta_{1,i}$ and $\theta_{2,i}$ satisfies $h_{1,i} (\theta_{1,i}) = h_{1,i}(\theta_0)$ and $h_{2,i} (\theta_{2,i}) = h_{2,i}(\theta_0)$, respectively.
Thus, if there exists $\theta^{*} \in \Theta$ such that $h_{1,i} (\theta^{*}) = h_{1,i}(\theta_0)$ and $h_{2,i} (\theta^{*}) = h_{2,i}(\theta_0)$ for all $i=1,2,\ldots,n$, then $\theta^{*}$ is the maximizer.
Suppose that there exists the maximizer $\theta ^{*}$ which is not the same as $\theta _0$.
Since
\begin{eqnarray*}
    h_{1,i}(\theta) &=& \omega_{1}^g + \gamma_{1}h_{1,i-1}(\theta) + \alpha_{1}^g{\lambda_1}^{-1}IV_{1,i-1} + \beta_{1+}^g \lambda_1^{-1}IJ_{1,i-1}^{+} + \beta_{1-}^g \lambda_1^{-1}IJ_{1,i-1}^{-} \cr && + \kappa_{1}^g(1 - \lambda_1)^{-1}OV_{1,i-1} + \alpha_{12}^g{\lambda_2}^{-1}IV_{2,i-1} + \beta_{12+}^g \lambda_2^{-1}IJ_{2,i-1}^{+} + \beta_{12-}^g \lambda_2^{-1}IJ_{2,i-1}^{-},\cr
    h_{2,i}(\theta) &=& \omega_{2}^g + \gamma_{2}h_{2,i-1}(\theta) + \alpha_{2}^g{\lambda_2}^{-1}IV_{2,i-1} + \beta_{2+}^g \lambda_2^{-1}IJ_{2,i-1}^{+} + \beta_{2-}^g \lambda_2^{-1}IJ_{2,i-1}^{-} \cr && + \kappa_{2}^g(1 - \lambda_2)^{-1}OV_{2,i-1} + \alpha_{21}^g{\lambda_1}^{-1}IV_{1,i} + \beta_{21+}^g \lambda_1^{-1}IJ_{1,i}^{+} + \beta_{21-}^g \lambda_1^{-1}IJ_{1,i}^{-},
\end{eqnarray*}
\begin{equation*}
    M_1 
    \begin{pmatrix}
        \omega_{01}^g - \omega_{1}^{*} \\
        \gamma_{01} - \gamma_{1}^{*} \\
        \alpha_{01}^g - \alpha_{1}^{*} \\
        \beta_{01+}^g - \beta_{1+}^{*} \\
        \beta_{01-}^g - \beta_{1-}^{*} \\
        \kappa_{01}^g - \kappa_{1}^{*} \\
        \alpha_{012}^g - \alpha_{12}^{*} \\
        \beta_{012+}^g - \beta_{12+}^{*} \\
        \beta_{012-}^g - \beta_{12-}^{*} \\
    \end{pmatrix}
    = 0 \quad \text{ and } \quad
    M_2 
    \begin{pmatrix}
        \omega_{02}^g - \omega_{2}^{*} \\
        \gamma_{02} - \gamma_{2}^{*} \\
        \alpha_{02}^g - \alpha_{2}^{*} \\
        \beta_{02+}^g - \beta_{2+}^{*} \\
        \beta_{02-}^g - \beta_{2-}^{*} \\
        \kappa_{02}^g - \kappa_{2}^{*} \\
        \alpha_{021}^g - \alpha_{21}^{*} \\
        \beta_{021+}^g - \beta_{21+}^{*} \\
        \beta_{021-}^g - \beta_{21-}^{*} \\
    \end{pmatrix}
    = 0 \quad  \text{a.s.},
\end{equation*}
where
\begin{equation*}
M_1 = 
\begin{pmatrix}
1       & h_{1,1}(\theta_0)     & \frac{IV_{1,1}}{\lambda_1}    & \frac{IJ_{1,1}^{+}}{\lambda_1}    & \frac{IJ_{1,1}^{-}}{\lambda_1}    & \frac{OV_{1,1}}{1-\lambda_1}      & \frac{IV_{2,1}}{\lambda_2}  & \frac{IJ_{2,1}^{+}}{\lambda_2}  & \frac{IJ_{2,1}^{-}}{\lambda_2}  \\
1       & h_{1,2}(\theta_0)     & \frac{IV_{1,2}}{\lambda_1}    & \frac{IJ_{1,2}^{+}}{\lambda_1}    & \frac{IJ_{1,2}^{-}}{\lambda_1}    & \frac{OV_{1,2}}{1-\lambda_1}      & \frac{IV_{2,2}}{\lambda_2}  & \frac{IJ_{2,2}^{+}}{\lambda_2}  & \frac{IJ_{2,2}^{-}}{\lambda_2}  \\
\vdots  & \vdots                & \vdots                & \vdots                & \vdots                            & \vdots                      & \vdots              & \vdots              \\
1       & h_{1,n-1}(\theta_0)   & \frac{IV_{1,n-1}}{\lambda_1}  & \frac{IJ_{1,n-1}^{+}}{\lambda_1}  & \frac{IJ_{1,n-1}^{-}}{\lambda_1}  & \frac{OV_{1,n-1}}{1-\lambda_1}    & \frac{IV_{2,n-1}}{\lambda_2}& \frac{IJ_{2,n-1}^{+}}{\lambda_2}& \frac{IJ_{2,n-1}^{-}}{\lambda_2}
\end{pmatrix}
\end{equation*}
and
\begin{equation*}
M_2 = 
\begin{pmatrix}
1       & h_{2,1}(\theta_0)     & \frac{IV_{2,1}}{\lambda_1}    & \frac{IJ_{2,1}^{+}}{\lambda_1}    & \frac{IJ_{2,1}^{-}}{\lambda_1}    & \frac{OV_{2,1}}{1-\lambda_1}      & \frac{IV_{1,1}}{\lambda_2}  & \frac{IJ_{1,1}^{+}}{\lambda_2}  & \frac{IJ_{1,1}^{-}}{\lambda_2}  \\
1       & h_{2,2}(\theta_0)     & \frac{IV_{2,2}}{\lambda_1}    & \frac{IJ_{2,2}^{+}}{\lambda_1}    & \frac{IJ_{2,2}^{-}}{\lambda_1}    & \frac{OV_{2,2}}{1-\lambda_1}      & \frac{IV_{1,2}}{\lambda_2}  & \frac{IJ_{1,2}^{+}}{\lambda_2}  & \frac{IJ_{1,2}^{-}}{\lambda_2}  \\
\vdots  & \vdots                & \vdots                & \vdots                & \vdots                            & \vdots                      & \vdots              & \vdots              \\
1       & h_{2,n-1}(\theta_0)   & \frac{IV_{2,n-1}}{\lambda_1}  & \frac{IJ_{2,n-1}^{+}}{\lambda_1}  & \frac{IJ_{1,n-1}^{-}}{\lambda_1}  & \frac{OV_{2,n-1}}{1-\lambda_1}    & \frac{IV_{1,n-1}}{\lambda_2}& \frac{IJ_{1,n-1}^{+}}{\lambda_2}& \frac{IJ_{1,n-1}^{-}}{\lambda_2}
\end{pmatrix}.
\end{equation*}
Since $IV_{1,i}$'s, $IV_{2,i}$'s, $OV_{1,i}$'s, $OV_{2,i}$'s, $IJ_{1,i}^{+}$'s, $IJ_{1,i}^{-}$'s, $IJ_{2,i}^{+}$'s, and $IJ_{2,i}^{-}$'s are nondegenerate random variables, we have
\begin{equation*}
    \begin{pmatrix}
        \omega_{01}^g - \omega_{1}^{*} \\
        \gamma_{01} - \gamma_{1}^{*} \\
        \alpha_{01}^g - \alpha_{1}^{*} \\
        \beta_{01+}^g - \beta_{1+}^{*} \\
        \beta_{01-}^g - \beta_{1-}^{*} \\
        \kappa_{01}^g - \kappa_{1}^{*} \\
        \alpha_{012}^g - \alpha_{12}^{*} \\
        \beta_{012+}^g - \beta_{12+}^{*} \\
        \beta_{012-}^g - \beta_{12-}^{*} \\
    \end{pmatrix}
    = 0 \quad \text{ and } \quad
    \begin{pmatrix}
        \omega_{02}^g - \omega_{2}^{*} \\
        \gamma_{02} - \gamma_{2}^{*} \\
        \alpha_{02}^g - \alpha_{2}^{*} \\
        \beta_{02+}^g - \beta_{2+}^{*} \\
        \beta_{02-}^g - \beta_{2-}^{*} \\
        \kappa_{02}^g - \kappa_{2}^{*} \\
        \alpha_{021}^g - \alpha_{21}^{*} \\
        \beta_{021+}^g - \beta_{21+}^{*} \\
        \beta_{021-}^g - \beta_{21-}^{*} \\
    \end{pmatrix}
    = 0 \quad a.s.,
\end{equation*}
which implies $\theta^{*} = \theta_0$ a.s.
Thus, there is a unique maximizer.
Now, $\hat{\theta} \xrightarrow{p} \theta_0$ is a consequence of Theorem 1 in \citet{xiu2010quasi}.
$\blacksquare$

\begin{lemma}\label{lemma3}
Under Assumption \ref{assumption-1}, we have
\begin{enumerate}
    \item [(a)] $-\nabla\psi_{n}(\theta_0)$ is a positive definite matrix for $n \geq 18$, and $-\nabla\psi_{n}(\theta_0)\xrightarrow{p}B$;
    \item [(b)] $\hat{\psi}_{n,m}(\theta_0) = O_p(n^{-1/2}) + O_p(m^{-1/4})$;
    \item [(c)] $-\sqrt{n}\hat{\psi}_{n}(\theta_0)=\frac{1}{\sqrt{n}}\left(\frac{1}{2}\sum_{i=1}^{n}\frac{\partial h_{1,i}(\theta)}{\partial\theta}h_{1,i}^{-1}\frac{D_{1,i}}{\lambda_{1}h_{1,i}} + \frac{\partial h_{2,i}(\theta)}{\partial\theta}h_{2,i}^{-1}\frac{D_{2,i}}{\lambda_{2}h_{2,i}}\right)\xrightarrow{d}N(0,A)$.
\end{enumerate}
\end{lemma}

\textbf{Proof of Lemma \ref{lemma3}.} 
Consider (a).
 Simple algebra shows
\begin{equation*}
    -\nabla \psi_n (\theta_0) = \frac{1}{2n} \sum_{l=1}^{2} \sum_{i=1}^{n} \frac{\partial h_{l,i}(\theta_0)}{\partial \theta} \frac{\partial h_{l,i}(\theta_0) ^\top}{\partial \theta} h_{l,i} (\theta_0) ^{-2} = \frac{1}{2n} \sum_{l=1}^{2} \sum_{i=1}^{n} h_{\theta_0,l,i} h_{\theta_0,l,i} ^\top,
\end{equation*}
where $h_{\theta_0,l,i} = \frac{\partial h_{l,i}(\theta_0)}{\partial \theta} h_{l,i} (\theta_0) ^{-1}$.
Suppose that $-\nabla \psi_n (\theta_0)$ is not a positive definite matrix.
Then, there exists non-zero $\phi \in \mathbb{R}^{18}$ such that $\frac{1}{2n} \sum_{l=1}^{2} \sum_{i=1}^{n} \phi^\top h_{\theta_0,l,i} h_{\theta_0,l,i} ^\top \phi = 0$.
This implies that
\begin{equation*}
    h_{\theta_0,l,i}^\top \phi = 0 \text{ a.s. for any } i = 1, \ldots , n \text{ and } l = 1,2 .
\end{equation*}
Similar to the proofs of Lemma 4 in \citet{kim2016unified}, we can show that $\phi = 0$ a.s., which is contradiction, using the fact that $h_{l,i}(\theta_0)$ stays away from zero and $(IV_{1,i}, IV_{2,i}, OV_{1,i},$ $OV_{2,i}, $ $IJ_{1,i}^{+}, IJ_{1,i}^{-}, IJ_{2,i}^{+}, IJ_{2,i}^{-})$'s are nondegenerate.
Therefore, $-\nabla \psi_n (\theta_0)$ is a positive definite matrix.
On the other hand, we have $-\nabla \psi_n (\theta_0) \xrightarrow{p} B$, by the ergodic theorem.

Consider (b).
 It is enough to show that
\begin{equation}\label{lemma3b-1}
    \hat{\psi}_{n,m}(\theta_0) - \hat{\psi}_{n}(\theta_0) = O_p (m^{-1/4})
\end{equation}
and
\begin{equation*}
    \hat{\psi}_{n}(\theta_0) = O_p(n^{-1/2}).
\end{equation*}
Simple algebra shows that
\begin{eqnarray}\label{lemma3b-1tri}
\hat{\psi}_{n,m}(\theta_0) - \hat{\psi}_{n}(\theta_0) &=& -\frac{1}{2n} \sum_{l=1}^{2} \sum_{i=1}^{n}    \hat{h}_{l,i}^{-1}(\theta_0) \frac{\partial \hat{h}_{l,i}(\theta_0)}{\partial \theta} - h_{l,i}^{-1}(\theta_0) \frac{\partial h_{l,i}(\theta_0)}{\partial \theta}  \cr
&& + \frac{1}{2n} \sum_{l=1}^{2} \sum_{i=1}^{n}  \frac{\partial \hat{h}_{l,i}(\theta_0)}{\partial \theta} \frac{RV_{l,i}/\lambda_l}{\hat{h}_{l,i}^2(\theta_0)} -  \frac{\partial h_{l,i}(\theta_0)}{\partial \theta} \frac{IV_{l,i}/\lambda_l}{h_{l,i}^2(\theta_0)}.
\end{eqnarray}
For the first term on the right hand side of \eqref{lemma3b-1tri}, we can apply Minkowski's inequality to show that for any $j \in \lbrace 1, \ldots , 18 \rbrace$,
\begin{eqnarray*}
&&\norm{-\frac{1}{2n} \sum_{l=1}^{2} \sum_{i=1}^{n}    \hat{h}_{l,i}^{-1}(\theta_0) \frac{\partial \hat{h}_{l,i}(\theta_0)}{\partial \theta_j} - h_{l,i}^{-1}(\theta_0) \frac{\partial h_{l,i}(\theta_0)}{\partial \theta_j} }_{L_1} \cr
&&\leq \norm{-\frac{1}{2n} \sum_{l=1}^{2} \sum_{i=1}^{n}  h_{l,i}^{-1}(\theta_0) \hat{h}_{l,i}^{-1}(\theta_0) \frac{\partial \hat{h}_{l,i}(\theta_0)}{\partial \theta_j} \Big( h_{l,i}(\theta_0) - \hat{h}_{l,i}(\theta_0) \Big) }_{L_1} \cr
&& \quad + \norm{\frac{1}{2n} \sum_{l=1}^{2} \sum_{i=1}^{n} h_{l,i}^{-1}(\theta_0) \Biggl( \frac{\partial \hat{h}_{l,i}(\theta_0)}{\partial \theta_j} - \frac{\partial h_{l,i}(\theta_0)}{\partial \theta_j} \Biggl)}_{L_1} \cr
&& \leq \frac{C}{n} \sum_{l=1}^{2} \sum_{i=1}^{n} \norm{\hat{h}_{l,i}^{-1}(\theta_0) \frac{\partial \hat{h}_{l,i}(\theta_0)}{\partial \theta_j}}_{L_2} \norm{\hat{h}_{l,i}(\theta_0) - h_{l,i}(\theta_0)}_{L_2} \cr
&& \quad + \frac{C}{n} \sum_{l=1}^{2} \sum_{i=1}^{n} \norm{\frac{\partial \hat{h}_{l,i}(\theta_0)}{\partial \theta_j} - \frac{\partial h_{l,i}(\theta_0)}{\partial \theta_j}}_{L_1} \\
&& \leq C m^{-1/4},
\end{eqnarray*}
where the second and last inequalities are due to the fact that $\hat{h}_{l,i}(\theta_0)$, $h_{l,i}(\theta_0)$ stay away from zero and H\"{o}lder's inequality, Lemma \ref{lemma1}(b) and (d), respectively.

For the second term on the right hand side of \eqref{lemma3b-1tri}, similar to the first term, we can show for any  $j \in \lbrace 1, \ldots , 18 \rbrace$,
\begin{eqnarray*}
&&\norm{\frac{1}{2n} \sum_{l=1}^{2} \sum_{i=1}^{n}  \frac{\partial \hat{h}_{l,i}(\theta_0)}{\partial \theta_j} \frac{RV_{l,i}/\lambda_l}{\hat{h}_{l,i}^2(\theta_0)} -  \frac{\partial h_{l,i}(\theta_0)}{\partial \theta_j} \frac{IV_{l,i}/\lambda_l}{h_{l,i}^2(\theta_0)}}_{L_1} \cr
&& = \Biggl\lVert \frac{1}{2n} \sum_{l=1}^{2} \sum_{i=1}^{n} \hat{h}_{l,i}^{-2}(\theta_0) \frac{\partial \hat{h}_{l,i}(\theta_0)}{\partial \theta_j} (RV_{l,i} - IV_{l,i}) + IV_{l,i} \hat{h}_{l,i}^{-2}(\theta_0) \Biggl( \frac{\partial \hat{h}_{l,i}(\theta_0)}{\partial \theta_j} - \frac{\partial h_{l,i}(\theta_0)}{\partial \theta_j} \Biggl) \cr
&& \quad \quad +  \frac{IV_{l,i}}{h_{l,i}(\theta_0)} h_{l,i}^{-1}(\theta_0) \frac{\partial h_{l,i}(\theta_0)}{\partial \theta_j} \Big(\frac{h_{l,i}^{2}(\theta_0)}{\hat{h}_{l,i}^{2}(\theta_0)} - 1 \Big)\Biggl\rVert_{L_1} \cr
&& \leq \Biggl\lVert \frac{1}{2n} \sum_{l=1}^{2} \sum_{i=1}^{n} \hat{h}_{l,i}^{-2}(\theta_0) \frac{\partial \hat{h}_{l,i}(\theta_0)}{\partial \theta_j} (RV_{l,i} - IV_{l,i}) \Biggl\rVert_{L_1} \cr
&& \quad \quad + \Biggl\lVert \frac{1}{2n} \sum_{l=1}^{2} \sum_{i=1}^{n} IV_{l,i} \hat{h}_{l,i}^{-2}(\theta_0) \Biggl( \frac{\partial \hat{h}_{l,i}(\theta_0)}{\partial \theta_j} - \frac{\partial h_{l,i}(\theta_0)}{\partial \theta_j} \Biggl) \Biggl\rVert_{L_1} \cr
&& \quad \quad + \Biggl\lVert \frac{1}{2n} \sum_{l=1}^{2} \sum_{i=1}^{n} \frac{IV_{l,i}}{h_{l,i}(\theta_0)} h_{l,i}^{-1}(\theta_0) \frac{\partial h_{l,i}(\theta_0)}{\partial \theta_j} \Big(\frac{h_{l,i}^{2}(\theta_0)}{\hat{h}_{l,i}^{2}(\theta_0)} - 1 \Big)\Biggl\rVert_{L_1} \cr
&& \leq C m^{-1/4} + \Biggl\lVert \frac{1}{2n} \sum_{l=1}^{2} \sum_{i=1}^{n} \frac{IV_{l,i}}{h_{l,i}(\theta_0)} h_{l,i}^{-1}(\theta_0) \frac{\partial h_{l,i}(\theta_0)}{\partial \theta_j} \left(h_{l,i}(\theta_0) - \hat{h}_{l,i}(\theta_0)\right) \frac{h_{l,i}(\theta_0) + \hat{h}_{l,i}(\theta_0)}{\hat{h}_{l,i}^{2}(\theta_0)} \Biggl\rVert_{L_1}.
\end{eqnarray*}
By the fact that $IV_{l,i} = \lambda_l h_{l,i}(\theta_0) + D_{l,i}$, Minkowski's inequality, and H\"{o}lder's inequality, we have
\begin{eqnarray*}
    && \Biggl\lVert \frac{1}{2n} \sum_{l=1}^{2} \sum_{i=1}^{n} \frac{IV_{l,i}}{h_{l,i}(\theta_0)} h_{l,i}^{-1}(\theta_0) \frac{\partial h_{l,i}(\theta_0)}{\partial \theta_j} \left(h_{l,i}(\theta_0) - \hat{h}_{l,i}(\theta_0)\right) \frac{h_{l,i}(\theta_0) + \hat{h}_{l,i}(\theta_0)}{\hat{h}_{l,i}^{2}(\theta_0)} \Biggl\rVert_{L_1} \cr
    && \leq \Biggl\lVert \frac{1}{2n} \sum_{l=1}^{2} \sum_{i=1}^{n} \lambda_l h_{l,i}^{-1}(\theta_0) \frac{\partial h_{l,i}(\theta_0)}{\partial \theta_j} \left(h_{l,i}(\theta_0) - \hat{h}_{l,i}(\theta_0)\right) \frac{h_{l,i}(\theta_0) + \hat{h}_{l,i}(\theta_0)}{\hat{h}_{l,i}^{2}(\theta_0)} \Biggl\rVert_{L_1} \cr
    && \qquad + \Biggl\lVert \frac{1}{2n} \sum_{l=1}^{2} \sum_{i=1}^{n} D_{l,i} h_{l,i}^{-1}(\theta_0) \frac{\partial h_{l,i}(\theta_0)}{\partial \theta_j} \left(h_{l,i}(\theta_0) - \hat{h}_{l,i}(\theta_0)\right) \frac{h_{l,i}(\theta_0) + \hat{h}_{l,i}(\theta_0)}{h_{l,i}(\theta_0) \hat{h}_{l,i}^{2}(\theta_0)} \Biggl\rVert_{L_1} \cr
    && \leq \Biggl\lVert \frac{C}{n} \sum_{l=1}^{2} \sum_{i=1}^{n} h_{l,i}^{-1}(\theta_0) \frac{\partial h_{l,i}(\theta_0)}{\partial \theta_j} \left(h_{l,i}(\theta_0) - \hat{h}_{l,i}(\theta_0)\right) h_{l,i}(\theta_0) \Biggl\rVert_{L_1} \cr
    && \qquad + \Biggl\lVert \frac{C}{n} \sum_{l=1}^{2} \sum_{i=1}^{n} h_{l,i}^{-1}(\theta_0) \frac{\partial h_{l,i}(\theta_0)}{\partial \theta_j} \left(h_{l,i}(\theta_0) - \hat{h}_{l,i}(\theta_0)\right) \Biggl\rVert_{L_1} \cr
    && \qquad + \Biggl\lVert \frac{C}{n} \sum_{l=1}^{2} \sum_{i=1}^{n} D_{l,i} h_{l,i}^{-1}(\theta_0) \frac{\partial h_{l,i}(\theta_0)}{\partial \theta_j} \left(h_{l,i}(\theta_0) - \hat{h}_{l,i}(\theta_0)\right) \Biggl\rVert_{L_1} \cr
    && \leq \frac{C}{n} \sum_{l=1}^{2} \sum_{i=1}^{n} \left \Vert h_{l,i}^{-1}(\theta_0) \frac{\partial h_{l,i}(\theta_0)}{\partial \theta_j} \right \Vert_{L_{\eta/(4+2\eta)}} \left \Vert \left(h_{l,i}(\theta_0) - \hat{h}_{l,i}(\theta_0)\right) \right \Vert_{L_2} \left\Vert h_{l,i}(\theta_0) \right \Vert_{L_{2+\eta}} \cr
    && \qquad + \frac{C}{n} \sum_{l=1}^{2} \sum_{i=1}^{n} \left \Vert h_{l,i}^{-1}(\theta_0) \frac{\partial h_{l,i}(\theta_0)}{\partial \theta_j} \right \Vert_{L_2} \left \Vert \left(h_{l,i}(\theta_0) - \hat{h}_{l,i}(\theta_0)\right) \right \Vert_{L_2} \cr
    && \qquad +  \frac{C}{n} \sum_{l=1}^{2} \sum_{i=1}^{n} \left \Vert D_{l,i} \right \Vert_{L_4} \left \Vert h_{l,i}^{-1}(\theta_0) \frac{\partial h_{l,i}(\theta_0)}{\partial \theta_j} \right \Vert_{L_4} \left \Vert \left(h_{l,i}(\theta_0) - \hat{h}_{l,i}(\theta_0)\right) \right \Vert_{L_2} \cr
    && \leq Cm^{-1/4},
\end{eqnarray*}
where the second and last inequalities are due to the fact that $\hat{h}_{l,i}(\theta)$ and $h_{l,i}(\theta)$ stay away from zero and Lemma \ref{lemma1}(b) and (d), respectively.
Thus, we have
\begin{equation*}
    \hat{\psi}_{n,m}(\theta_0) - \hat{\psi}_{n}(\theta_0) = O_p (m^{-1/4}).
\end{equation*}

In case of $\hat{\psi}_{n}(\theta_0) = O_p(n^{-1/2})$, we have
\begin{equation*}
\hat{\psi}_{n}(\theta_0) = \frac{1}{2n} \sum_{l=1}^{2}\sum_{i=1}^{n} \frac{\partial h_{l,i}(\theta_0)}{\partial \theta_j} \frac{D_{l,i}}{\lambda_l h_{l,i}^{2}(\theta_0)}.
\end{equation*}
By the tower property and Lemma \ref{lemma1}(b), we can show  for any $j \in \lbrace 1, \ldots , 18 \rbrace$,
\begin{eqnarray}\label{bound-Ed2}
&& E\left[ \left( \frac{1}{2n} \sum_{l=1}^{2}\sum_{i=1}^{n} \frac{\partial h_{l,i}(\theta_0)}{\partial \theta_j} \frac{D_{l,i}}{\lambda_l h_{l,i}^{2}(\theta_0)} \right)^2 \right] \cr
&& = \frac{1}{4n^2} \sum_{l=1}^{2}\sum_{i=1}^{n} E \left[ \left( \frac{\partial h_{l,i}(\theta_0)}{\partial \theta_j} \right)^2 h_{l,i}^{-2}(\theta_0) \frac{E\left[D_{l,i}^2 | \mathcal{F}_{i-1}\right]}{\lambda_l^2 h_{l,i}^{2}(\theta_0)} \right] \cr
&& \leq \frac{C}{n^2}\sum_{l=1}^{2}\sum_{i=1}^{n} E \left[ \left( \frac{\partial h_{l,i}(\theta_0)}{\partial \theta_j} \right)^2 h_{l,i}^{-2}(\theta_0) \frac{1}{\lambda_l^2 h_{l,i}^{2}(\theta_0)} \right] \cr
&& \leq \frac{C}{n^2} \sum_{l=1}^{2}\sum_{i=1}^{n} E \left[ \left( \frac{\partial h_{l,i}(\theta_0)}{\partial \theta_j} \right)^2 h_{l,i}^{-2}(\theta_0)  \right] \leq C n^{-1}.
\end{eqnarray}
where the second inequality is due to the fact that $\hat{h}_{l,i}(\theta)$ and $h_{l,i}(\theta)$ stay away from zero.
Thus, we have
 $$\hat{\psi}_{n}(\theta_0) = O_p(n^{-1/2}).$$
Finally, we establish $$\hat{\psi}_{n,m}(\theta_0) = O_p(n^{-1/2}) + O_p(m^{-1/4}).$$

Consider  (c).
 For any $\pi \in \mathbb{R}^{18}$, let 
$$d_i = \pi^{\top}\left[\frac{\partial h_{1,1}(\theta_0)}{\partial\theta}h_{1,1}^{-1}(\theta_0)\frac{D_{1,1}}{\lambda_{1}h_{1,1}(\theta_0)} + \frac{\partial h_{2,1}(\theta_0)}{\partial\theta}h_{2,1}^{-1}(\theta_0)\frac{D_{2,1}}{\lambda_{2}h_{2,1}(\theta_0)}\right].
$$
Then, $d_{i}$ is a martingale difference, and similar to the proof of \eqref{bound-Ed2}, we can show that $E(d_{i}^2) < \infty$.
$(D_{1,i}, D_{2,i}, IV_{1,i}, IV_{2,i}, OV_{1,i}, OV_{2,i}, IJ_{1,i}^{+},$ $ IJ_{2,i}^{+},$ $IJ_{1,i}^{-}, IJ_{2,i}^{-})$'s are stationary and ergodic processes, thus, $d_{i}$ is stationary and ergodic.
Applying the martingale central limit theorem, we obtain $\frac{1}{\sqrt{n}}\sum_{i=1}^{n}d_{i}\xrightarrow{d}N(0,E(d_{i}^2))$.
Using Cram\'er-Wold device, we can show that
\begin{equation*}
    -\sqrt{n}\hat{\psi}_{n}(\theta_0)=\frac{1}{\sqrt{n}}\left(\frac{1}{2}\sum_{i=1}^{n}\frac{\partial h_{1,i}(\theta)}{\partial\theta}h_{1,i}^{-1}\frac{D_{1,i}}{\lambda_{1}h_{1,i}} + \frac{\partial h_{2,i}(\theta)}{\partial\theta}h_{2,i}^{-1}\frac{D_{2,i}}{\lambda_{2}h_{2,i}}\right)\xrightarrow{d}N(0,A).    
\end{equation*}
$\blacksquare$

\textbf{Proof of Theorem \ref{theorem2}.}
By the mean value theorem, there exists a $\theta^*$ between $\theta_0$ and $\htheta$ such that
\begin{equation*}
    \hat{\psi}_{n,m}(\htheta) - \hat{\psi}_{n,m}(\theta_0) = -\hat{\psi}_{n,m}(\theta_0) = \nabla\hat{\psi}_{n,m}(\theta^*)(\htheta-\theta_0).
\end{equation*}
If $-\nabla\hat{\psi}_{n,m}(\theta^*) \xrightarrow{p}-\nabla\psi_n (\theta_0)$ which is a positive definite matrix by Lemma \ref{lemma3}(a), the convergence rate of $\norm{\htheta - \theta_0}_{max}$ is the same as that of $\hat{\psi}_{n,m}(\theta_0)$.
Similar to the proofs of Theorem 2 \citep{kim2016unified}, we can show $$\norm{\nabla\hat{\psi}_{n,m}(\theta^*) - \nabla\psi_n (\theta_0)}_{max} = o_p(1),$$ by using the result of Lemma \ref{lemma1}(c), Lemma \ref{lemma3}(a) and Theorem \ref{theorem2}.
We can show that
\begin{eqnarray*}
    \sqrt{n}(\hat{\theta}-\theta_0) = -\sqrt{n}B^{-1}\hat{\psi}_{n,m}(\theta_0) = -\sqrt{n}B^{-1}\hat{\psi}_{n}(\theta_0) + O_p(n^{1/2}m^{-1/4}),
\end{eqnarray*}
where the first and last equality is due to Lemma \ref{lemma3}(a) and \eqref{lemma3b-1}, respectively.
By Lemma \ref{lemma3}(c), we conclude $\sqrt{n}(\htheta-\theta_0)\xrightarrow{d}N(0,B^{-1}AB^{-1})$.
$\blacksquare$

\subsection{Proof of Theorem \ref{theorem3}}
\textbf{Proof of Theorem \ref{theorem3}.}
Note that $\hat{L}_{1,n_1,m}(\theta)$ and $\hat{L}_{2,n_2,m}(\theta)$ are the same, except for index. Therefore, all statements for $\htheta_{1}$ and $\htheta_{2}$ in the proof of Theorem \ref{theorem2} are satisfied.
We first show that
\begin{equation*}
    \sqrt{n_1}
    \begin{pmatrix}
    \htheta_1 - \theta_{1} \\
    \htheta_2 - \theta_{2}
    \end{pmatrix} \xrightarrow{d} N(0,V),
\end{equation*}
where $V =
\begin{pmatrix}
B_1^{-1}A_1B_1^{-1} & 0 \\
0 & rB_2^{-1}A_2B_2^{-1}
\end{pmatrix}.$ By Theorem \ref{theorem2}, it is enough to show that the covariance terms between $\htheta_{1}$ and $\htheta_{2}$ converge to zero as $n,m\rightarrow\infty$.  
For any $l_1, l_2 \in \lbrace1,2\rbrace^2$, $j_1,j_2 \in \lbrace1,2, \ldots ,18\rbrace^2$, $i_1 \leq 0$, and $i_2 > 0$, by the tower property, we can show that
\begin{eqnarray*}
    &&E\left[\frac{\partial h_{l_1,i_1}(\theta_0)}{\partial\theta_{j_1}}h_{l_1,i_1}^{-2}(\theta_0)D_{l_1,i_1}\frac{\partial h_{l_2,i_2}(\theta_0)}{\partial\theta_{j_2}}h_{l_2,i_2}^{-2}(\theta_0)D_{l_2,i_2}\right]\\
    &&= E\left[E\left[\frac{\partial h_{l_1,i_1}(\theta_0)}{\partial\theta_{j_1}}h_{l_1,i_1}^{-2}(\theta_0)D_{l_1,i_1}\frac{\partial h_{l_2,i_2}(\theta_0)}{\partial\theta_{j_2}}h_{l_2,i_2}^{-2}(\theta_0)D_{l_2,i_2}|\mathcal{F}_{i_2-1}\right]\right]\\
    &&= E\left[\frac{\partial h_{l_1,i_1}(\theta_0)}{\partial\theta_{j_1}}h_{l_1,i_1}^{-2}(\theta_0)D_{l_1,i_1}\frac{\partial h_{l_2,i_2}(\theta_0)}{\partial\theta_{j_2}}h_{l_2,i_2}^{-2}(\theta_0)E\left[D_{l_2,i_2}|\mathcal{F}_{i_2-1}\right]\right]\\
    &&=0,
\end{eqnarray*}
which implies the covariance terms between $\htheta_{1}$ and $\htheta_{2}$ converge to zero as $n,m\rightarrow\infty$.
By Slutsky's theorem, we have $\sqrt{n_1}\left(\htheta_1-\htheta_2 - \delta \right)\xrightarrow{d} N(0,B_1^{-1}A_1B_1^{-1}+r B_2^{-1}A_2B_2^{-1})$.
$\blacksquare$

\subsection{Proof of Proposition \ref{proposition1}}
\begin{lemma}\label{lemma4}
    Let $P_{m,i}$ and $Q_{m,i}$, $m,i \in \mathbb{N}$ are random variables such that $\sup_{i} \norm{P_{i,m}}_{L_1} \leq C$ and $\sup_{i} \norm{Q_{i,m}}_{L_1} \leq C m^{-1/4}$ as $m \rightarrow \infty$.
    We have as $m,n \rightarrow \infty$, $\left | \frac{1}{n}\sum_{i=1}^{n} P_{m,i}Q_{m,i} \right | \xrightarrow{p} 0$ if $n^{2}m^{-1} \rightarrow 0$.
\end{lemma}

\textbf{Proof of Lemma \ref{lemma4}.}
We can apply Markov's inequality and H\"{o}lder's inequality to show that for given $\epsilon > 0$,
\begin{eqnarray*}
    P \left ( \left | \frac{1}{n}\sum_{i=1}^{n} P_{m,i}Q_{m,i} \right | \geq \epsilon  \right) &\leq& P \left ( \frac{1}{n}\sum_{i=1}^{n} \left | P_{m,i}Q_{m,i} \right | \geq \epsilon \right) \\
    & \leq & \sum_{i=1}^{n} P \left ( \left | P_{m,i}Q_{m,i} \right | \geq \epsilon \right)\\
    & = & \sum_{i=1}^{n} P \left ( \left | P_{m,i}Q_{m,i} \right |^{1/2} \geq \epsilon^{1/2} \right)\\
    & \leq & \epsilon^{-1/2} \sum_{i=1}^{n} E\left[ \left | P_{m,i}Q_{m,i} \right |^{1/2} \right]\\
    & \leq & \epsilon^{-1/2} \sum_{i=1}^{n} \norm{P_{m,i}}_{L_1}^{2} \norm{Q_{m,i}}_{L_1}^{2} \\
    & \leq & C n m^{-1/2}.
\end{eqnarray*}
Therefore, $\left | \frac{1}{n}\sum_{i=1}^{n} P_{m,i}Q_{m,i} \right | \xrightarrow{p} 0$ as $n^{2}m^{-1} \rightarrow 0$.
$\blacksquare$

\textbf{Proof of Proposition \ref{proposition1}.}
If $\hat{A}_{1}$, $\hat{A}_{2}$, $\hat{B}_{1}$, and $\hat{B}_{2}$  are consistent estimators, then we can show the consistency of the estimator for asymptotic variance by the continuous mapping theorem and Theorem \ref{theorem2}.
Then, by the Slutsky's theorem, we can show the statement. 
Thus, it is enough to show that $\hat{A}_{1}\xrightarrow{p}A_{1}$, $\hat{A}_{2}\xrightarrow{p}A_{2}$, $\hat{B}_{1}\xrightarrow{p}B_{1}$, and $\hat{B}_{2}\xrightarrow{p}B_{2}$.
Consider $\hat{B}_{2}\xrightarrow{p}B_{2}$.
We first define
\begin{flalign*}
    B_2(\theta) &=  \frac{1}{4} E\left[ \sum_{l=1}^{2} \frac{ \partial h_{l,1}(\theta)}{\partial\theta}\frac{\partial h_{l,1}(\theta)}{\partial\theta ^{\top}}h_{l,1}^{-2}(\theta )\right] \quad \text{and} \\
    \hat{B}_2(\theta) &=  \frac{1}{4n_2} \sum_{i=1}^{n_2} \sum_{l=1}^{2} \frac{ \partial \hat{h}_{l,i}(\theta)}{\partial\theta}\frac{\partial \hat{h}_{l,i}(\theta)}{\partial\theta ^{\top}}h_{l,i}^{-2}(\theta).
\end{flalign*}
By H\"{o}lder's inequality, we have
\begin{eqnarray*}
    E\left[ \sum_{l=1}^{2} \frac{ \partial h_{l,1}(\theta)}{\partial\theta_j}\frac{\partial h_{l,1}(\theta)}{\partial\theta_k}h_{l,1}^{-2}(\theta)\right] &\leq& C \sum_{l=1}^{2} \norm{\frac{ \partial h_{l,1}(\theta)}{\partial\theta_j} h_{l,i}^{-1}(\theta)}_{L_2} \norm{\frac{ \partial h_{l,1}(\theta)}{\partial\theta_k} h_{l,i}^{-1}(\theta)}_{L_2} \\
    &\leq& C
\end{eqnarray*}
for $j,k \in \{1,\ldots, 18\}$, where the last inequality is due to Lemma \ref{lemma1}(b).
Since for all $\theta \in \Theta$, $u_{ijk}(\theta) =  \sum_{l=1}^{2} \frac{ \partial h_{l,1}(\theta)}{\partial\theta_j}\frac{\partial h_{l,1}(\theta)}{\partial\theta_k}h_{l,1}^{-2}(\theta)$ is stationary and ergodic with finite expectation, we have
\begin{equation*}
    \frac{1}{2n_2}\sum_{i=1}^{n_2}\sum_{l=1}^{2} \frac{ \partial h_{l,1}(\theta)}{\partial\theta_j}\frac{\partial h_{l,1}(\theta)}{\partial\theta_k}h_{l,1}^{-2}(\theta) \xrightarrow{p} B_{2 jk} (\theta) \quad \text{for all} \quad \theta \in \Theta,
\end{equation*}
by the martingale convergence theorem.
By the Lemma \ref{lemma1}(b), we can easily show that for any $q \in \{1, \ldots, 18\}$, $E\left[ \frac{\partial u_{ijk}(\theta)}{\partial \theta_q} \right] < \infty$, which implies that it is stochastic equicontinuous.
Thus, by Theorem 3 in \citet{andrews1992generic}, we have
\begin{equation}\label{eq11}
    \frac{1}{2n_2}\sum_{i=1}^{n_2}\sum_{l=1}^{2} \frac{ \partial h_{l,1}(\theta)}{\partial\theta_j}\frac{\partial h_{l,1}(\theta)}{\partial\theta_k}h_{l,1}^{-2}(\theta) \xrightarrow{p} B_{2 jk} (\theta) \quad \text{uniformly} \quad \theta \in \Theta
\end{equation}
and  $B_2(\theta)$ is continuous in $\Theta$.

We now show that $\sup_{\theta \in \Theta} \left \| \hat{B}_2(\theta) - B_2(\theta)  \right \|_{\max} \xrightarrow{p} 0$.
For any $l \in \{1,2\}$ and $j,k \in \{1, \ldots, 18\}$, we have
\begin{eqnarray}\label{eq-final}
    && \frac{1}{n} \sum_{i=1}^{n} \left[ \frac{ \partial \hat{h}_{l,1}(\theta)}{\partial\theta_j}\frac{\partial \hat{h}_{l,1}(\theta)}{\partial\theta_k}\hat{h}_{l,1}^{-2}(\theta) - \frac{ \partial h_{l,1}(\theta)}{\partial\theta_j}\frac{\partial h_{l,1}(\theta)}{\partial\theta_k}h_{l,1}^{-2}(\theta) \right ] \cr
    && \quad = \frac{1}{n} \sum_{i=1}^{n} \left( \frac{ \partial \hat{h}_{l,1}(\theta)}{\partial\theta_j} - \frac{ \partial h_{l,1}(\theta)}{\partial\theta_j} \right)\frac{\partial \hat{h}_{l,1}(\theta)}{\partial\theta_k}\hat{h}_{l,1}^{-2}(\theta)\cr
    && \qquad + \frac{1}{n} \sum_{i=1}^{n} \frac{ \partial h_{l,1}(\theta)}{\partial\theta_j}\frac{\partial \hat{h}_{l,1}(\theta)}{\partial\theta_k}\hat{h}_{l,1}^{-2}(\theta) h_{l,1}^{-1}(\theta) \left( \hat{h}_{l,1}(\theta) - h_{l,1}(\theta) \right)\cr
    && \qquad + \frac{1}{n} \sum_{i=1}^{n} \left( \frac{ \partial \hat{h}_{l,1}(\theta)}{\partial\theta_k} - \frac{ \partial h_{l,1}(\theta)}{\partial\theta_k} \right)\frac{\partial h_{l,1}(\theta)}{\partial\theta_k}\hat{h}_{l,1}^{-1}(\theta) h_{l,1}^{-1}(\theta)\cr
    && \qquad + \frac{1}{n} \sum_{i=1}^{n} \frac{ \partial h_{l,1}(\theta)}{\partial\theta_j}\frac{\partial h_{l,1}(\theta)}{\partial\theta_k}\hat{h}_{l,1}^{-1}(\theta) h_{l,1}^{-2}(\theta) \left( \hat{h}_{l,1}(\theta) - h_{l,1}(\theta) \right).
\end{eqnarray}
By Lemma \ref{lemma1}(b) and (d) and the fact that $\hat{h}_{l,i}(\theta)$ and $h_{l,i}(\theta)$ stay away from zero, we have
\begin{flalign*}
    \sup_{i \in \mathbb{N}}\norm{\sup_{\theta \in \Theta}\frac{ \partial \hat{h}_{l,1}(\theta)}{\partial\theta_j} - \frac{ \partial h_{l,1}(\theta)}{\partial\theta_j}}_{L_1} \leq C m^{-1/4} \quad \text{and} \quad \sup_{i \in \mathbb{N}}\norm{\sup_{\theta \in \Theta} \frac{\partial \hat{h}_{l,1}(\theta)}{\partial\theta_k}\hat{h}_{l,1}^{-2}(\theta)}_{L_1} \leq C.
\end{flalign*}
Thus, by Lemma \ref{lemma4}, we have
\begin{equation*}
    \left| \frac{1}{n} \sum_{i=1}^{n} \sup_{\theta \in \Theta} \left( \frac{ \partial \hat{h}_{l,1}(\theta)}{\partial\theta_j} - \frac{ \partial h_{l,1}(\theta)}{\partial\theta_j} \right)\frac{\partial \hat{h}_{l,1}(\theta)}{\partial\theta_k}\hat{h}_{l,1}^{-2}(\theta) \right| \xrightarrow{p} 0.
\end{equation*}
Similarly, we can show that the other terms in \eqref{eq-final} converges to zero in probability.
Therefore, by \eqref{eq11}, we have
\begin{equation*}
    \sup_{\theta \in \Theta} \left \| \hat{B}_2(\theta) - B_2(\theta)  \right \|_{\max} \xrightarrow{p} 0.   
\end{equation*}
Since $B_2(\theta)$ is continuous in $\Theta$, $\sup_{\theta \in \Theta} \left \| \hat{B}_2(\theta) - B_2(\theta)  \right \|_{\max} \xrightarrow{p} 0$, and $\htheta \xrightarrow{p} \theta_0$, we have
\begin{eqnarray*}
	\| \hat{B}_2(\htheta)-  B_2(\theta_0) \|_{\max} &\leq&  \sup_{\theta} \| \hat{B}_2(\theta)-  B_2(\theta) \|_{\max} + \| B_2(\htheta)-  B_2(\theta_0) \|_{\max} \cr
	&=& o_p(1).
\end{eqnarray*}
Similarly, we can show that $\hat{A}_1$, $\hat{A}_2$, and $\hat{B}_1$ are consistent estimators for $A_1$, $A_2$, and $B_1$, respectively.
$\blacksquare$

\subsection{Proof of Proposition \ref{proposition2}}
\textbf{Proof of Proposition \ref{proposition2}.}
Proposition \ref{proposition2} is an immediate consequence of Proposition \ref{proposition1}.
$\blacksquare$

\end{spacing}
\end{document}